\documentclass[twocolumn,showpacs,preprintnumbers,amsmath,amssymb,prb]{revtex4}

\usepackage{graphicx}
\usepackage{dcolumn}
\usepackage{bm}

\newcommand{\bol}[1]{\boldsymbol #1}
\def\rnum#1{\expandafter{%
\romannumeral #1}}
\def\Rnum#1{\uppercase\expandafter{%
\romannumeral #1}}

\begin{document}

\title{$\bol N$-leg integer-spin ladders and tubes in commensurate external
fields:\\ Nonlinear sigma model approach}

\author{Masahiro Sato}
\affiliation{Synchrotron Radiation Research Center, Japan Atomic Energy
Research Institute, Mikazuki, Sayo, Hyogo 679-5148, Japan and CREST JST}
\date{\today}

\begin{abstract}
We investigate the low-energy properties, especially the low-energy 
excitation structures, of $N$-leg integer-spin ladders and tubes 
with an antiferromagnetic (AF) intrachain coupling. 
In the odd-leg tubes, the AF rung coupling 
causes the frustration. 
To treat all ladders and tubes systematically, 
we apply S\'en\'echal's method [Phys. Rev. B {\bf 52}, 15319 (1995)], 
based on the nonlinear sigma model, 
together with a saddle-point approximation. This strategy is valid in 
the weak interchain (rung) coupling regime. 
We show that all frustrated tubes possess six-fold degenerate 
spin-$1$ magnon bands, as the lowest excitations, 
while other ladders and tubes have a standard 
triply degenerate bands. We also consider effects of four kinds of 
Zeeman terms: uniform, staggered only along 
the rung, only along the chain, or both directions.
The above prediction of the no-field case implies 
that a sufficiently strong uniform field yields a two-component 
Tomonaga-Luttinger liquid (TLL) due to the condensation of doubly 
degenerate lowest magnons in frustrated tubes. 
In contrast, the field induces 
a standard one-component TLL in all other systems. 
This is supported by symmetry and bosonization arguments based 
on the Ginzburg-Landau theory. The bosonization also suggests 
that the two-component TLL vanishes and a one-component TLL appears, 
when the uniform field becomes larger for the second lowest magnon 
bands to touch the zero-energy line. This transition could be 
observed as a cusp singularity in the magnetization process. 
When the field is staggered only along the rung direction, it
is implied that the lowest doubly-degenerate bands fall down with the 
field increasing in all systems. 
For final two cases where the fields are staggered along the chain, 
it is showed that at least in the weak rung-coupling region, 
the lowest-excitation gap grows with the field increasing,
and no critical phenomena occurs.
Furthermore, for the ladders of the final two cases, 
we predict that the inhomogeneous magnetization along the rung 
occurs, and the frustration between the field and the rung coupling can 
induce the magnetization pointing to the opposite direction to the 
field. All the analyses suggest that the emergence of the doubly 
degenerate transverse magnons and the single longitudinal one is 
universal for the one-dimensional AF spin systems with a 
weak staggered field. 

\end{abstract}


\maketitle

\section{\label{sec1}Introduction}
Low dimensional quantum spin systems have provided much interest
for a long time. In particular, the understanding of one-dimensional
(1D) spin-$\frac{1}{2}$ systems has shown a significant progress.
Recently, quasi-1D systems, such as ladders and tubes, 
have been among the central issues. 
Here, spin tubes means cylinder-type spin systems, i.e., 
spin ladders with the periodic boundary condition along the rung
(interchain) direction. 

In the spin ladders with an antiferromagnetic (AF) 
intrachain coupling, one of the most dramatic properties is
the following ``even-odd'' nature, which is an extension of the Haldane
conjecture~\cite{NLS-Hal,LSM,Aff-Lieb} for the single AF spin chain. 
For odd-leg and half-integer-spin cases,
there exist massless excitations above the ground state (GS), 
the spin correlation functions decay algebraically, and 
the low-energy physics is described by a one-component Tomonaga-Luttinger
liquid (TLL),~\cite{Gogo} which is equal to a conformal field theory
(CFT)~\cite{CFT} with the central charge $c=1$. Meanwhile, for other 
(even-leg or integer-spin) cases, the system is gapful, and the decay 
of spin correlations is an exponential type. This even-odd property 
has been established by both numerical~\cite{Dag,White,Dag2,Fri} and 
analytical~\cite{Rojo,Dell,Sier} works. 
Moreover several experiments~\cite{Hiroi,Azuma,Chab,DT-TTF} also 
support it.

Both theoretical and experimental studies of spin tubes 
are not as active as those of spin ladders. 
As far as we know, there are only two spin-tube-like
materials~\cite{Nojiri,Mill} even now.
However, odd-leg tubes with an AF rung
coupling~\cite{Shul,Taka,Cab,Ichi,Cit,Cit2,oddlegs1/2} have
attracted considerable interest at least theoretically, because
such tubes possess the frustration along the rung. 
It is known that at least for the strong rung-coupling regime, 
odd-leg AF-rung spin-$\frac{1}{2}$ tubes take doubly
degenerate and gapful GSs with the one-site translational symmetry along the
chain breaking.
Namely, the rung frustration induces the break down of the 
even-odd prediction.

As powerful theoretical tools to treat these ladders and tubes, 
there are nonlinear sigma model (NLSM)
approaches.~\cite{Sier,Dell,NLS-Hal,Aff-Lec,Man,Frad,Auer}
A standard NLSM technique, which has an ability to derive 
the above even-odd nature, assumes the development of a sufficient 
short-range order to all spatial directions.
Thus, it is not applicable for frustrated odd-leg tubes.   
However, if we first map a single AF spin chain to a NLSM, and next take
into account the rung coupling perturbatively, we can deal with
frustrated tubes as well as other non-frustrated systems within the NLSM
framework. In this paper, following this idea, we revisit and investigate 
the low-energy physics of spin ladders and tubes systematically, in the
weak rung-coupling regime. 
Note that the perturbative treatment of the rung coupling was already
proposed by S\'en\'echal, who applied it to 2-leg ladders.
Therefore, our method discussed below will be 
regarded as a natural extension of his work. 
As well known, the NLSM method for half-integer-spin chains
bears a topological term (Berry phase). Because 
(as S\'en\'echal mentioned in Ref.~\onlinecite{Sene}) 
it is difficult to treat such a term and the rung coupling concurrently, 
we concentrate on integer-spin cases only in this paper.

Our target is the following Hamiltonian for $N$-leg spin systems:
\begin{eqnarray}
\label{1-1}
\hat{\cal H} &=& J \sum_{l=1}^{N}\sum_{j}\vec S_{l,j}\cdot\vec S_{l,j+1}
+ J_{\perp}\sum_{l=1}^{\tilde{N}}
\sum_{j}\vec S_{l,j}\cdot\vec S_{l+1,j},\,\,\,\,\,\,\,\,\,
\end{eqnarray}
where $\vec S_{l,j}$ is the integer-spin-$S$ operator on site $(l,j)$, 
$J(>0)$ is the intrachain coupling, and $J_\perp$ is the rung one. 
In the rung-coupling term, ladders take $\tilde{N}=N-1$,
while $\tilde{N}=N$ and $\vec S_{N+1,j}=\vec S_{1,j}$ in tubes ($N\geq 3$). 

We further study external-field effects for the model~(\ref{1-1}).
In this paper, we consider following four kinds of the Zeeman terms:
\begin{subequations}
\label{1-2}
\begin{eqnarray}
\hat{\cal H}_{[0,0]}&=&-H\sum_{l,j}S_{l,j}^z,\label{1-2a}\\
\hat{\cal H}_{[0,\pi]}&=&-H\sum_{l,j}(-1)^{l+1}S_{l,j}^z,\label{1-2b}\\
\hat{\cal H}_{[\pi,0]}&=&-H\sum_{l,j}(-1)^{j}S_{l,j}^z,\label{1-2c}\\
\hat{\cal H}_{[\pi,\pi]}&=&-H\sum_{l,j}(-1)^{l+j+1}S_{l,j}^z,\label{1-2d}
\end{eqnarray}
\end{subequations}
where $H(>0)$ is the strength of the external field.
The first term $\hat{\cal H}_{[0,0]}$ is a standard uniform-field Zeeman
term. External fields of other terms have an alternation. We call the
fields in (\ref{1-2a})-(\ref{1-2d}) as a $[0,0]$ (uniform), a $[0,\pi]$
(staggered along the rung), a $[\pi,0]$ (staggered along the chain) and 
a $[\pi,\pi]$ (staggered along both directions) fields, respectively.
Staggered magnetic fields have been investigated 
recently.~\cite{O-A,A-O,Ess,E-T,Nomu,Z-R-M,Mas-Zhe,Lou,Erco,Erco2,Cap,Wang,M-O,Masa}
Actually such fields are present in real 
magnets,~\cite{Z-R-M,Mas-Zhe,Cu-Den,Cu-A1,Cu-A2,PMCu-ex,YbAs-O,YbAs-S,Yb4As3} 
and their origins have been explained.~\cite{A-O,Wang} 
One will see that these four terms are congenial to the NLSM method.

The organization of this paper is as follows.
First, we review the NLSM approach for single AF integer-spin chains in 
Sec.~\ref{sec2}. 
It provides an underlying effective theory for spin ladders and tubes. 
Section~\ref{sec3} presents our main results, in which
we treat spin ladders and tubes in quite detail.
Sections~\ref{sec3a}, \ref{sec3b}, \ref{sec3c}, and~\ref{sec3d} are
devoted to investigate the no-field, uniform($[0,0]$)-field, 
$[0,\pi]$-field, and $[\pi,0]$ or $[\pi,\pi]$-field cases,
respectively. One will see several new even-odd natures in spin ladder
and tube systems. Particularly, in the no-field and $[0,0]$-field cases, 
we find qualitative differences between the low-energy excitation in
even-leg tubes and that in odd-leg (frustrated) ones. 
Because these results in the two cases are supported by symmetry
arguments, we believe that they are not merely approximate results, 
and true.
In Sec.~\ref{sec5}, we summarize all the results and touch some
related topics. We write down the properties of some simple matrices 
in Appendix~\ref{app1}. Moreover, Appendix~\ref{app2} gives a 
review of Green's functional treatment of the staggered field along the chain.
These Appendices are useful for the calculations in Sec.~\ref{sec3}.

\section{\label{sec2}Review of single-chain cases}
We present a review of the NLSM approach for single integer-spin AF chains
and the saddle-point approximation (SPA) in this section, which mainly
follows Refs.~\onlinecite{Erco}, \onlinecite{Erco2}, and \onlinecite{Ven}.
This will be the basis of Sec.~\ref{sec3}. Readers who are familiar with
the approach can skip this section, especially Sec.~\ref{sec2a}.

\subsection{\label{sec2a}No-field case}
This subsection discusses the integer-spin-$S$ isotropic Heisenberg AF
chain without external fields,
\begin{eqnarray}
\label{2a-1}
\hat {\cal H}_{\rm chain}&=& J\sum_{j}\vec S_j\cdot \vec S_{j+1}.
\end{eqnarray}
There exist two celebrated ways to obtain the low-energy effective
theory for the chain (\ref{2a-1}), a NLSM: the operator
formalism~\cite{Aff-Lec, Sier}
and the path-integral one.~\cite{Man,Frad,Auer} 
We use the latter here. In the latter
formalism, the Euclidean action of the chain~(\ref{2a-1}) is given by 
$A_{\rm E}=\int_0^{\beta}d\tau {\cal H}[\vec \Omega(\tau)]$, 
where $\tau$ is the imaginary time, and $\beta=1/(k_{\rm B}T)$ is the inverse
of temperature. The quantity ${\cal H}[\vec\Omega(\tau)]$ is 
the ``classical'' Hamiltonian in which the spin operator $\vec S_j/S$ is
exchanged into a three-component unit vector $\vec \Omega_j(\tau)$.


Following Haldane's idea,~\cite{NLS-Hal} we take the spatially continuum limit 
$\vec\Omega_j(\tau)\to\vec\Omega(x_j,\tau)$ [$x_j=j\times a$: $a$ is the
lattice constant], and decompose $\vec\Omega(x,\tau)$ into the uniform 
fluctuation $\vec l(x,\tau)$ and the AF fluctuation $\vec n(x,\tau)$ as
follows:
\begin{eqnarray}
\label{2a-4}
\vec\Omega(x_j) &\approx& (-1)^j\vec n(x_j)\sqrt{1-a^2\vec l(x_j)^2}
+a\vec l(x_j),
\end{eqnarray}
where two new constraints $\vec n^2=1$ and $\vec n\cdot\vec l=0$ are
imposed to maintain the original constraint $\vec\Omega^2=1$ up to $O(a)$.
The approximation (\ref{2a-4}) is called the Haldane mapping, and it assumes
that there exist the low-energy and slowly-moving modes around both the
wave numbers $k=0$ and $k=\pi/a$. 
It is hence expected that the more the GS approaches a N\'eel
ordered one (i.e., the classical limit $S\to\infty$), 
the more the mapping is reliable. 
Through a few procedures
[(\rnum{1}) substituting Eq.~(\ref{2a-4}) to $A_{\rm E}$, (\rnum{2}) 
a gradient expansion for $A_{\rm E}$, and (\rnum{3})
integrating out the uniform part $\vec l$ or replacing $\vec l$ with its
classical solution $\vec l_{\rm{cl}}=\frac{i}{4SJa}
(\vec n \times\partial_\tau\vec n)$, which is defined by 
$\delta A_{\rm E}/\delta \vec l=0$],
we can finally obtain the effective model 
for the AF fluctuation $\vec n$,
\begin{subequations}
\label{2a-5}
\begin{eqnarray}
Z &\approx& \int{\cal D}\vec n{\cal D}\lambda \,\,
\exp(-{\cal S}_{\rm E}[\vec n,\lambda]),\label{2a-5-1}\\
{\cal S}_{\rm E} &=& \int d{\bf x} \,\,{\cal L}_{\rm E}(\vec n({\bf
 x}),\lambda({\bf x})),
\label{2a-5-2}\\
{\cal L}_{\rm E} &=& -\frac{1}{2g}\vec n\cdot
\Big[\frac{1}{c}\partial_\tau^2+ c\partial_x^2\Big]\vec n
-i\lambda(\vec n^2-1),\label{2a-5-3}
\end{eqnarray}
\end{subequations}
where $Z$, ${\cal S}_{\rm E}$, and ${\cal L}_{\rm E}$ are the partition function,
the Euclidean action, and the Lagrangian density, respectively. 
The symbol ${\bf x}$ means $(x,\tau)$, 
$g=2/S$ is the bare coupling constant, 
$c=2SJa$ is the bare spin-wave velocity, and $\lambda({\bf x})$ 
is the auxiliary field for the constraint $\vec n^2=1$. 
The model~(\ref{2a-5}) is nothing but an O(3) NLSM. 
In this framework, the spin operator is approximated as 
\begin{eqnarray}
\label{2a-spin}
\vec S_j  &\approx & 
(-1)^jS\vec n +  \frac{i}{4J}(\vec n\times\dot{\vec n}). 
\,\,\,\,\,\,\,\,(\dot{\vec n}\equiv\partial_\tau\vec n)
\end{eqnarray}

Using this model, 
let us consider the low-lying band structures in integer-spin chains.
It has been known well that the (1+1)D O(3) NLSM is integrable:~\cite{Zam} 
the system is gapful and the first excitation bands consist of the O(3)
triplet particles. 
However, the integrability method can not be extended to the case of
ladders and tubes~(\ref{1-1}). In this paper, 
we utilize a saddle-point approximation (SPA) instead to reveal 
the low-energy properties qualitatively, in the simplest manner. (As one
will see later, the SPA is available in ladders and tubes.) 
Integrating out the field $\vec n({\bf x})$ in $Z$, we obtain the
effective action including only the field $\lambda({\bf x})$, 
${\cal S}_{\rm E}[\lambda]$, which
is defined by $Z=\int{\cal D}\lambda e^{-{\cal S}_{\rm E}[\lambda]}$.
The SPA in the present work is given by replacing $\lambda({\bf x})$
with $\lambda_{\rm sp}$ (a constant independent of $x$ and $\tau$) 
which is the solution of the saddle-point equation (SPE)
$\partial {\cal S}_{\rm E}[\lambda_{\rm sp}]/\partial \lambda_{\rm sp}=0$.
To obtain the explicit form of $S_{\rm E}[\lambda_{\rm sp}]$, we introduce 
the Fourier transformation for $n^\alpha({\bf x})$ [$\alpha=x,y,z$] as
\begin{subequations}
\label{2a-7}
\begin{eqnarray}
\tilde{n}^\alpha(\omega_n,k)&=&\frac{1}{\sqrt{L\beta}}\int dx\int_0^\beta d\tau
\,\,e^{-ikx+i\omega_n\tau}n^\alpha({\bf x}),\,\,\,\,\,\,\,\,\,\,\,\,\label{2a-7-1}\\
n^\alpha({\bf x})&=&\frac{1}{\sqrt{L\beta}}\sum_{\omega_n,k}
e^{ikx-i\omega_n\tau}\tilde{n}^\alpha(\omega_n,k),\label{2a-7-2}
\end{eqnarray}
\end{subequations}
where $L=Ma$ is the system length ($M$ is the total site number), 
$k=2\pi m/L$ is the wave number, and $\omega_n=2\pi n/\beta$ is the 
Matsubara bosonic frequency ($m,n\in \mathbb{Z}$). 
Hereafter, we will often use a new symbol ${\bf k}=(\omega_n,k)$.
Because $\vec n({\bf x})$ is real, 
$\tilde{n}^\alpha({\bf k})^*=\tilde{n}^\alpha(-{\bf k})$.
Performing the Gaussian integral of the field 
$\tilde{n}^\alpha({\bf k})$ in $Z$, we obtain
\begin{eqnarray}
\label{2a-8}
{\cal S}_{\rm E}[\lambda_{\rm sp}]&=&\frac{3}{2}
\sum_{\bf k}\ln\left[\frac{1}{2gc}(\omega_n^2+c^2k^2)
-i\lambda_{\rm sp}\right] \nonumber\\
&&+i L\beta\lambda_{\rm sp}.
\end{eqnarray}
Therefore, the SPE is evaluated as
\begin{eqnarray}
\label{2a-9}
\frac{3gc}{2\pi}\int_0^\Lambda \frac{dk}{\epsilon(k)}
\coth\Big(\frac{\beta}{2}\epsilon(k)\Big) &=& 1,
\end{eqnarray}
where $\epsilon(k)=c\sqrt{k^2+\xi^{-2}}$, 
$\xi^{-2}\equiv-i2g\lambda_{\rm sp}/c$ and
$\Lambda$ is the ultraviolet cut off. We performed the sum
of $\omega_n$ using the standard prescription with the residue 
theorem,~\cite{Mahan}
and then took the continuous limit $\sum_k\rightarrow \int\frac{dk}{2\pi}$.
In the limit $T\rightarrow 0$, Eq.~(\ref{2a-9}) is reduced to 
\begin{eqnarray}
\label{2a-10}
\frac{3gc}{2\pi}\int_0^\Lambda \frac{dk}{\epsilon(k)}=
\frac{3g}{2\pi}\ln \left[\Lambda\xi+\sqrt{1+(\Lambda\xi)^2}\right]
= 1.
\end{eqnarray}
Equations~(\ref{2a-9}) or (\ref{2a-10}) fix $\lambda_{\rm sp}$ and
$\xi$. They also suggest that $\xi$ is real and $\lambda_{\rm sp}$ is purely 
imaginary. To complete the SPA, we must
determine the unknown parameter $\Lambda$. In other words, the present SPA 
provides a one-parameter-fitting theory.
Of course, other quantities such as $g$ and $c$ can be adopted as the
fitting parameter. However, we will always take $\Lambda$ as it
throughout this paper.

The SPA transforms the constraint to a mass term for the bosons $n^{\alpha}$. 
Therefore, after the SPA, the field $\vec n({\bf x})$ stands for 
the triply degenerate massive bosons with dispersion $\epsilon(k)$. 
This is consistent with the exact solution of the NLSM. 
The bosons should be regarded as the spin-1 magnon excitations of
integer-spin chains (\ref{2a-1}). The gap 
$\epsilon(0)=c\xi^{-1}$ hence corresponds to the Haldane gap $\Delta$. 
From Eq.~(\ref{2a-10}), we obtain
\begin{eqnarray}
\label{2a-11}
\epsilon(0) &=& c\Lambda/\sinh\left(S\pi/3\right).
\end{eqnarray}
It is remarkable that the Haldane gap depends on the spin magnitude $S$
in an exponential fashion. It corresponds to the fact 
that the conventional spin-wave theory ($1/S$ expansion) can not explain the
Haldane gap. The accurate values of $\Delta$ in 
spin-1, 2 and 3 AF chains are found by numerical works.~\cite{Todo-Kato} 
Therefore, we can determine the cut off
$\Lambda$ from the relation $\Delta=\epsilon (0)$, 
completing the SPA. The NLSM plus SPA scheme further 
leads to $\langle S_j^\alpha S_0^\alpha\rangle\approx 
(-1)^j S^2\langle n^\alpha(x_j)n^\alpha(0)\rangle+\dots$, 
$\langle\vec n^2\rangle=1$ and $\langle n^{x,y,z}\rangle=0$ 
($\langle\cdots\rangle$ stands for the expectation value).
The first result means that $\xi$ is interpreted as the spin correlation
length. The second is trivial from the SPE, 
$0=\partial {\cal S}_{\rm{E}}/\partial\lambda_{\rm{sp}}
\propto\langle\vec n^2-1\rangle$. The third is also trivial due to 
invariance of the action~(\ref{2a-5}) under $\vec n\to -\vec n$.

\begin{table*}
\caption{\label{tab1}Haldane gaps, spin-spin correlation
 lengths and spin-wave velocities in spin-1, 2, and 3 AF chains
 (\ref{2a-1}). The QMC data of Haldane gaps and correlation lengths are
 quoted from Ref.~\onlinecite{Todo-Kato}. The velocity $c$ of numerics
 is found in Refs.~\onlinecite{Qin}, \onlinecite{S-A} and \onlinecite{Yama}.}
\begin{ruledtabular}
\begin{tabular}{ccccccc}
& Haldane gap $\Delta$ &    & \multicolumn{2}{c}{correlation length $\xi$}
& \multicolumn{2}{c}{spin-wave velocity $c$} \\
Spin & (QMC data)     &  optimized cut off $\Lambda$   & QMC data         
& SPA data  & numerics  &   bare value \\ 
\hline
1 & $0.410\times J$   & $0.0816\times \pi/a$    & $6.015\times a$  &
 $4.872\times a$      & $\approx 2.5\times Ja$  &    $2\times Ja$  \\
2 & $0.0892\times J$  & $0.02837\times \pi/a$   & $49.49\times a$  &  
 $44.858\times a$     & $\approx4.65\times Ja$  &  $4\times Ja$    \\
3 & $0.0100\times J$  & $0.006139\times \pi/a$  & $637\times a$    &  
 $598.80\times a$     & $?$                     &  $6\times Ja$    \\
\end{tabular}
\end{ruledtabular}
\end{table*}

Table~\ref{tab1} provides the numerical data [quantum Monte Carlo
(QMC) simulation, exact diagonalization and density-matrix
renormalization-group (DMRG) method]
and the above SPA results for the spin-1, 2 and 3 chains.
From this, the SPA is expected to work well even 
in the minimum-integer-spin (spin-1) case.
The larger the spin magnitude $S$ becomes, the more the SPA correlation 
length $\xi$ approaches its correct value. This is consistent with the fact
that the NLSM method is considered as an expansion from the classical
limit ($S\to\infty$), i.e., a N\'eel state. 
On the other hand, the effective Brillouin zone 
(or the cut off $\Lambda$) rapidly becomes smaller with increasing $S$.
This implies that the SPA is efficient only for extremely low
temperatures; $k_B T\ll\Delta$. 
Of course, one can continue more precise analyses of the
NLSM (\ref{2a-5}) beyond the SPA (for example, using its exact solution,
renormalization group, large-$N$ expansion, the improvement of the magnon
dispersion, etc).~\cite{Frad,Zam,H-A,Ess2}

\subsection{\label{sec2b}Uniform-field case}
We consider integer-spin chains (\ref{2a-1}) with the uniform Zeeman
term, $-H\sum_jS_j^z$. Recalling that the boson $\vec n$ in the NLSM
represents the triply degenerate spin-1 magnons in the no-field case, 
one can immediately conclude that the uniform field splits the
degenerate bands into three ones, which have $S^z=1$, $0$, and $-1$,
respectively. In this subsection, we verify that the NLSM and the SPA
can reproduce this Zeeman splitting.

Within the NLSM formalism, the uniform field
$\vec H=(0,0,H)$ couples to the uniform fluctuation $\vec l$.~\cite{Ven}
In this case, the classical solution of $\vec l$ becomes
\begin{eqnarray}
\label{2b-1}
\vec l_{\rm cl}&=& \frac{i}{4SJa}(\vec n\times\dot{\vec n})
+\frac{1}{4SJa}(\vec H+i\tilde{\lambda}\vec n),
\end{eqnarray}
where the second term in the right-hand side originates from the field
$\vec H$, and $\tilde{\lambda}$ is the auxiliary field for the constraint 
$\vec l\cdot\vec n=0$. We fix $\tilde{\lambda}$ to $i\vec H\cdot \vec n$
which is the solution of $\vec l_{\rm cl}\cdot\vec n=0$. Substituting
$\vec l_{\rm cl}$ to the low-energy action, we obtain the action of
$\vec n$,
\begin{eqnarray}
\label{2b-2}
{\cal S}_{\rm E}[\vec n,\lambda] &=& \int d{\bf x} \Big[{\cal L}_{\rm E}
-\frac{1}{8Ja}\left\{\vec H^2-(\vec H\cdot\vec n)^2\right\}\nonumber\\
&&-\frac{i}{4Ja}\vec H\cdot(\vec n\times \dot{\vec n})\Big],
\end{eqnarray}
where ${\cal L}_{\rm E}$ is the same as Eq.~(\ref{2a-5-3}), and
remaining two terms are induced by the uniform field. 
Since the action is also quadratic in the field $\vec n$, 
it is possible to integrate out it in $Z$. 
As a result, the SPE for $\lambda$ is 
\begin{eqnarray}
\label{2b-3}
\frac{gc}{2\pi}\int_0^{\Lambda} \frac{dk}{\epsilon^0(k)}
\sum_{z=0,+,-}\coth\Big(\frac{\beta}{2}\epsilon^z(k)\Big)&=&1,
\end{eqnarray}
where $\epsilon^0(k)=c\sqrt{k^2+\xi^{-2}+H^2/c^2}$ and 
$\epsilon^\pm(k)=\epsilon^0(k)\mp H$. We use the same cut off $\Lambda$
in the no-field case. Observing the Fourier-space representation of the
action~(\ref{2b-2}) or calculating the two-point correlation functions
of $\vec n$, one finds that $\epsilon^{0,+,-}(k)$ are regarded as 
the magnon dispersions.
Therefore, the field $H$ induces the band splitting
$\epsilon(k)\to\epsilon^{0,+,-}(k)$.
At $T= 0$, Eq.~(\ref{2b-3}) is re-expressed as
\begin{eqnarray}
\label{2b-4}
\frac{3gc}{2\pi}\int_0^{\Lambda} \frac{dk}{\epsilon^0(k)}&=& 1.
\end{eqnarray}
The comparison between Eqs.~(\ref{2a-10}) and (\ref{2b-4}) shows
that $\xi(H=0)^{-2}=\xi(H)^{-2}+H^2/c^2$ and $\epsilon(k)=\epsilon^0(k)$
are realized at $T=0$. These two relations tell us that the SPA reproduces 
the Zeeman splitting of the spin-1 magnon modes at $T=0$. 
Modes $\epsilon^0(k)$, $\epsilon^+(k)$ and $\epsilon^-(k)$ 
can be regarded as $S^z=0$, $+1$ and $-1$ magnons, respectively.

Similarly to the preceding subsection, the SPA derives 
$\langle\vec n^2\rangle=1$ and $\langle n^{x,y,z}\rangle=0$. However, it
also does an incorrect result 
$\langle S_j^z\rangle\propto\langle l_{\rm{cl}}^z\rangle\neq 0$.
It would be because the SPA and the Haldane mapping do not take care of the
spin uniform part $\vec l$ sufficiently, in comparison with 
the staggered part $\vec n$.

\subsection{\label{sec2c}Staggered-field case}
Let us next discuss integer-spin chains (\ref{2a-1}) with 
the staggered Zeeman term, $-H\sum_j(-1)^j S_j^z$, 
in which the staggered field $(-1)^j\vec H$ directly 
couples to the AF fluctuation $\vec n$. 
The staggered magnetization $m_{\rm s}^z=(-1)^j\langle S_j^z\rangle$, 
magnetic susceptibilities and transverse excitations 
can be evaluated by applying the SPA in sufficiently low temperatures, 
like Secs.~\ref{sec2a} and \ref{sec2b}. 
However, it has been shown in Ref.~\onlinecite{Erco} that in addition to the
SPA, the Green's function method is necessary for a
quantitative estimation of longitudinal excitations.
``Transverse'' (``longitudinal'') means the components of $\vec n$ 
which are perpendicular (parallel) to the staggered field. 
Here, we provide only main results of the Green's function 
method in Ref.~\onlinecite{Erco}. Its brief explanation is in
Appendix~\ref{app2}, which is applied in Sec.~\ref{sec3c}. 
For more details, see Ref.~\onlinecite{Erco}.

\begin{figure}
\scalebox{0.4}{\includegraphics{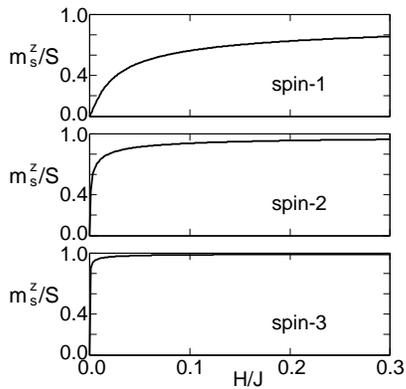}}
\caption{\label{fig1}Normalized staggered magnetizations $m_{\rm s}^z/S$ in 
spin-1, 2, and 3 chains with the staggered field at $T=0$.}
\end{figure}

\begin{figure}
\scalebox{0.3}{\includegraphics{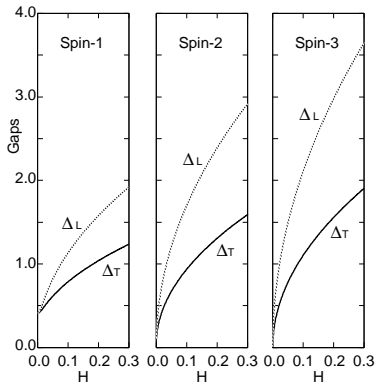}}
\caption{\label{fig2}Transverse and longitudinal gaps in 
spin-1, 2, and 3 chains with the staggered field. We set $J=1$.}
\end{figure}

The normalized magnetization $m_{\rm s}^z(H)/S$ is fixed by
Eq.~(\ref{2-11}). We draw it in Fig.~\ref{fig1}, which explains that 
as $S$ becomes larger, $m_{\rm s}^z(H)/S$ grows extremely rapidly.
The Green's function method (plus SPA) derives two-fold degenerate
transverse magnons with the dispersion $\epsilon_T(k)$ and
a nondegenerate longitudinal magnon with $\epsilon_L(k)$.
Of course, these two bands return to $\epsilon(k)$ as $H=0$.
The transverse gap (lowest excitation energy) $\Delta_T=\epsilon_T(0)$
and the longitudinal one $\Delta_L=\epsilon_L(0)$ are determined from
the SPE (\ref{2-10}) and the relation (\ref{2-20}). 
Figure~\ref{fig2} represents two gaps $\Delta_T$ and $\Delta_L$. 
It shows that the gap grows more rapidly as a function of $H$ for 
larger $S$. The relation $\Delta_T<\Delta_L<2\Delta_T$ always holds within the 
present Green's function method.~\cite{Erco}

In Ref.~\onlinecite{Erco}, it has been confirmed that
in the spin-1 chain, 
$m_{\rm s}^z$, $\Delta_T$ and $\Delta_L$ excellently agree with those
determined from DMRG method 
within the weak-field regime ($H\ll J$).~\cite{note0} 
Recalling again that the NLSM is a semiclassical approach, 
we can guess that the above quantities of the spin-2 and 3 chains, 
determined by the NLSM, are also consistent with their correct values at
least in the regime $H\ll J$.

\section{\label{sec3}Ladders and tubes}
Based on the NLSM method for the single-chain problems in the preceding section, 
we investigate our targets, $N$-leg integer-spin ladders and tubes
(\ref{1-1}) in $T=0$. 

\subsection{\label{sec3a}No-field case}
This subsection considers $N$-leg integer-spin ladders and tubes without
external fields. (Here we remark that the 2-leg spin-1 case has
already been analyzed by the NLSM,~\cite{Sene} 
a non-Abelian bosonization,~\cite{Allen} and the
QMC simulation.~\cite{Todo}) 
Following the method in Ref.~\onlinecite{Sene}, 
we treat the rung-coupling term as a perturbation on 
$N$ decoupled chains, each of which can be mapped to an NLSM. 
Namely, within the NLSM framework, we approximate it as follows:
\begin{eqnarray}
\label{3a-1}
J_\perp\sum_{l=1}^{\tilde{N}}\sum_j\vec S_{l,j}\cdot\vec S_{l+1,j}
&\to & \frac{S^2J_\perp}{a}\sum_{l=1}^{\tilde{N}}\int dx\,\, 
\vec n_l\cdot\vec n_{l+1}\nonumber\\
&&+\cdots,
\end{eqnarray}
where $\vec n_l$ is the AF fluctuation field of the $l$-th chain ($\vec
n_{N+1}\equiv \vec n_1$).
This prescription enables us to deal with any rung-coupling
terms even including frustrations, although it would be valid only in
the weak rung-coupling regime; $J\gg |J_\perp|$. 
However, as a price, 
we have to take into account $N$ constraints: $\vec n_l^2=1$ ($l=1,\dots,N$). 
(As well known, there is only one constraint in the standard NLSM method.) 
Here, as a crude approximation, 
we replace these constraints with an averaged one, 
\begin{eqnarray}
\label{3a-2}
\sum_l\vec n_l^2 &=& N.
\end{eqnarray}
We will discuss the validity of Eqs.~(\ref{3a-1}) and (\ref{3a-2}) 
in the final part of this subsection. (As one will see there, we can
predict that these two approximations are allowed in the any-leg
weak-rung-coupling systems, at least within the qualitative level.)
Under the approximations~(\ref{3a-1}) and (\ref{3a-2}), 
the total action of ladders or tubes is described as
\begin{eqnarray}
\label{3a-3}
S_{\rm E}[\{\vec n_l\},\lambda]&=&\int d{\bf x}\,\Big[{}^T {\cal N}_\alpha 
{\bol A} {\cal N}_\alpha +i N\lambda\Big], 
\end{eqnarray}
where the subscript ${}^T$ means transpose, 
${\cal N}_\alpha={}^T(n_1^\alpha,\dots,n_N^\alpha)$,
$\lambda$ is the auxiliary field for the constraint (\ref{3a-2}), and
the $N\times N$ matrix $\bol A$ is defined as
\begin{subequations}
\label{3a-4}
\begin{eqnarray}
{\bol A}&=& 
\left(
\begin{array}{cccccccc}
a_1        & a_2     &            &         &  a_0     \\
a_2        & a_1     &   \ddots   &         &          \\
           & \ddots  &   \ddots   &  \ddots &         \\
           &         &   \ddots   &  \ddots &  a_2     \\
a_0        &         &            & a_2     &  a_1     \\
\end{array}
\right),\label{3a-4-1}\\
a_1&=&-\frac{1}{2gc}\left(\partial_{\tau}^2+c^2\partial_{x}^2\right)-i\lambda,
\label{3a-4-2}\\
a_2 &=& S^2 J_\perp/(2a), \label{3a-4-3}
\end{eqnarray}
\end{subequations}
where $a_0=0$ ($a_2$) for ladders (tubes). 
It is noteworthy that the action of an AF-rung
ladder (an even-leg AF-rung tube) can be transformed to that of the
ferromagnetic (FM)-rung ladder (the FM-rung tube) through the unitary
transformation, $\vec n_{l={\rm even}}\to -\vec n_{l={\rm even}}$.
Therefore, both AF- and FM-rung ladders (both even-leg AF- and FM-rung
tubes) with the same leg number have the same low-energy excitation
structure within the present scheme. 
Indeed, this property has been partially observed in a QMC study
of the 2-leg spin-1 ladder
with $|J_\perp|\ll J$.~\cite{Todo} On the other hand, we also emphasize
that there are no unitary transformations connecting an odd-leg AF-rung
(frustrated) tube and the FM-rung one.

Because the action (\ref{3a-3}) is quadratic in the fields $\vec n_l$
like the chain cases, 
we can integrate out $\vec n_l$ and derive the SPE for $\lambda$. 
After diagonalizing $\bol A$ (see Appendix~\ref{app1}) 
and performing the Fourier transformations 
for $\vec n_l({\bf x})$, the action becomes 
\begin{eqnarray}
\label{3a-4}
S_{\rm E}&=& \sum_{\bf k}\sum_r (\omega_n^2+\epsilon_r(k)^2)
\tilde{m}_r^\alpha({\bf k})^* \tilde{m}_r^\alpha({\bf k})\nonumber\\
&&+ iL\beta N \lambda_{\rm sp},\,\,\,\,\,\,\,\,
\end{eqnarray}
where 
\begin{subequations}
\label{3a-4-2}
\begin{eqnarray}
\epsilon_r(k) &=& c\sqrt{k^2+\xi_r^{-2}},\label{3a-4-2-1}\\
\xi_r^{-2} &\equiv & \xi^{-2}+2J_\perp\cos k_r/(Ja^2),\label{3a-4-2-2}
\end{eqnarray}
\end{subequations}
and $k_r=\frac{\pi r}{N+1}$ ($\frac{2\pi r}{N}$) and 
$r=1,\dots,N$ ($-[\frac{N}{2}]< r \leq [\frac{N}{2}]$) for ladders
(tubes: $N\geq 3$).
The symbol $[v]$ means the maximum integer $u$ satisfying $u\leq v$.
The new field $\tilde{m}_r^\alpha({\bf k})$ is defined by 
$\tilde{m}_r^\alpha({\bf k})=U_{rl}\tilde{n}_l^\alpha({\bf k})$, where
$U_{rl}$ is the unitary matrix diagonalizing $\bol A$. 
In the derivation of Eq.~(\ref{3a-4}), we 
performed the replacement $\lambda\to\lambda_{\rm sp}$ 
(a constant), and assumed that $\lambda_{\rm sp}$ is purely imaginary. 
In tubes, $k_r$ means the wave number for the rung direction. 
Following the SPA prescription similar to that in
Sec.~\ref{sec2}, we easily estimate the SPE, $\partial S_{\rm
E}[\lambda_{\rm sp}]/\partial \lambda_{\rm sp}=0$, where 
$S_{\rm E}[\lambda]=-\ln(\int \prod_l{\cal D}\vec n_l 
e^{-S_{\rm E}[\{\vec n_l\},\lambda]})$, as follows:
\begin{eqnarray}
\label{3a-5}
\frac{3g}{2\pi}\sum_r\int_0^\Lambda \frac{dk}{\epsilon_r(k)}
\coth\Big(\frac{\beta}{2}\epsilon_r(k)\Big)&=& N,
\end{eqnarray}
where we use the same cut off $\Lambda$ as that of chains. 
For $J_\perp=0$, Eq.~(\ref{3a-5}) reduces 
to the SPE for chains (\ref{2a-9}).

\begin{figure}
\scalebox{0.35}{\includegraphics{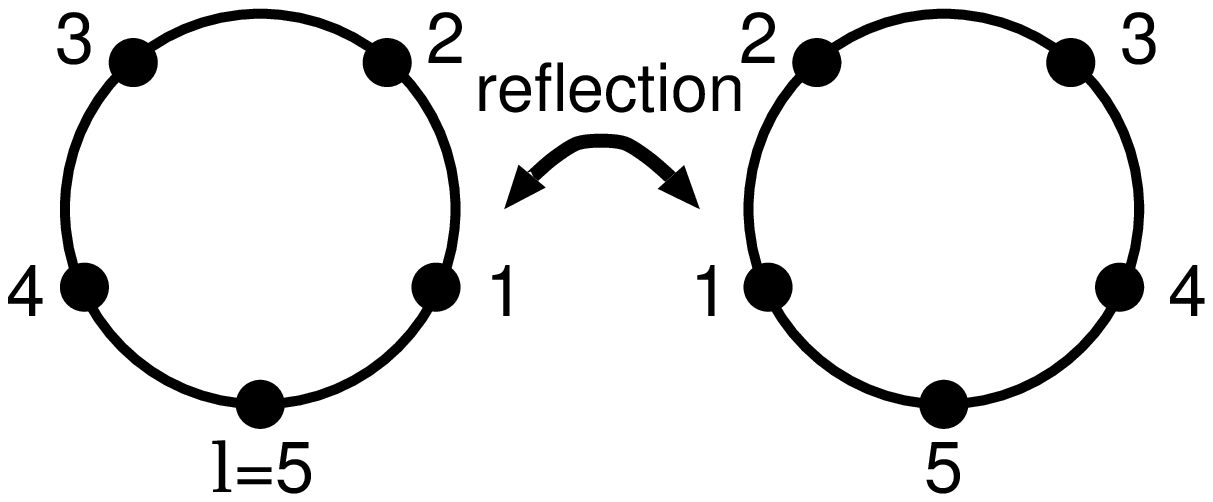}}
\caption{\label{fig3}Cross section of the 5-leg tube and the reflecting
 operation for the rung direction.}
\end{figure}

\begin{figure}
\scalebox{0.3}{\includegraphics{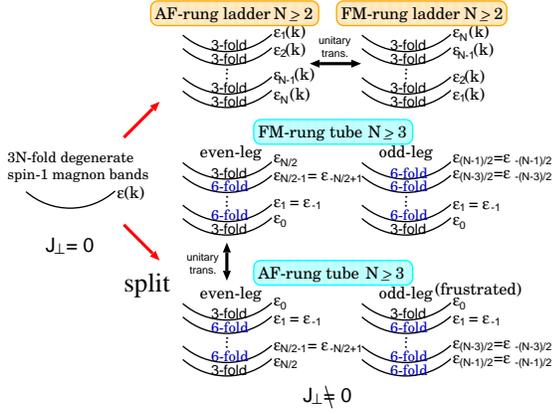}}
\caption{\label{fig4}Band splitting induced by the rung 
coupling $J_\perp$. ``Three- or sixfold'' means the degree of the band degeneracy.}
\end{figure}

The representation (\ref{3a-4}) tells us the following several
low-energy properties of ladders and tubes. (\rnum{1}) 
One can interpret $\epsilon_r(k)$ as a spin-1 magnon dispersion. 
The band splitting $\epsilon(k)\to\epsilon_r(k)$ is due to the
hybridization effect of the rung coupling. 
Each band $\epsilon_r(k)$ is triply
degenerate correspondingly to $S^z=1$, $0$ and $-1$.
(\rnum{2}) The $3N$-band crossing occurs at $J_\perp=0$.
Obtaining this level-crossing picture is an advantage of the NLSM
approach.~\cite{Sene} Other methods, such as the QMC 
simulation~\cite{Todo} and the non-Abelian
bosonization,~\cite{Tsv,Kita-Nomu,Allen} cannot derive it.   
(\rnum{3}) In addition to the triple degeneracy of $S^z$, 
tubes (not ladders) have an extra degeneracy
$\epsilon_r(k)=\epsilon_{-r}(k)$.
Namely, in tubes, only two bands $\epsilon_0(k)$ and 
$\epsilon_{\frac{N}{2}}(k)$ are triply degenerate, while all other bands 
have a sixfold degeneracy. Here, note that
the original tube (\ref{1-1}) is invariant under reflection including
(or $\pi$ rotation with respect to) the central axis of the cross
section of tubes (Fig.~\ref{fig3}).
Because this operation causes $k_r\to -k_r$,   
we can conclude that the degeneracy $\epsilon_r(k)=\epsilon_{-r}(k)$ comes
from the reflection symmetry and it must be a correct result independent
of our approximation scheme. 
(\rnum{4}) Noticing the contents of (\rnum{1})-(\rnum{3}) and the form of
$\epsilon_r(k)$, we can show the band splitting as in Fig.~\ref{fig4}.
For any $r$, $\epsilon_r(k)$ has the minimum at $k=0$. Thus we define
the gap of each band, $\Delta_r\equiv \epsilon_r(0)$. The true gap 
$\Delta_{\rm min}$ of the system would be the smallest among $\Delta_r$.
For the nonfrustrated systems, 
$\Delta_{\rm min}$ is always carried by a triply degenerate band.
Those of AF-rung ladders, FM-rung ladders, FM-rung
tubes, and even-leg AF-rung tubes are given by $\Delta_N$,
$\Delta_1$, $\Delta_0$, and $\Delta_{\frac{N}{2}}$, respectively.
On the other hand, the gap of frustrated tubes is carried by 
a six-fold degenerate band with 
$\epsilon_{\frac{N-1}{2}}(k)=\epsilon_{-\frac{N-1}{2}}(k)$.
This is a {\em new even-odd property in the AF-rung
spin tubes}. 
This interesting phenomena can be intuitively understood as
follows. The GS in all AF-rung tubes would tend to take a
short-range N\'eel order along the rung. 
Therefore, we guess that the lowest excitations
on such a GS are around the wave number $k_r=\pi$. 
Actually, those in nonfrustrated tubes always have $k_r=\pi$. 
However, the wave number $k_r=\pi$ can not be admitted in frustrated
tubes. Instead, the lowest bands in $N$-leg frustrated tubes hence have   
$k_r=\pi\pm \frac{2\pi}{N}$, which is closest to $\pi$ in all the wave
numbers. These bands are just sixfold degenerate. On the other hand, 
the lowest band in the FM-rung case always has $k_r=0$ because of the
similar reason; FM-rung tubes tend to develop a FM short-range order
along the rung. All the band with $k_r=0$ are not degenerate, except for
the degeneracy of the spin-1 triplet.

\begin{figure}
\scalebox{0.3}{\includegraphics{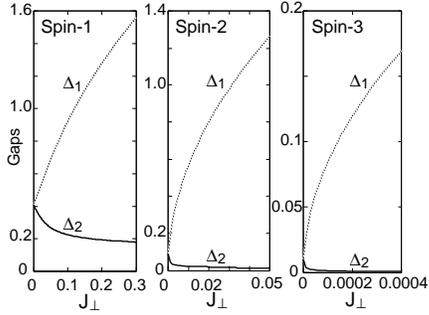}}
\caption{\label{fig5}Gaps $\Delta_1$ and $\Delta_2=\Delta_{\rm min}$ 
in 2-leg AF-rung spin-1, -2, or -3 ladders with $J=1$.}
\end{figure}

\begin{figure}
\scalebox{0.3}{\includegraphics{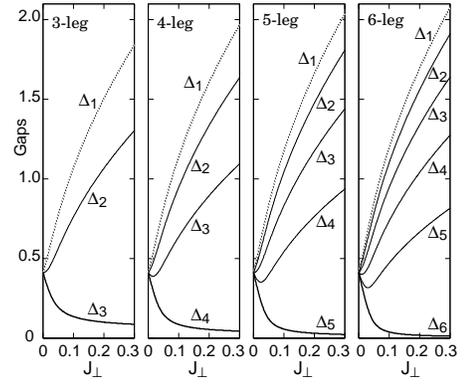}}
\caption{\label{fig6}All band gaps $\Delta_r$ in $N$-leg AF-rung 
spin-1 ladders with $J=1$.}
\end{figure}

\begin{figure}
\scalebox{0.3}{\includegraphics{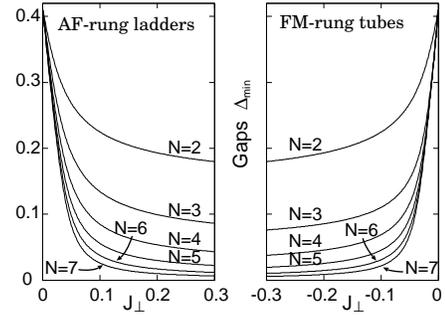}}
\caption{\label{fig7}Lowest magnon gaps $\Delta_{\rm min}$ in $N$-leg 
AF-rung spin-1 ladders and FM-rung tubes (non-frustrated systems) 
with $J=1$. The gap-reduction speed of the $N$-leg tube is larger a 
little than that of the ladder. The 2-leg ``tube'' means the 2-leg ladder.}
\end{figure}

\begin{figure}
\scalebox{0.3}{\includegraphics{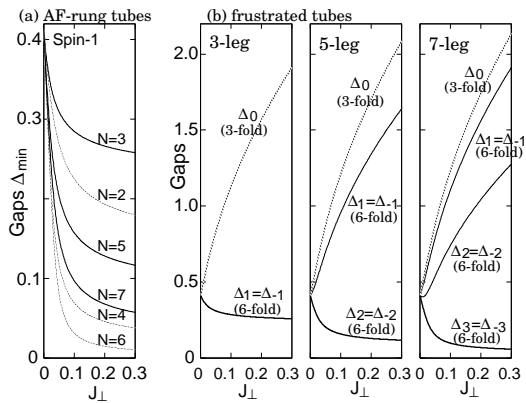}}
\caption{\label{fig8} Lowest magnon gaps $\Delta_{\rm min}$ and band 
gaps $\Delta_r$ in $N$-leg AF-rung spin-1 tubes with $J=1$.}
\end{figure}

\begin{figure}
\scalebox{0.3}{\includegraphics{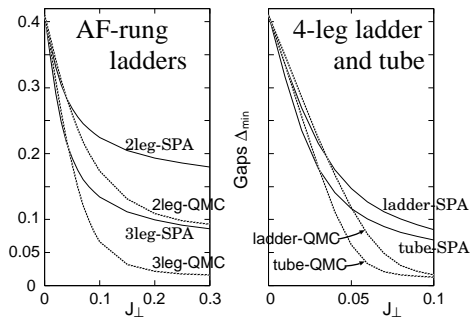}}
\caption{\label{fig9}Gaps $\Delta_{\rm min}$ of spin-1 ladders and 
tubes with $J=1$. Marks ``SPA'' and ``QMC'' mean ``derived by the SPA'' and 
``derived by the QMC simulation'' respectively. 
The QMC data, which all achieve their
thermodynamic-limit values, are calculated by Munehisa 
Matsumoto.~\cite{MMcom} The method of determining the gap and the 
correlation length in the QMC simulation is, for example, explained in 
Refs.~\onlinecite{Todo-Kato} adn \onlinecite{Todo}.}
\end{figure}

\begin{figure}
\scalebox{0.3}{\includegraphics{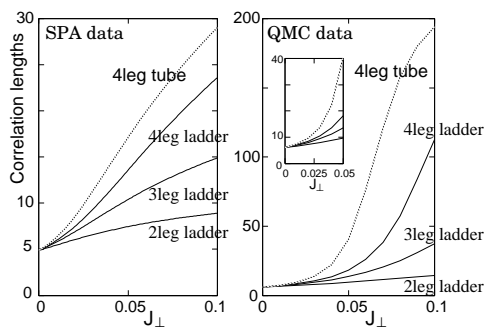}}
\caption{\label{fig10}Spin correlation lengths in spin-1 ladders and 
tubes determined by the SPA or the QMC simulation. The QMC data is 
provided by Munehisa Matsumoto.~\cite{MMcom}}
\end{figure}

We investigate the low-energy excitations more quantitatively by
solving the SPE (\ref{3a-5}). As we will see in Figs.~\ref{fig5}-\ref{fig10}, 
all ladders and tubes have a positive-$\xi_r$ solution, 
which means that all magnon bands $\epsilon_r(k)$ are well-defined. 
First, we discuss nonfrustrated systems. 
Figure~\ref{fig5} shows the rung-coupling and spin-magnitude
dependence of gaps in 2-leg ladders. Figure~\ref{fig6}
represents the rung-coupling dependence of gaps in $N$-leg
spin-1 ladders. Moreover, Fig.~\ref{fig7} provides the lowest gap
$\Delta_{\rm min}$ in $N$-leg spin-1 ladders and tubes.
(Since the band structure of a nonfrustrated system with
$J_\perp$ is equivalent to that of the system with $-J_\perp$ in the
present scheme, the same result applies to the FM-rung ladders, and to
even-leg AF-rung tubes.)
The former two figures indicate that the rung coupling induces rapid
rises of bands except for the lowest bands. It supports the known result 
that the standard NLSM approach for nonfrustrated systems,~\cite{Sier,Dell} 
which extracts only the lowest bands, 
captures the low-energy physics in $|J_\perp| \sim J$. 
Figure~\ref{fig6} shows that the increase of $N$
gradually enlarges splitting-magnon-band width.  
One finds from Figs.~\ref{fig5}
and \ref{fig7} that the more $N$ or $S$ increase, the larger the
decreasing speed of gaps $\Delta_{\rm min}$ becomes 
around the decoupling point $J_\perp =0$. Particularly, in
Fig.~\ref{fig5}, it is remarkable 
that once one attaches an awfully weak rung coupling for a ladder with
$S>1$, the gap $\Delta_{\rm min}$ sharply  approaches zero. 
For example, when we set $(J,J_\perp,T)=(1,0.05,0)$, the SPA predicts
$(\Delta_{\rm min},\frac{\Delta_{\rm min}}{\Delta})=(0.268,0.653)$ for $S=1$, 
$(0.0143,0.160)$ for $S=2$, and $(0.000459,0.0458)$ for $S=3$
($\Delta$ is the Haldane gap of single chains in Table~\ref{tab1}). 
These gap reductions are naturally expected from the consideration that 
the growths of $S$ and $N$ help the GS (a massive spin-liquid state) 
be close to a N\'eel state, which has a massless Nambu-Goldstone mode.

We next focus on the frustrated spin tubes.
Figure~\ref{fig8} is the gap structures of AF-rung spin-1 tubes. 
The panel (a) shows that gap reductions of frustrated (odd-leg) tubes are
considerably slower than those of even-leg tubes. 
It must reflect that the rung frustration obstructs 
the rise of the AF short-range order unlike in the nonfrustrated
systems. While, similarly to nonfrustrated
systems, the growth of $N$ prompts the gap reduction in the frustrated
tubes. It would be a relaxation effect of the frustration.    
We see from the panels (b) that all magnon bands, except for lowest one, 
quickly rise together with increasing $J_\perp$ even in frustrated tubes.
This may suggest the possibility to construct an effective theory for 
frustrated tubes, which includes only lowest sixfold degenerate bands.

Finally, we discuss the validity of our strategy in this subsection. 
In order to investigate the integer-spin ladders and tubes, we took
approximations~(\ref{3a-1}) and (\ref{3a-2}).
First, we consider the validity of Eq.~(\ref{3a-2}). If we 
adopt the original constraint $\vec n_l^2=1$ ($l=1,\dots,N$) 
instead of the averaged one~(\ref{3a-2}), the action corresponding to
the model~(\ref{1-1}) is given by 
\begin{eqnarray}
\label{originalNLSM}
S_{\text{E}}'[\{\vec n_l\},\{\lambda_l\}] = \int\, d{\bf x}\Big\{
\sum_{l=1}^N\Big[\vec n_l
\Big(-\frac{1}{2gc}\left(\partial_{\tau}^2+c^2\partial_{x}^2\right)
\nonumber\\
-i\lambda_l\Big)\vec n_l+i\lambda_l \Big]
+\frac{S^2J_\perp}{a}\sum_{l=1}^{\tilde N}\vec n_l\cdot\vec n_{l+1}\Big\},
\,\,\,\,\,\,\,\,
\end{eqnarray}
where $\lambda_l$ is the auxiliary field for the constraint $\vec n_l^2=1$.
At least for the 2-leg ladder system, the SPEs for the original
constraint $\vec n_1^2=\vec n_2^2=1$ turn out to be identical with 
that for the averaged one.~\cite{Sene} 
The deviation between original and averaged
constraints could appear in 3-leg or higher-leg systems within the SPA. 
Actually, one can easily check that the SPEs for the original and
averaged constraints provide different solutions in the 3-leg ladder. 
Therefore, averaging~(\ref{3a-2}) is expected to be 
invalid for large-$N$ systems. 
Here, notice that 
the action~(\ref{originalNLSM}) is invariant under the
transformation $\vec n_{l}\to\vec n_{l+{\cal Q}}$ and 
$\lambda_l\to\lambda_{l+{\cal Q}}$ [${\cal Q}\in \mathbb{Z}$] 
($\vec n_l\to\vec n_{N+1-l}$ and 
$\lambda_l\to\lambda_{N+1-l}$) in the tube (ladder) systems.
These transformations of course correspond to the translational
operation along the
rung in tubes ($\vec S_{l,j}\to \vec S_{l+{\cal Q},j}$) and reflection 
around the central axis along the chain in ladders 
($\vec S_{l,j}\to \vec S_{N+1-l,j}$), respectively. 
Let us discuss the form of the SPEs, using these symmetries.  
Each original SPE is written as 
$\partial S'_{\rm{E}}[\{\lambda_m\}]/\partial \lambda_l
\propto \langle \vec n_l^2-1\rangle=0$,
where $S'[\{\lambda_m\}]=-\ln(\int \prod_m{\cal D}\vec n_m 
e^{-S_{\text{E}}'[\{\vec n_m\},\{\lambda_m\}]})$. The expectation value 
$\langle \vec n_l^2\rangle$ is a function of $\{\lambda_m\}$; 
$\langle \vec n_l^2\rangle=f(\lambda_1,\cdots,\lambda_N)$. In the tube
systems, these facts and the above symmetry of the
action~(\ref{originalNLSM}) lead to
the identity $1=\langle \vec n_l^2\rangle=f(\lambda_1,\cdots,\lambda_N)
=\langle \vec n_{l+1}^2\rangle
=f(\lambda_2,\cdots,\lambda_N,\lambda_1)=\cdots$. Therefore, we find
that $\partial S'_{\rm{E}}[\{\lambda_m\}]/\partial 
\lambda_l|_{\lambda_1=\cdots=\lambda_N}$ is independent of all the chain
indices $l$ in tubes. On the other hand, in tubes, the SPE for the averaged
constraint concerns those for the original ones as follows: 
\begin{eqnarray}
\label{OriAve}
0=\frac{\partial S_{\rm{E}}[\lambda]}{\partial \lambda}
&=& \sum_{l=1}^N 
\frac{\partial S_{\rm{E}}'[\{\lambda_m\}]}{\partial \lambda_l}
\Big|_{\lambda_1=\cdots=\lambda_N=\lambda}\nonumber\\
&= & N\frac{\partial S_{\rm{E}}'[\{\lambda_m\}]}{\partial \lambda_1}
\Big|_{\lambda_1=\cdots=\lambda_N=\lambda},
\end{eqnarray}
where the final equal sign is thanks to the above property of the function
$f$. Because we have already known that 
$\partial S_{\rm{E}}[\lambda]/\partial \lambda=0$ has a physically
suitable solution $\lambda_{\rm{sp}}$, Eq.~(\ref{OriAve}) indicates that
the original SPEs in tubes can take the same solution, 
$\lambda_1=\cdots=\lambda_N=\lambda_{\rm{sp}}$. 
Thus, the averaging of the constraints, Eq.~(\ref{3a-2}), should be
valid on the symmetric solutions of the original SPEs,
$\lambda_1=\cdots=\lambda_N$, although other
possible solutions would not be covered. 
Meanwhile, in ladder systems, the similar argument can not lead to the
validity of Eq.~(\ref{3a-2}). If the effective
theory~(\ref{originalNLSM}) for ladders possesses the reflection
symmetry ($\vec S_{l,j}\to \vec S_{N+1-l,j}$), 
the original SPEs should have a solution with
$\lambda_l=\lambda_{N+1-l}$. This solution does not contradict the
symmetric one. Therefore, we expect that the
averaging~(\ref{3a-2}) is admitted even in ladders.

Subsequently, we discuss the perturbative treatment of the rung 
coupling~(\ref{3a-1}). As mentioned before, the approximation~(\ref{3a-1}) 
would be justified only in the weak rung-coupling regime $|J_\perp|\ll J$. 
As we see from
Figs.~\ref{fig5}-\ref{fig8}, the present scheme always predicts that 
the gap $\Delta_{\rm min}$ monotonically decreases with $|J_\perp|$
increasing in all systems. 
However, it has been known, 
from the QMC simulation,~\cite{Todo} that in the 2-leg
spin-1 AF-rung ladder, the gap reaches its minimum value
at a finite $J_\perp$, and then grows and approaches the gap for 
the 2-spin problem in the single rung with increasing $J_\perp$.
(The simulation also shows that the gap monotonically
decreases for FM-rung side.) 
In addition, the standard NLSM method for nonfrustrated 
systems,~\cite{Sier,Dell,Sene} which is reliable for the case 
$|J_\perp|\sim J$, shows that the gap is a monotonically increasing
function of $J_\perp$ in all the AF-rung cases. 
These gap growths cannot be explained in our
weak rung-coupling framework. (Inversely, it also means that the
standard NLSM approach breaks down in the weak rung-coupling regime.) 
Figure~\ref{fig9} displays gaps 
$\Delta_{\rm min}$ of spin-1 ladders and tubes with a few leg numbers, 
which are determined from the SPA and the QMC method (note that the 
QMC method cannot be applicable for the frustrated tubes doe to the
negative-sign problem). As expected, 
the SPA gap is semiquantitatively identical to the QMC one within
the sufficiently weak rung-coupling regime 
$|J_\perp| \lesssim 0.05\times J$, outside which the deviation between 
SPA and QMC gaps becomes clear. The good agreement between the SPA and
QMC methods is observed in 2, 3, and 4-leg systems. It encourages
and allows us to apply the present SPA scheme to large-$N$ systems. 
The SPA gap is always larger than the QMC one in the region 
$|J_\perp| \gtrsim 0.05\times J$. 
It might be because the SPA does not take into account 
the quantum fluctuation effects enough.
In Fig.~\ref{fig10}, we draw the spin-spin correlation lengths of small-$N$
systems, which are determined by the SPA and the QMC methods. There, we
define the SPA correlation length as 
$\xi_{\rm corr}\equiv c\Delta_{\rm min}^{-1}$.  
Since our scheme optimizes the gaps and not the correlation lengths, 
the deviation of the SPA and QMC data already emerges 
at the starting point $J_\perp=0$.    
However, like the gap, the behavior of the SPA correlation lengths 
is similar to that of the QMC ones within 
$|J_\perp| \lesssim 0.05\times J$.

We believe that our results are qualitatively correct 
in all ladders and tubes with a weak rung coupling, and in
particular, the predicted band structures in Fig.~\ref{fig4} are true. 
The band degeneracy is strongly protected by the symmetry argument. 
Like the final statements in Sec.~\ref{sec2a}, 
one can imagine several modifications of approximations~(\ref{3a-1})
and (\ref{3a-2}).

\subsection{\label{sec3b}$[0,0]$-field case}
In this subsection, we investigate the ladders and tubes with the
uniform Zeeman term (\ref{1-2a}). 
To this end, one first had better notice the following general aspects 
in 1D spin systems. (\rnum{1}) The uniform field always splits
triply degenerate spin-1 magnon states into $S^z=1$, $0$ and $-1$
states. (\rnum{2}) When a magnon band crosses the zero energy and 
the magnon condensation occurs in a 1D U(1)-symmetric spin system, 
the GS is usually regarded as a $c=1$ one-component
TLL.~\cite{Masa,S-A,Tsv,Aff-magcond,Aff-bose,S-T,Kon,Fath} 
The combination of these statements and the band structures in
Fig.~\ref{fig4} leads to the conjecture that 
as a sufficiently strong uniform field is applied,
nonfrustrated systems enter in a standard one-component TLL phase ($c=1$),
whereas the GSs of frustrated tubes become a two-component TLL ($c=1+1$). 
This is {\em another new even-odd property in AF-rung tubes}.

However, as multimagnon modes are
condensed simultaneously, interactions among the resulting massless
excitations (TLLs) can be present in general.
They could induce the hybridizations and gaps 
in a part of the massless modes. [For example, 
on the $c=1+1$ critical GS of two independent spin-$\frac{1}{2}$ (or
spin-1) AF chains with a uniform field,~\cite{Masa,Cabra,Selton,Fu-Zh} 
a rung coupling brings the hybridization of massless modes. As a result, 
a massless mode becomes gapful and only the other one remains being massless.]
While, all magnon excitations have to possess a wave number $k_r$ 
in the (frustrated) tubes, due to the translational symmetry along the
rung. Moreover, the tubes also have the reflection symmetry
(Fig.~\ref{fig3}). Therefore, we expect that when the lowest
doubly degenerate magnons are condensed in frustrated tubes, 
these two symmetries 
strongly restrict possible interactions and
hybridizations between the low-energy excitations.  
Consequently, it would be natural that two kinds of 
massless excitations are present after the lowest two magnon bands
are condensed. 
It has been suggested in Ref.~\onlinecite{Cit} that a strong
uniform field brings a $c=1+1$ phase in the 3-leg AF-rung 
``spin-$\frac{1}{2}$'' tube. (Note that this massless phase of
the spin-$\frac{1}{2}$ tube is located just under the saturated state. 
While, the massless phase in the integer-spin tube, expected here, 
appears just when an infinitesimal magnetization occurs.)
Therefore, the emergence of a GS with two bosonic massless modes 
might be universal for odd-leg AF-rung (frustrated) spin tubes 
with any spin magnitudes.

We investigate the above expectation of the $c=1+1$ phase in more
detail, below. 
First, we demonstrate that our NLSM plus SPA method for 
ladders and tubes can reproduce the Zeeman splitting in
Sec.~\ref{sec3b-1}.
Subsequently, in Secs.~\ref{sec3b-2} and \ref{sec3b-3}, 
we discuss the magnon condensed state in frustrated tubes using
a heuristic bosonization method. The discussion there takes into account
the translational symmetry along the rung and the reflection one
more carefully. 
The results further support the presence of the $c=1+1$ state.


\subsubsection{\label{sec3b-1}Zeeman splitting in ladders and tubes} 
From Eqs.~(\ref{2b-2}) and (\ref{3a-3}), we can represent the Euclidean
action of the [0,0]-field case, in the Fourier space, as follows:
\begin{eqnarray}
\label{3b-1}
S_{\rm E}^{[0,0]}[\{\vec n_l\},\lambda_{\rm sp}]&=& \sum_{\bf k}\Big[
\tilde{\cal N}_\perp^\dag
\tilde{\bol A}_\perp
\tilde{\cal N}_\perp 
\nonumber\\
&&+ \tilde{\cal N}_z^\dag \tilde{\bol A}_z\tilde{\cal N}_z\Big] 
 +i N\beta L\lambda_{\rm sp},
\end{eqnarray}
where the mark $^\dag$ means Hermitian conjugate, 
$\tilde{\cal N}_\alpha({\bf k})
=(\tilde{n}_1^\alpha({\bf k}),\dots,\tilde{n}_N^\alpha({\bf k}))$, 
$\tilde{\cal N}_\perp({\bf k})
=({}^T\tilde{\cal N}_x({\bf k}),{}^T\tilde{\cal N}_y({\bf k}))$. 
The $2N\times 2N$ matrix $\tilde{\bol A}_\perp$ and $N\times N$ one 
$\tilde{\bol A}_z$ are defined by
\begin{subequations}
\label{3b-2}
\begin{eqnarray}
\tilde{\bol A}_\perp ({\bf k})&=&
\left(
\begin{array}{ccc}
{\bol A}_{\bf k} &  {\bol E}_{\omega_n}  \\
 -{\bol E}_{\omega_n}&{\bol A}_{\bf k}
\end{array}
\right),\label{3b-2-1}\\
{\bol A}_{\bf k}&=& 
\left(
\begin{array}{cccccc}
\tilde{a}_1   & \tilde{a}_2   &             &               & \tilde{a}_0   \\
\tilde{a}_2   & \tilde{a}_1   & \ddots      &               &               \\
              & \ddots        & \ddots      & \ddots        &               \\
              &               & \ddots      & \ddots        & \tilde{a}_2   \\
\tilde{a}_0   &               &             & \tilde{a}_2   & \tilde{a}_1 
\end{array}
\right),\label{3b-2-2}\\
{\bol E}_{\omega_n}&=& -\frac{H\omega_n}{gc} \hat{\bol 1},\label{3b-2-3}\\
\tilde{\bol A}_z({\bf k}) &=& {\bol A}_{\bf k}
+\frac{H^2}{2gc} \hat{\bol 1},\label{3b-2-4}
\end{eqnarray}
\end{subequations}
where $\tilde{a}_1=(\omega_n^2+\epsilon(k)^2)/(2gc)$, $\tilde{a}_2=a_2$, 
$\tilde{a}_0=0$ [$\tilde{a}_2$] for ladders [tubes],  
and $\hat{\bol 1}$ is an $N\times N$ unit matrix.
We stress that $\tilde{\bol A}_\perp$ is not Hermite but normal 
($\tilde{\bol A}_\perp^\dag\tilde{\bol A}_\perp=
\tilde{\bol A}_\perp\tilde{\bol A}_\perp^\dag$), so it
can be diagonalized by a unitary matrix (hence, we do not need to consider 
the Jacobian generated from the diagonalization in the 
path-integral formalism). 
From Eqs.~(\ref{ap1-2})-(\ref{ap1-4}), 
the eigenvalues of $\tilde{\bol A}_\perp$ [$\tilde{\bol A}_z$] are 
$A_r^\pm=(\omega_n^2+\epsilon_r(k)^2 \pm 2i H\omega_n)/(2gc)$
[$A_r^z=(\omega_n^2+\epsilon_r(k)^2+H^2)/(2gc)$]. 
Using these, we can integrate out $\vec n_l$ in $Z$ and derive the SPE,
\begin{eqnarray}
\label{3b-3}
\frac{gc}{2\pi}\sum_r\int_0^\Lambda \frac{dk}{\epsilon_r^0(k)}
\sum_{z=0,+,-}\coth\Big(\frac{\beta}{2}\epsilon_r^z(k)\Big)&=&N,
\end{eqnarray}
where $\epsilon_r^0(k)=c\sqrt{k^2+\xi_r^{-2}+H^2/c^2}$ and 
$\epsilon_r^\pm(k)=\epsilon_r^0(k)\mp H$. The quantities 
$\epsilon_r^{0,+,-}(k)$ can be considered as the magnon dispersions. 
The comparison between Eqs.~(\ref{3b-3}) and (\ref{3a-5}) elucidates the
relation $\xi_r(H=0)^{-2}=\xi_r(H)^{-2}+H^2/c^2$ at $T=0$.
Thus, one can see that new dispersions $\epsilon_r^z(k)$ restore the
Zeeman splitting at $T=0$. 
The magnons with $\epsilon_r^0(k)$, $\epsilon_r^+(k)$ and
$\epsilon_r^-(k)$ take $S^z=0$, $1$ and $-1$, respectively.

Strictly speaking, the above SPA cannot handle the strong-field case
($H\sim J$), where some magnons are condensed, since the magnon 
condensation and the finite uniform magnetization are not taken into 
account in the Haldane mapping (\ref{2a-4}).    
However, as will be explained below, under certain approximations, 
the above consideration can be
extended to the situation where the magnons are condensed.


\subsubsection{\label{sec3b-2}Magnon condensed state in frustrated tubes} 
Referring the arguments in Refs.~\onlinecite{Aff-magcond} and 
\onlinecite{Kon}, 
we try to construct the effective theory for 
the magnon condensed state in frustrated tubes. 
After integrating out the magnon fields except for the degenerate lowest
ones in the action~(\ref{3b-1}), the effective action would be given as 
\begin{eqnarray}
\label{3b-2-1}
S_{\rm{eff}}[\vec m_{\pm p}]&=&
\sum_{q=\pm p}\int d{\bf x}\Big\{\frac{1}{2gc}
\Big[(\partial_\tau\vec m_q)^2 +c^2(\partial_x\vec m_q)^2\Big]\nonumber\\
&&-\frac{i}{gc}\vec H\cdot(\vec m_q\times\dot{\vec m}_q)+V(\vec m_q)\Big\}, 
\end{eqnarray}
where $p=\frac{N-1}{2}$ ($k_p=\frac{N-1}{N}\pi$) and 
\begin{subequations}
\label{3b-2-2}
\begin{eqnarray}
\vec m_r({\bf x})&=&U_{rl}\vec n_l({\bf x})=\sqrt{\frac{2}{N}}\sum_{l=1}^N
\sin(k_rl+\pi/4)\vec n_l,\,\,\,\,\,\,\,\,\label{3b-2-2a}\\
V(\vec m_q)&=&\frac{\Delta_q^2-H^2}{2gc}\vec m_q^2+\frac{H^2}{2gc}(m_q^z)^2
+u|\vec m_q|^4.\label{3b-2-2b}
\end{eqnarray}
\end{subequations}
In the potential $V(\vec m_q)$, $\Delta_q=\Delta_{\rm{min}}$ 
($\Delta_{\rm{min}}$ is defined in the no-field case), 
$c\xi_q^{-1}=\frac{\Delta_q^2-H^2}{2gc}$, and we introduced the
biquadratic term $u|\vec m_q|^4$ to ensure the stability of the magnon
condensed state [one can interpret that it originates from the large-$N$
expansion for the O($N$) NLSM]. 
We perform the Ginzburg-Landau (GL) analysis (i.e., a mean-field theory) 
for the action~(\ref{3b-2-1}).
The minimum of the GL potential $V(\vec m_q)$ is at $\vec m_q=0$ 
in the weak uniform-field regime $H<\Delta_q$. On the other hand, 
as $H$ exceeds $\Delta_q$,
the minimum is located in the field configuration, 
\begin{eqnarray}
\label{3b-2-3}
(m_q^x)^2 + (m_q^y)^2=\frac{H^2-\Delta_q^2}{4ugc}\equiv \bar m^2,
&&m^z=0.
\end{eqnarray}
Namely, $H=\Delta_q$ is just the critical point between the condensed
phase with a finite uniform magnetization and the noncondensed phase, 
within the GL analysis. For the condensed case 
$H>\Delta_q$, let us introduce the new field parameterization, 
\begin{eqnarray}
\label{3b-2-4}
m_q^{\pm} \equiv m_q^x \pm i m_q^y &=& (\bar m_q+\bar m)
e^{\pm i\sqrt{\pi}\theta_q}.
\end{eqnarray}
Substituting Eq.~(\ref{3b-2-4}) in the action~(\ref{3b-2-1}), we obtain 
\begin{eqnarray}
\label{3b-2-5}
S_{\rm{eff}}&=&
\sum_{q=\pm p}\int d{\bf x}\Big\{\frac{1}{2gc}
\Big[(\partial_\tau m_q^z)^2 +c^2(\partial_x m_q^z)^2\Big]\nonumber\\
&&+\frac{1}{2gc}\Big[(\partial_\tau \bar m_q)^2 
+c^2(\partial_x \bar m_q)^2\Big]\nonumber\\
&&+\frac{H^2}{2gc}(m_q^z)^2
+\frac{H^2-\Delta_q^2}{gc}\bar m_q^2 \nonumber\\
&&-i\frac{\sqrt{\pi}H}{gc}(2\bar m \bar m_q+\bar m^2)\partial_\tau\theta_q
\nonumber\\
&&+\frac{\pi\bar m^2}{2gc}\Big[(\partial_\tau \theta_q)^2
+c^2(\partial_x \theta_q)^2\Big]+\dots\Big\},
\end{eqnarray}
up to the quadratic order of the fields. The action~(\ref{3b-2-5})
indicates that fields $m_q^z$ and $\bar m_q$ are massive, and it means
that the low-energy limit of the magnon condensed state 
is described by the two phase fields $\theta_{q=\pm p}$. 
If one integrates out the massive fields neglecting the
third or higher derivatives of all fields in the action~(\ref{3b-2-5}),
the resultant effective Euclidean Lagrangian density is 
\begin{eqnarray}
\label{3b-2-6}
{\cal L}_{\rm{E}}(\theta_p,\theta_{-p}) &=& \sum_{q=\pm p}
\frac{K}{2v}\Big[
(\partial_\tau \theta_q)^2 + v^2(\partial_x \theta_q)^2\Big]
\nonumber\\
&&-ih_1 \partial_\tau\theta_q,\,\,\,\,
\end{eqnarray}
where $K=\frac{\pi\bar m^2}{g}\sqrt{h_2}$, 
$v=c/\sqrt{h_2}$, $h_1=\sqrt{\pi}H\bar m^2/(gc)$, and 
$h_2=(3H^2-\Delta_q^2)/(H^2-\Delta_q^2)$. 
This is just the Lagrangian density for a two-component 
TLL:~\cite{CFT,Gogo} $K$ and $v$
correspond to the TLL parameter and the Fermi velocity, respectively. 
These two quantities are renormalized, from their values of the GL theory, 
by the interactions neglected here (we will discuss the effects of such
interactions below). 
It has been known that in the vicinity of the lower or upper critical
fields, the TLL parameter for a gapless AF spin chain 
preserving the $z$ component of the total spin 
is generally approximated as unity corresponding to the 
free fermion.~\cite{Aff-magcond,Kon,Hal2} 
Under the assumption that this property also holds in tubes, 
$K$ in Eq.~(\ref{3b-2-6}) is close to unity as $H\to \Delta_q+0$.
Here, imitating $\vec m_r=U_{rl}\vec n_l$, we define the 
Fourier transformation of the spin operator as 
$\vec T_r\equiv U_{rl}\vec S_{l,j}\approx 
\vec T_{r,\rm{u}}+(-1)^j\vec T_{r,\rm{st}}$.
Equations~(\ref{3b-2-4}), (\ref{2a-spin}), and (\ref{2b-1}) serve
\begin{subequations}
\label{3b-2-7}
\begin{eqnarray}
T_{0,\rm{u}}^z &=&\frac{1}{\sqrt{N}}\sum_j\frac{i}{4J}
(\vec n_j\times\dot{\vec n}_j)^z+\frac{H}{4J}[1-(n_j^z)^2]\nonumber\\
&=&\frac{1}{\sqrt{N}}\sum_r\frac{i}{4J}(\vec m_r\times\dot{\vec m}_r)^z
+\dots \nonumber\\
&\approx&\frac{1}{\sqrt{N}}
\Big[\frac{i\bar m^2}{4J}(\partial_\tau\theta_p+\partial_\tau\theta_{-p})
+\dots\Big], \label{3b-2-7a}\\
T_{q,\rm{st}}^+&\approx&(\bar m_q+\bar m)e^{i\sqrt{\pi}\theta_q}+\dots,
\label{3b-2-7b}
\end{eqnarray}
\end{subequations}
where the latter equation suggests that the radius of $\theta_q$ is 
$1/\sqrt{\pi}$, and it also means that the radius of the dual field
$\phi_q$ is $1/\sqrt{4\pi}$ in our notation:~\cite{Gogo} the equal-time
commutation relation between $\theta_q$ and $\phi_q$ is defined as 
$[\phi_q(x),\theta_q(y)]=i\Theta_{\rm s}(y-x)$,~\cite{Shank,Delft} where 
$\Theta_{\rm s}$ is the Heaviside's step function. We expect that 
$\partial_\tau\theta_q$ ($e^{i\sqrt{\pi}\theta_q}$) in
Eq.~(\ref{3b-2-7}) is the most relevant bosonic term in the effective
representation of $T_{0,\rm{u}}^z$ ($T_{q,\rm{st}}^+$). 
However, it is difficult to determine the forms of 
second relevant or more irrelevant terms 
within only the above NLSM plus GL analysis.

In order to examine a more proper bosonic representation of $\vec T_r$, 
let us once review the low-energy physics of a spin-1 AF chain
with a uniform field. 
In addition to the NLSM approach in Sec.~\ref{sec2}, there is another
low-energy effective theory for the spin-1 AF chains. 
The latter approximates the spin-1 chain without external fields as 
three copies of massive real fermions, each of which is equivalent to 
an off-critical transverse Ising chain.~\cite{Masa,Tsv,Allen,Kita-Nomu} 
In this scheme, the spin operator is written as 
\begin{eqnarray} 
\label{3b-2-8}
S_j^\alpha &\approx & S_{\rm{u}}^\alpha+(-1)^jS_{\rm{st}}^\alpha,\,\,\,\,
(\alpha=1,2,3 \,\rm{or}\,x,y,z),\,\,\,\,\nonumber\\
S_{\rm{u}}^\alpha/a&=& -i(\xi_L^{\alpha+1}\xi_L^{\alpha+2}+
\xi_R^{\alpha+1}\xi_R^{\alpha+2}),\nonumber\\
S_{\rm{st}}^\alpha/a &=& C_{\rm{st}}\sigma_{\alpha}\mu_{\alpha+1}\mu_{\alpha+2},
\,\,\,\,(\alpha+3=\alpha),
\end{eqnarray}
where $\xi_{L(R)}^\alpha$ is the left (right) mover of the $\alpha$th
real fermion, $\sigma_\alpha$ ($\mu_\alpha$) stands for the order (disorder)
field in the $\alpha$-th Ising system, 
and $C_{\rm{st}}$ is a nonuniversal constant. 
The GS in the fermion picture
corresponds to the disorder phase in the Ising picture: 
$\langle \mu_\alpha\rangle \neq 0$. When the uniform field exceeds the 
lower critical value, the low-energy physics is governed by a TLL
with a scalar boson $\phi$ and its dual $\theta$. In this case, 
the representation of the spin operator~(\ref{3b-2-8}) should
be changed by using these two fields. Equation~(\ref{3b-2-8}) and the
known results of the 2-leg spin-$\frac{1}{2}$ AF ladder with a uniform
field~\cite{Selton,Fu-Zh} provide a desirable representation,
\begin{subequations}
\label{3b-2-9}
\begin{eqnarray} 
\frac{S_{\rm{u}}^z}{a} &\approx& \frac{\partial_x\phi }{\sqrt{\pi}} 
+ \frac{M_u}{a} + C_1\cos\Big[\sqrt{4\pi}\phi+\frac{2\pi M_ux}{a}\Big],
\,\,\,\,\,\,\,\,\label{3b-2-9a}\\
\frac{S_{\rm{st}}^z}{a} &\approx& C_2
\sigma_3\cos\Big[\sqrt{\pi}\phi+\frac{\pi M_ux}{a}\Big],\label{3b-2-9b}\\
\frac{S_{\rm{st}}^+}{a} &\approx& C_3
\mu_3 e^{i\sqrt{\pi}\theta}\Big\{1   \nonumber\\
&&+C_4 \sin\Big[\sqrt{4\pi}\phi+\frac{2\pi M_ux}{a}\Big]\Big\},
\,\,\,\,\,\,\,\,\,\,\label{3b-2-9c}
\end{eqnarray}
\end{subequations}
where $M_u=\langle S_j^z\rangle$, and $C_{1\text{-}4}$ are a constant.
The third fermion system is still massive  (it
corresponds to $\langle\mu_3\rangle\neq 0$) and 
the fermion $\xi_{L,R}^3$ stands for the magnon with $S^z=0$.
The formula~(\ref{3b-2-9}) is valid in $M_u\ll 1$.~\cite{Fu-Zh}
Equation~(\ref{3b-2-9c}) means that the radius of
$\theta$ is $1/\sqrt{\pi}$. While, Eq.~(\ref{3b-2-9b})
also suggests that the radius of $\phi$ is the same value,
$1/\sqrt{\pi}$. This apparently contradicts 
the framework of the standard TLL theory.
Actually, if we allow the presence of both 
$e^{i\sqrt{\pi}\theta}$ and $e^{i\sqrt{\pi}\phi}$, the commutation
relation between these two is nonlocal: 
$e^{i\sqrt{\pi}\phi(x)}e^{i\sqrt{\pi}\theta(y)}
 =e^{-\pi[\phi(x),\theta(y)]}e^{i\sqrt{\pi}\theta(y)}e^{i\sqrt{\pi}\phi(x)}
 ={\rm sgn}(y-x)e^{i\sqrt{\pi}\theta(y)}e^{i\sqrt{\pi}\phi(x)}$ 
$(x\neq y)$, where ${\rm sgn}(y-x)$ is the sign function. 
The nonlocal property is inconsistent with the fact that the original
spin-1 operators are mutually local (i.e., two spins on
different sites commute with each other). However, observing 
Eq.~(\ref{3b-2-9}) carefully, one can find that 
the nonlocality between $\sigma_3$ and $\mu_3$ and 
that between $\cos(\sqrt{\pi}\phi+\pi M_ux/a)$ and 
$e^{i\sqrt{\pi}\theta}$ cooperatively restore the locality among the
spin operators $S_j^z$ and $S_k^\pm$. Therefore, we believe that the 
formula~(\ref{3b-2-9}) is valid and the radius of $\theta$ ($\phi$) 
may be defined as $1/\sqrt{\pi}$ ($1/\sqrt{4\pi}$). 
Here, further notice that the effective Hamiltonian 
in the spin-$1$ chain has to be constructed by the sum of 
terms being locally related with each other, because the original chain 
is a {\em locally interacting} system. This statement must hold in the
effective theories of the ladders and tubes~(\ref{1-1}).

Now, we go back to the frustrated tubes. 
Following Refs.~\onlinecite{Aff-magcond} and \onlinecite{Kon}, 
one can see that the $\vec m_p$ (or $\vec m_{-p}$) part of 
the effective action~(\ref{3b-2-1}) [or (\ref{3b-2-5})] is the same form
as the effective one for the spin-1 chain with a uniform field. 
Moreover, the bosonic representation~(\ref{3b-2-7}) is
very similar to that of the spin-1 chain, Eq.~(\ref{3b-2-9}).
From these facts, it is expected that 
Eq.~(\ref{3b-2-9}) helps us improve the imperfect formula~(\ref{3b-2-7}).
We propose the following new bosonic representation of $\vec T_r$: 
\begin{subequations}
\label{3b-2-10}
\begin{eqnarray} 
\sqrt{N}T_{0,\rm{u}}^z&\approx&\frac{a}{\sqrt{\pi}}
(\partial_x\phi_p+\partial_x\phi_{-p})+2M_t\nonumber\\
&&+C_{t1}\Big\{\cos\Big[\sqrt{4\pi}\phi_p+\frac{2\pi M_tx}{a}\Big]\nonumber\\
&&+\cos\Big[\sqrt{4\pi}\phi_{-p}+\frac{2\pi M_tx}{a}\Big]\Big\},
\label{3b-2-10a}\\
T_{q,\rm{st}}^z &\approx&
C_{t2}\cos\Big[\sqrt{\pi}\phi_q+\frac{\pi M_tx}{a}\Big],\label{3b-2-10b}\\
T_{q,\rm{st}}^+&\approx& C_{t3}e^{i\sqrt{\pi}\theta_q}
\Big\{1\nonumber\\
&&+C_{t4} \sin\Big[\sqrt{4\pi}\phi_q+\frac{2\pi M_tx}{a}\Big]\Big\},
\,\,\,\,\,\,\,\,\,\,\,
\label{3b-2-10c}
\end{eqnarray}
\end{subequations}
where $\phi_q$ is the dual of $\theta_q$, and $C_{t1\text{-}t4}$ are a
constant. The first term in Eq.~(\ref{3b-2-10a}) would be acceptable
from the real-time operator identity
$\frac{1}{v}\partial_t\theta_q=\partial_x\phi_q$. The parameter $M_t$
can be fixed by the magnetization per site 
$\langle S_{l,j}^z\rangle=2M_t/N$. (Determining the correct value of
$M_t$ is difficult within the GL theory.) To preserve the locality among
the spin operators $T_q^\alpha$, we should regard that $C_{t2}$ and
$C_{t3}$ contain massive fields such as $\sigma_3$ and $\mu_3$ in 
Eq.~(\ref{3b-2-9}). However, since we actually cannot determine what
massive fields the constants $C_{t2,t3}$ contain within the present
heuristic approach, Eq.~(\ref{3b-2-10b}) might be
somewhat doubtful.

We proceed to the discussion 
employing the formula~(\ref{3b-2-10}).
So far, we have omitted the interactions generated from the higher-order
terms of $\vec m_q$ and the trace out of the massive fields. 
We hence study their effects towards the two-component TLL. 
The interactions can induce terms involving $\phi_q$
and $\theta_q$ in the effective Hamiltonian for the TLL. 
Let us concentrate on the situation near $H\to \Delta_q$, 
and assume that $K$, the radius of
$\phi_q$, and that of $\theta_q$ are approximated as unity, 
$1/\sqrt{4\pi}$, and $1/\sqrt{\pi}$, respectively. In such a case,
the relevant or marginal vertex
operators are restricted to 
$e^{\pm i\sqrt{4\pi}\phi_q}$, $e^{\pm i\sqrt{\pi}n\theta_q}$,  
$e^{\pm i\sqrt{4\pi}\Phi_\pm}$, and $e^{\pm i\sqrt{\pi}n\Theta_\pm}$ 
($n=1$ or $2$), where we defined 
$\Phi_\pm=\phi_p\pm\phi_{-p}$ and $\Theta_\pm= \theta_p\pm\theta_{-p}$. 
[Note that the scaling dimension of a vertex operator 
$e^{\pm i\sqrt{A}\phi_q(\theta_q)}$ is $A/4\pi$ in the
Lagrangian~(\ref{3b-2-6}) with $K=1$ in our notation.~\cite{Gogo}]    
It is sufficient to investigate whether these terms can be allowed or
not in the low-energy effective theory, 
in order to know how critical state appears in the frustrated tubes.~\cite{relevant} 
To this end, we utilize several symmetries in the spin tube systems.

From Eq.~(\ref{3b-2-4}), the U(1) spin rotation around the $z$ axis 
corresponds to the transformation
$\theta_q\to\theta_q+\text{constant}$.~\cite{OYA} 
Since the spin-tube Hamiltonian should be invariant under the U(1) rotation,
the effective theory does not have any interaction terms 
with $e^{\pm i\sqrt{\pi}n\theta_q}$ and $e^{\pm i\sqrt{\pi}n\Theta_+}$. 
Equation~(\ref{3b-2-10}) shows that the one-site translation along the chain
is identified with $\phi_q\to \phi_q+\sqrt{\pi}(M_t\pm 1)$ and 
$\theta_q\to\theta_q\pm\sqrt{\pi}$.~\cite{OYA} Thus, the appearance of 
$e^{\pm i\sqrt{4\pi}\phi_q}$ and $e^{\pm i\sqrt{4\pi}\Phi_+}$ 
is also prohibited as far as $M_t$ is not equal to a special
commensurate value. (For the nonfrustrated tubes or ladders, 
the restriction from the above two symmetries 
is sufficient to confirm the $c=1$ state.)

To further restrict the possible terms of vertex operators, 
we consider the symmetries with respect to the rung direction. 
All the tubes have the reflection symmetry illustrated in Fig.~\ref{fig3}: 
the corresponding transformations are $\vec S_{l,j}\to \vec S_{N-l,j}$, 
$\vec n_l\to \vec n_{N-l}$ ($\vec S_{N,j}$ and $\vec n_N$ are fixed), 
$\vec T_r\to\vec T_{-r}$, and $\vec m_r\to\vec m_{-r}$. From
Eq.~(\ref{3b-2-10}), these obviously require the effective theory to be 
invariant under the mapping 
$(\phi_q,\theta_q)\to(\phi_{-q},\theta_{-q})$. This prohibits 
the sine-type operators, $\sin(\beta\Theta_-)$ and 
$\sin(\beta\Phi_-)$ [$\beta\in {\mathbb R}$], 
since they change their signs under the mapping. 
To discuss the remaining relevant terms $\cos(\sqrt{4\pi}\Phi_-)$ and 
$\cos(\sqrt{\pi}l_1\Theta_-)$, we further examine 
the invariance under the translation along the rung: 
$\vec S_{l,j}\to\vec S_{l+{\cal Q},j}$ and 
$\vec n_l\to\vec n_{l+{\cal Q}}$ [$l$ mod $N$, and 
${\cal Q}\in {\mathbb Z}$]. 
From Eq.~(\ref{3b-2-2a}) and the definition of $\vec T_r$, 
these operations cause
\begin{subequations}
\label{3b-2-11}
\begin{eqnarray}
\vec T_r  &\to &  \cos ({\cal Q}k_r)\vec T_r-\sin({\cal Q}k_r) \vec T_{-r},
\\
\label{3b-2-11-1}
\vec m_r  &\to &  \cos ({\cal Q}k_r)\vec m_r-\sin({\cal Q}k_r) \vec m_{-r}.
\label{3b-2-11-1}
\end{eqnarray}
\end{subequations}
Comparing Eqs.~(\ref{3b-2-11}) and (\ref{3b-2-10}), 
we propose the following transformation for the vertex operators:
\begin{subequations}
\label{3b-2-12}
\begin{eqnarray}
e^{\pm i\sqrt{\pi}\theta_q} &\to& \cos({\cal Q}k_q) 
e^{\pm i\sqrt{\pi}\theta_q}
-\sin({\cal Q}k_q)e^{\pm i\sqrt{\pi}\theta_{-q}},
\,\,\,\,\,\,\,\,\,\,\,\,\,\,\label{3b-2-12a}\\
e^{\pm i\sqrt{\pi}\phi_q} &\to& \cos({\cal Q}k_q)e^{\pm i\sqrt{\pi}\phi_q}
-\sin({\cal Q}k_q) e^{\pm i\sqrt{\pi}\phi_{-q}}.
\,\,\,\,\,\,\,\,\,\,\,\,\,\,\label{3b-2-12b}
\end{eqnarray}
\end{subequations}
In fact, as far as one focuses on the most relevant term in
Eqs.~(\ref{3b-2-10b}) and (\ref{3b-2-10c}), Eq.~(\ref{3b-2-12}) is
consistent with the transformation~(\ref{3b-2-11}). 
[We will discuss the second relevant term in Eq.~(\ref{3b-2-10c}) later.]
The transformation~(\ref{3b-2-12}) leads to 
\begin{eqnarray}
\label{3b-2-13}
e^{\pm i\sqrt{\pi}\Theta_-} &\to& \cos^2({\cal Q}k_q) 
e^{\pm i\sqrt{\pi}\Theta_-}
-\sin^2({\cal Q}k_q)  e^{\mp i\sqrt{\pi}\Theta_-}\nonumber\\
&&+{\cal O}_\pm,
\end{eqnarray}
where ${\cal O}_\pm=2\sin (2{\cal Q}k_p) 
(e^{\pm i\sqrt{\pi}\Theta_+}e^{\mp i\sqrt{\pi}\Theta_+}-
e^{\pm i\sqrt{\pi}\Theta_-}e^{\mp i\sqrt{\pi}\Theta_-})$.
If ${\cal O}_\pm$ can be negligible in the sense of the
point-splitting technique,~\cite{CFT,Gogo,Cardy} 
Eq.~(\ref{3b-2-13}) provides 
\begin{eqnarray}
\cos(\sqrt{\pi}\Theta_-) &\to& \cos(2{\cal Q}k_p)\cos(\sqrt{\pi}\Theta_-).
\,\,\,\,\,\,\label{3b-2-14}
\end{eqnarray}
Due to $\cos(2{\cal Q}k_p)\neq 1$, 
$\cos(\sqrt{\pi}\Theta_-)$ has to be absent in the effective Hamiltonian. 
Furthermore, using Eq.~(\ref{3b-2-13}), one can obtain
\begin{eqnarray}
\label{3b-2-15}
\cos(2\sqrt{\pi}\Theta_-) &\to& [\cos^4({\cal Q}k_p)+\sin^4({\cal Q}k_p)]
\cos(2\sqrt{\pi}\Theta_-)\nonumber\\ &&-\tilde{\cal O},
\end{eqnarray}
where $\tilde{\cal O}=2\sin^2({\cal Q} k_p)
[(e^{i\sqrt{\pi}\Theta_-}e^{-i\sqrt{\pi}\Theta_-}
+e^{-i\sqrt{\pi}\Theta_-}e^{i\sqrt{\pi}\Theta_-})+(\text{h.c})]$. 
By using the point splitting, $\tilde{\cal O}$ may be replaced with a constant. 
Since $k_p$ can not satisfy $\cos^4({\cal Q}k_p)+\sin^4({\cal Q}k_p)=1$ and 
$\sin({\cal Q} k_p)=0$ simultaneously, 
the marginal terms $\cos(2\sqrt{\pi}\Theta_-)$ is also forbidden. 
From Eq.~(\ref{3b-2-12b}), 
the similar argument from Eq.~(\ref{3b-2-13}) to (\ref{3b-2-15}), of
course, can be adopted to the vertex operators with $\phi_q$.

From these arguments, 
we can say that the symmetries of tubes make all the 
relevant or marginal operators absent in the effective theory. 
Namely, the above bosonization argument
supports the presence of the $c=1+1$ state. 

Here, we had better think again the proposal~(\ref{3b-2-12}).
The reader will immediately (or already) find that the final term in 
Eq.~(\ref{3b-2-10c}) does not obey the desirable transformation
corresponding to Eq.~(\ref{3b-2-11}).
Therefore, it is expected that 
either the term in Eq.~(\ref{3b-2-10c}) or the
proposal~(\ref{3b-2-12b}) is invalid.
In the latter case, one cannot
forbid the existence of $\cos(2\sqrt{\pi}\Phi_-)$. Then, instead of the
discussion on the symmetries, let us  count on the known result: 
in the single integer-spin-$S$ AF chain, 
the TLL parameter $K$ monotonically increases together with the growth
of the magnetization $\langle S_j^z\rangle$ within the region 
$\langle S_j^z\rangle\ll S$.~\cite{Aff-magcond,Kon,Fath} 
Provided there exists the same nature in
the integer-spin frustrated tubes, the scaling dimension of 
$\cos(2\sqrt{\pi}\Phi_-)$, $2K$, is larger than 2 (i.e., irrelevant) 
for the small-$M_t$ case. 
Therefore, we can predict again the presence of the 
$c=1+1$ state.

Now, are there any vertex terms which survive from the restriction of 
symmetries of the reflection $\vec T_r\to\vec T_{-r}$ and
translation~(\ref{3b-2-11})? 
We can find that the following four terms:
\begin{eqnarray}
\label{3b-2-16}
\cos(\sqrt{4\pi}\phi_p)+\cos(\sqrt{4\pi}\phi_{-p}),\nonumber\\
\sin(\sqrt{4\pi}\phi_p)+\sin(\sqrt{4\pi}\phi_{-p}),\nonumber\\
\cos(\sqrt{4\pi}\theta_p)+\cos(\sqrt{4\pi}\theta_{-p}),\nonumber\\
\sin(\sqrt{4\pi}\theta_p)+\sin(\sqrt{4\pi}\theta_{-p}),
\end{eqnarray}
are invariant under these two operations. 
This is consistent with 
the presence of the final term in Eq.~(\ref{3b-2-10a}).

\subsubsection{\label{sec3b-3}Stronger uniform-field case in the
frustrated tubes (quantum phase transition)} 
We next discuss the frustrated tubes with a stronger uniform field,
where the second lowest magnons are condensed as well as the
lowest ones. 
  
First, we consider the 3-leg tube. The second lowest magnon corresponds
to the field $\vec m_0$. When its condensation takes place, the new
phase field $\theta_0$ and its dual $\phi_0$ would emerge 
from the field $\vec m_0$, like Eq.~(\ref{3b-2-4}). 
Because $\vec m_0$ is invariant under the
reflection $\vec n_l\to\vec n_{N-l}$ and the translation along the 
rung~(\ref{3b-2-11}), the symmetries do not at all restrict the
form of interaction terms with $\theta_0$ and $\phi_0$ in the effective
theory. On the other hand, other symmetries of the U(1) rotation and the
translation along the chain demand the invariance under 
$\theta_0\to\theta_0+{\text{constant}}$ and $(\phi_0,\theta_0)\to
(\phi_0+\sqrt{\pi}(M_0\pm 1),\theta_0\pm\sqrt{\pi})$ [$M_0\neq M_t$]. 
Therefore, all the vertex operators including $\phi_0$ or $\theta_0$
alone are prohibited. Are there any vertex operators with 
$\phi_0$, $\theta_0$, $\phi_q$, and $\theta_q$ which are invariant under
all symmetry operations? Employing the term~(\ref{3b-2-16}), one can
find the following terms permitted for all symmetries: 
\begin{eqnarray} 
\label{3b-2-17}
\cos[\sqrt{4\pi}(\theta_p-\theta_0)]
+\cos[\sqrt{4\pi}(\theta_{-p}-\theta_0)],\nonumber\\
\sin[\sqrt{4\pi}(\theta_p-\theta_0)]+\sin[\sqrt{4\pi}(\theta_{-p}-\theta_0)].
\end{eqnarray}
Notice that the same type of terms as Eq.~(\ref{3b-2-17}), where 
$\theta_{q,0}$ are replaced with $\phi_{q,0}$, 
are not permitted because $M_0\neq M_t$.
Relying again on the known result that the TLL parameter $K$ ($K_0$) for
$\theta_q$ ($\theta_0$) are larger than (close to) unity, we can
expect that the scaling dimension of terms~(\ref{3b-2-17}), 
$1/K+1/K_0$, is smaller than two, and they must be relevant. 
Introducing the new parameterization:~\cite{Cabra} 
$\Theta_0=(\theta_p+\theta_0+\theta_{-p})/\sqrt{3}$, 
$\Theta_1=(\theta_p-\theta_{-p})/\sqrt{2}$ and 
$\Theta_2=(\theta_p+\theta_{-p}-2\theta_0)/\sqrt{6}$, one sees that 
the relevant terms~(\ref{3b-2-17}) can be rewritten by the two fields 
$\Theta_{1,2}$ and do not contain the field $\Theta_0$. As a result, the
field $\Theta_0$ provides a one-component TLL, whereas the remaining 
two fields $\Theta_{1,2}$ carry a gapful excitation.
Thus, we can predict that the $c=1+1$ state in the 3-leg tube are
broken down to a $c=1$ one once the condensate of the $k_r=0$ magnon occurs. 
(Although the above new parameterization makes the Gaussian part of three
TLLs be a nondiagonal form, it would not influence the prediction of
the $c=1$ state.)

\begin{figure}[h]
\begin{center}
\scalebox{0.5}{\includegraphics{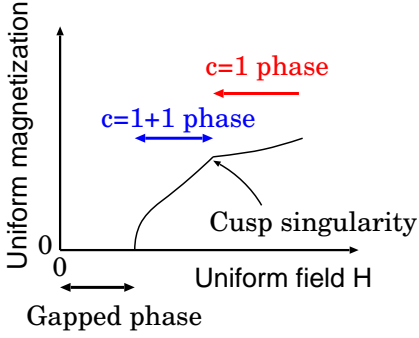}}
\caption{Expected magnetization curves in frustrated integer-spin tubes.}
\label{cusp}
\end{center}
\end{figure}

\begin{figure}[h]
\begin{center}
\scalebox{0.5}{\includegraphics{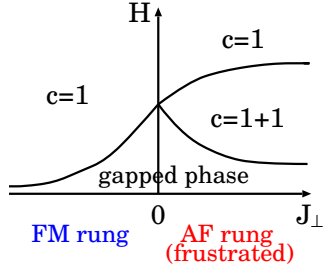}}
\caption{Expected GS phase diagram of odd-leg integer-spin tubes around 
$(J_\perp,H)=(0,0)$.}
\label{phase-frust}
\end{center}
\end{figure}

The similar argument also holds in other frustrated ($N\geq 5$) tubes. 
In these cases, the second lowest bands are twice degenerate: 
the bands with the wave numbers $k_{p-1}$ and $k_{-p+1}$. 
Thus, as the field is applied enough, two
pairs of phase fields $(\phi_{\pm(p-1)},\theta_{\pm(p-1)})$ appear
correspondingly to the condensations of $\vec m_{\pm(p-1)}$.
Similarly to Eq.~(\ref{3b-2-17}), we can find the following 
interaction terms, which are permitted from all symmetry
operations:
\begin{eqnarray} 
\label{3b-2-18}
\cos[\sqrt{4\pi}(\theta_p-\theta_{p-1})]
+\cos[\sqrt{4\pi}(\theta_{-p}-\theta_{p-1})]\nonumber\\
+\cos[\sqrt{4\pi}(\theta_p-\theta_{-p+1})]
+\cos[\sqrt{4\pi}(\theta_{-p}-\theta_{-p+1})],\nonumber\\
\sin[\sqrt{4\pi}(\theta_p-\theta_{p-1})]
+\sin[\sqrt{4\pi}(\theta_{-p}-\theta_{p-1})]\nonumber\\
+\sin[\sqrt{4\pi}(\theta_p-\theta_{-p+1})]
+\sin[\sqrt{4\pi}(\theta_{-p}-\theta_{-p+1})].
\end{eqnarray}
These are expected to be relevant. As in the 3-leg case, if we introduce
the new fields~\cite{Cabra}
$\tilde{\Theta}_0=(\theta_p+\theta_{-p}+\theta_{p-1}+\theta_{-p+1})/\sqrt{4}$, 
$\tilde{\Theta}_1=(\theta_p-\theta_{-p}+\theta_{p-1}-\theta_{-p+1})/\sqrt{4}$, 
$\tilde{\Theta}_2=(\theta_p-\theta_{p-1})/\sqrt{2}$ and 
$\tilde{\Theta}_3=(\theta_{-p}-\theta_{-p+1})/\sqrt{2}$, 
the terms~(\ref{3b-2-18}) are 
re-expressed by using only three fields $\tilde{\Theta}_{1,2,3}$. 
Consequently, a $c=1$ state with the scalar field $\tilde{\Theta}_0$ would
appear, and other three fields have a massive spectrum.
Thus, we can finally arrive at the general prediction that 
a $c=1$ state emerges instead of the $c=1+1$ one as the second lowest
bands crosses the zero-energy line in all the frustrated tubes. 
The quantum phase transition between these
two critical ($c=1+1$ and $c=1$) states would be observed as 
a cusp singularity in the uniform-field magnetization curve as in
Fig.~\ref{cusp}, because the uniform susceptibility is generally
proportional to the number of massless modes in 1D spin systems. 
Moreover, the GS phase diagram for the frustrated tubes is drawn as in
Fig.~\ref{phase-frust}.
(Applying the same argument to
nonfrustrated systems, it is found that the $c=1$ state lasts out 
even when the condensate of the second lowest magnon occurs.)
Besides the above scenario of the magnetization cusp, other cusp
singularities have already been found in theoretical studies of 
1D quantum systems.~\cite{Fra-Kre,Kawa1,Kawa2,Oku1,Oku2,Yamamoto} 
It is known that around such cusp points, 
the left or right derivatives of the magnetization (i.e.,
susceptibilities) always diverge. However, such a singular phenomena is
expected not to occur in our cusp mechanism. Thus, we may insist 
that the cusp in the frustrated tubes is a new type.

Finally, we remark the validity and the stability of the bosonization
arguments in Secs.~\ref{sec3b-2} and \ref{sec3b-3}. 
The proposal~(\ref{3b-2-12}) is a strange
form: other symmetry operations correspond to a transformation for the
phase fields own, but Eq.~(\ref{3b-2-12}) is the
transformation for the vertex operators. 
In fact, as stated already, Eq.~(\ref{3b-2-12}) is 
incompatible with the final term in Eq.~(\ref{3b-2-10c}). 
There might be a more natural transformation than 
Eq.~(\ref{3b-2-12}). 
As far as we know, it has never been elucidated whether the GL
theory is efficient or not even when multimagnon bands are condensed. 
Furthermore, the coupling constants of the
phase-fields interactions [for example, Eqs.~(\ref{3b-2-17}), 
(\ref{3b-2-18}), and other irrelevant interactions with phase fields] 
might be as large as the order of $J$. 
In such a case, the perturbative treatment of them and the biquadratic
term $u|\vec m|^4$ would be dangerous. Thus, there is still a
little possibility that the interactions break down the $c=1+1$ state.
On the other hand, 
we essentially use only four symmetries of the spin tubes in
order to lead to the $c=1+1$ state. Thus, if our strategy based on the
bosonization and the GL theory is admitted, 
the $c=1+1$ phase must be stabilized against several small perturbations 
preserving those symmetries 
(e.g., $XXZ$-type anisotropy, single-ion anisotropy
$D\sum_{l,j}(S_{l,j}^z)^2$, next-nearest-neighbor coupling, etc.).

\subsection{\label{sec3c}$[0,\pi]$-field case}
This subsection discusses ladders and tubes with the $[0,\pi]$-field term
(\ref{1-2b}). Odd-leg tubes, which include frustrated tubes, 
are not admitted in this case. 

Imitating the argument in Sec.~\ref{sec3b-1}, we can also express the
action with the quadratic form of bosons $\tilde{n}_l^\alpha$, in which
${\bol E}_{\omega_n}$ is replaced with 
${\bol F}_{\omega_n}=\frac{H\omega_n}{gc}{\rm diag}(-1,1,\dots,(-1)^N)$. 
In the action, the $2N\times 2N$ matrix held between 
$\tilde{\cal N}_\perp^\dag$ and $\tilde{\cal N}_\perp$ are not normal
due to ${\bol F}_{\omega_n}$. 
Its diagonalization hence generates a nontrivial Jacobian differently from
the $[0,0]$-field case. To avoid this difficulty, 
we turn to the {\em real-time} formalism, even though 
it restricts our consideration to the zero-temperature case. 
The partition function 
$Z=\int {\cal D}\vec n_l{\cal D}\lambda e^{-S_{\rm E}^{[0,\pi]}}$ 
is associated with the real-time vacuum-vacuum amplitude 
$Z_{\rm v}=\int {\cal D}\vec n_l{\cal D}\lambda e^{iS_{\rm R}^{[0,\pi]}}$ via 
$S_{\rm R}^{[0,\pi]}[t]=i S_{\rm E}^{[0,\pi]}[\tau= it]$. 
In the Fourier space (note that the frequency $\omega$ 
is a real number, and not the Matsubara one $\omega_n$), the real-time
action $S_{\rm R}^{[0,\pi]}$ can be represented as 
\begin{eqnarray}
\label{3c-1}
S_{\rm R}^{[0,\pi]}[\{\vec n_l\},\lambda_{\rm sp}]&=& \sum_{\bar{\bf k}}
\Big[
\tilde{\cal N}_\perp^{R\,\dag}
\tilde{\bol A}_\perp^R
\tilde{\cal N}_\perp^R 
\nonumber\\
&&+ \tilde{\cal N}_z^{R\,\dag} \tilde{\bol A}_z^R\tilde{\cal N}_z^R\Big] 
 -N{\cal T}L\lambda_{\rm sp},
\end{eqnarray}
where $\bar{\bf k}\equiv(\omega,k)$, $\tilde{\cal N}_\alpha^R
=(n_1^\alpha(\bar{\bf k}),\dots,n_N^\alpha(\bar{\bf k}))$, 
$\tilde{\cal N}_\perp^R=({}^T\tilde{\cal N}_x^R,{}^T\tilde{\cal N}_y^R)$, 
$\cal T$ is the time distance from the initial vacuum to the final one, 
and we performed a unitary transformation $n_l^y\to (-1)^l n_l^y$ 
for convenience. Two matrices $\tilde{\bol A}_\perp^R$ 
and $\tilde{\bol A}_z^R$ are defined as
\begin{subequations}
\label{3c-2}
\begin{eqnarray}
\tilde{\bol A}_\perp^R (\bar{\bf k})&=&
\left(
\begin{array}{ccc}
{\bol A}_{\bar{\bf k}}  &     {\bol E}_\omega  \\
 -{\bol E}_\omega    &     {\bol B}_{\bar{\bf k}}
\end{array}
\right),\label{3c-2-1}\\
\tilde{\bol A}_z^R(\bar{\bf k}) &=& 
{\bol A}_{\bar{\bf k}} -\frac{H^2}{2gc}\hat{\bol 1},
\label{3c-2-2}\\
{\bol E}_\omega &=& -i\frac{H\omega}{gc}\hat{\bol 1},\label{3c-2-3}
\end{eqnarray}
\end{subequations}
where ${\bol A}_{\bar{\bf k}}$ [${\bol B}_{\bar{\bf k}}$] is the same
form as ${\bol A}_{\bf k}$, in which 
$\tilde{a}_1\to (\omega^2-c^2k^2+2gc\lambda_{\rm sp})/(2gc)$ and  
$\tilde{a}_{0,2}\to -\tilde{a}_{0,2}$ [$\tilde{a}_{0,2}\to \tilde{a}_{0,2}$].
Since $\tilde{\bol A}_\perp^R$ and $\tilde{\bol A}_z^R$ 
are Hermite and can be diagonalized by a
unitary matrix, we can apply the SPA prescription 
as in the preceding subsections. 
Eigenvalues of $\tilde{\bol A}_\perp^R$ 
and $\tilde{\bol A}_z^R$ are, respectively, 
$\bar A_r^{\pm}=(\omega^2-\bar\epsilon(k)^2)/(2gc)\pm f_r(\bar {\bf k})$
and $\bar A_r^{z}=(\omega^2-\bar\epsilon_r(k)^2-H^2)/(2gc)$, 
where we defined 
\begin{subequations}
\label{3c-3}
\begin{eqnarray}
\bar\epsilon(k)=c\sqrt{k^2+\bar\xi^{-2}},\,\,\,\,\,
\bar\xi^{-2} \equiv -2g\lambda_{\rm sp}/c, \nonumber\\
\bar\epsilon_r(k)=c\sqrt{k^2+\bar\xi_r^{-2}}, \,\,\,\,\,
\bar\xi_r^{-2}\equiv \bar\xi^{-2}+\frac{2J_\perp}{Ja^2}\cos k_r,\nonumber\\
f_r(\bar{\bf k})=\sqrt{\left(\frac{cJ_\perp}{gJa^2}\right)^2\cos^2k_r
+\left(\frac{H\omega}{gc}\right)^2}.\nonumber
\end{eqnarray}
\end{subequations}
Employing these eigenvalues, one can trace out $\vec n_l$ in $Z_{\rm v}$, 
and then obtain the effective action $S_{\rm R}^{[0,\pi]}[\lambda_{\rm sp}]$. 
The SPE 
$\partial S_{\rm R}^{[0,\pi]}[\lambda_{\rm sp}]/\partial \lambda_{\rm sp}=0$
can be calculated as
\begin{widetext}
\begin{eqnarray}
\label{3c-4}
\frac{gc}{2\pi}\sum_r\int_0^\Lambda dk \Bigg[\sum_{z=0,+,-}
\frac{1}{\bar\epsilon_r^z(k)}
\,\,\,+\frac{2H^2}{h_r(k)}
\left(\frac{1}{\bar\epsilon_r^{+}(k)}-\frac{1}{\bar\epsilon_r^{-}(k)}\right)
\Bigg]   &=&  N,
\end{eqnarray}
where 
\begin{subequations}
\label{3c-5}
\begin{eqnarray}
\bar\epsilon_r^0(k)= \sqrt{\bar\epsilon_r(k)^2 + H^2},
\,\,\,\,\,\,\,\,\,\,\,\,\,\,\,
\bar\epsilon_r^\pm(k)= \sqrt{ \bar\epsilon(k)^2  +  2H^2
\pm h_r(k)},\label{3c-5-1}\\
h_r(k) = \sqrt{4H^4+4H^2 \bar\epsilon(k)^2
+c^4\left(\frac{2J_\perp}{Ja^2}\cos k_r\right)^2}.\label{3c-5-2}
\end{eqnarray}
\end{subequations}
\end{widetext}
Here, we used the so-called $i\varepsilon$-prescription~\cite{Wein} 
in the derivation of the SPE~(\ref{3c-4}). 
The SPE indicates that $\lambda_{\rm sp}$ is real (not imaginary) and 
negative in contrast to the imaginary-time schemes. 
It is verified that as $H\to 0$ [$J_\perp\to 0$], Eq.~(\ref{3c-4}) is 
reduced to the SPE (\ref{3a-5}) [(\ref{2b-3})] of the zero-temperature case, 
where $(\bar\xi,\lambda_{\rm sp})$ corresponds to $(\xi, i\lambda_{\rm sp})$.

\begin{figure}
\scalebox{0.3}{\includegraphics{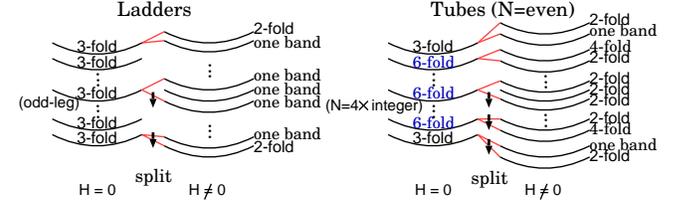}}
\caption{\label{fig11}Predicted band splitting in the $[0,\pi]$-field case.}
\end{figure}

One can regard $\bar\epsilon_r^{0,\pm}(k)$ as the magnon
band dispersions. At $H=0$, $\bar\epsilon_r^0(k)$, $\bar\epsilon_r^+(k)$,
and $\bar\epsilon_r^-(k)$ coincide with $\epsilon_r(k)$ in Eq.~(\ref{3a-4}), 
$\epsilon_{|r|}(k)$ and $\epsilon_{N+1-|r|}(k)$ 
[$\epsilon_{[\frac{N}{2}]+1-|r|}(k)$], 
respectively, for ladders [tubes: $N\geq 3$]. 
Although it is hard to solve the transcendental
equation (\ref{3c-4}) for $\lambda_{\rm sp}$, we can
extract some features of the band splitting induced by the $[0,\pi]$
field, from Eqs.~(\ref{3c-4}) and (\ref{3c-5}). The inequalities 
$\bar\epsilon_r^+(k)>\bar\epsilon_r^-(k)$ and $h_r(k)> 2H^2\geq 0$ show 
that the final term in the left-hand side of Eq.~(\ref{3c-4}) 
is negative or zero. Therefore, $\bar\xi^{-2}$ decreases and the bands 
$\bar\epsilon_r^-(k)$ fall down as $H$ is applied. 
The form of the dispersion~(\ref{3c-5-1}) indicates that as $H$ is
applied, the triply degenerate bands $\epsilon_r(k)$ with $\cos k_r>0$
($<0$) are split into doubly degenerate upper (lower) bands 
and a nondegenerate lower (upper) one. 
The former two are the transverse modes, and the latter is the
longitudinal one. 
The bands with $k_r=\pm\frac{\pi}{2}$, which are present only in
odd-leg ladders and $(4\times Q)$-leg tubes [$Q\in \mathbb{Z}$], 
are divided into three bands. 
From these considerations, we can illustrate the band splitting as
in Fig.~\ref{fig11}.
[Remember that tubes have a sixfold degeneracy in the
no-field case (Fig.~\ref{fig4}).]
The figure tells us three remarkable aspects. 
(\rnum{1}) 
Any band $\bar \epsilon_r^{0,+,-}$ split by the $[0,\pi]$ field tends
not to approach the other neighboring bands (to avoid the band
crossings). This contrasts with the Zeeman splitting in the 
uniform ($[0,0]$)-field case, where the crossings among magnon bands
with different indices $k_r$ occur.  
(\rnum{2}) The lowest bands are {\em doubly degenerate} in all systems. 
It might imply that a sufficiently strong $[0,\pi]$ field always
engenders a $c=1+1$ phase.   
(\rnum{3}) The FM-rung coupling competes with the $[0,\pi]$
field. However, the figure suggests that any qualitative differences 
between the competitive and noncompetitive cases do not emerge at least 
in the weak rung-coupling regime.

We believe that the band structure in Fig.~\ref{fig11} is
qualitatively valid. However, its details would strongly depend upon
the SPA strategy. Particularly, we mind that the symmetry
corresponding to the remaining double degeneracy of transverse modes 
cannot be found. Therefore, the degeneracy and the prediction 
(\rnum{2}) would be ruined in more quantitative approaches.

\subsection{\label{sec3d}$[\pi,0]$- and $[\pi,\pi]$-field cases}
This subsection addresses the
$[\pi,0]$- and $[\pi,\pi]$-field cases, in which the external field
is alternated along the chain. 
Utilizing Eqs.~(\ref{2-1}) and (\ref{3a-3}), 
we can describe their low-energy action as follows:
\begin{eqnarray}
\label{3d-1}
S_{\rm E}[\{\vec n_l\},\lambda]&=& \int d{\bf x}
\Big[{}^T {\cal N}_\alpha 
{\bol A} {\cal N}_\alpha - S\sum_l\vec K_l\cdot\vec n_l \nonumber\\
&&+i N\lambda({\bf x})\Big],  
\end{eqnarray}
where $\vec K_l=\vec H/a$ ($\vec K_l=(-1)^{l+1}\vec H/a$) for the
$[\pi,0]$-field ($[\pi,\pi]$-field) case [$\vec H=(0,0,H)$]. 
Notice that odd-leg tubes are not permitted
for the $[\pi,\pi]$-field case. From Eq.~(\ref{3d-1}), it is
found that a unitary transformation 
$\vec n_{l={\rm even}}\to -\vec n_{l={\rm even}}$ exchanges 
the action of the AF (FM)-rung system with the $[\pi,\pi]$ field 
into that of the FM (AF)-rung one with the $[\pi,0]$ field. 
Therefore, it is enough to investigate only either $[\pi,0]$-field case or
the $[\pi,\pi]$-field case. We analyze the former case below. 

After diagonalizing the quadratic part of $\vec n_l$ in Eq.~(\ref{3d-1}), 
the action are arranged as 
\begin{eqnarray}
\label{3d-2}
S_{\rm E}^{[\pi,0]}[\{\vec n_l\},\lambda]&=& \int d{\bf x}\Big[\sum_r 
\Big\{\vec m_r({\bf x}) A_r({\bf x}) \vec m_r({\bf x})\nonumber\\
&&-S\vec J_r \cdot\vec m_r\Big\} 
+iN\lambda({\bf x})\Big], 
\end{eqnarray}
where 
\begin{widetext}
\begin{subequations}
\label{3d-3}
\begin{eqnarray}
A_r({\bf x})=-\frac{1}{gc}\Big[\partial_\tau^2+c^2\partial_x^2
+\xi_r({\bf x})^{-2}\Big],\hspace{1cm}\label{3d-3-1}
\xi_r({\bf x})^{-2}=\frac{-2ig\lambda({\bf x})}{c} +\frac{2J_\perp}{Ja^2}\cos k_r,
\hspace{1cm}
\\
\vec J_r = 
\left\{
\begin{array}{cc}
\delta_{r,{\rm odd}}\sqrt{\frac{2}{N+1}}
\left[\tan\left(\frac{r\pi}{2(N+1)}\right)\right]^{-1}\frac{\vec H}{a} & 
({\rm ladders})\\
\delta_{r,0}\sqrt{N}\frac{\vec H}{a}  &({\rm tubes}:\,\,N\geq 3),
\end{array}
\right.\,\,\,\,\label{3d-3-2}
\end{eqnarray}
\end{subequations}
\end{widetext}
and the field $\vec m_r({\bf x})=U_{rl}\vec n_l$ [see Eq.~(\ref{3b-2-2a})].
The action (\ref{3d-2}) can be considered as a 
generalization of that of the single chain with a staggered field
(\ref{2-1}). Thus, we can apply the Green's function method in
Appendix~\ref{app2} to the present $[\pi,0]$-field case.
(We would like the reader to refer Appendix~\ref{app2} or 
Ref.~\onlinecite{Erco} before proceeding below.)

Following Appendix~\ref{app2}, let us define Green's functions: 
\begin{subequations}
\label{3d-4}
\begin{eqnarray}
A_r({\bf x})G_r^0({\bf x}-{\bf x'})&=&\delta^2({\bf x}-{\bf x'}),
\label{3d-4-1}\\
G_r^T({\bf x}-{\bf x'})&=&
\langle {\cal T}_\tau m_r^{x(y)}({\bf x})m_r^{x(y)}({\bf x'})\rangle_c,
\label{3d-4-2}\\
G_r^L({\bf x}-{\bf x'})&=&
\langle {\cal T}_\tau m_r^z({\bf x})m_r^z({\bf x'})\rangle_c,
\label{3d-4-3}
\end{eqnarray}
\end{subequations}
where the subscript ${}_c$ means ``connected,'' ${\cal T}_\tau$ denotes
imaginary-time ordered product [see Eq.~(\ref{2-6})], and 
$\delta^2({\bf x}-{\bf x'})=\delta(x-x')\delta(\tau-\tau')$. 
In anticipation of
removing the space-time dependence of $\xi_r$ via the SPA process, we
already assumed that the above Green's functions depend only on the 
distance between ${\bf x}$ and ${\bf x'}$.
The magnon dispersions of transverse and
longitudinal modes can be determined from $G_r^T({\bf x}-{\bf x'})$ and 
$G_r^L({\bf x}-{\bf x'})$, respectively. 
As in Eq.~(\ref{2-9}), the Fourier transformation of $G_r^0$ is
estimated as
\begin{eqnarray}
\label{3d-5}
\tilde G_r^0 ({\bf k}) &=& \frac{gc}{\omega_n^2+c^2k^2+c^2\xi_r^{-2}}, 
\end{eqnarray}
where $\xi_r^{-2}=
-\frac{2ig\lambda{\rm sp}}{c}+\frac{2J_\perp}{Ja^2}\cos k_r$ 
and $\lambda_{\rm sp}$ is the saddle-point value of $\lambda({\bf x})$.
Using $\tilde G_r^0$ and referring the way deriving Eqs.~(\ref{2-4}) and
(\ref{2-10}), we obtain the following 
SPE determining $\lambda_{\rm sp}$ and $\xi_r$,
\begin{eqnarray}
\label{3d-6}
\frac{3gc}{2\pi}\sum_r\int \frac{dk}{\epsilon_r(k)}
\coth\left(\frac{\beta}{2}\epsilon_r(k)\right)\nonumber\\
= N-S^2 \sum_r {\vec J_r}^2 \left(\frac{g}{c}\right)^2\xi_r^4,
\end{eqnarray}
where the final term in the right-hand side represents the $[\pi,0]$-field 
effect. As $J_\perp\to 0$, Eq.~(\ref{3d-6}) returns to
Eq.~(\ref{2-10}). Applying Eqs.~(\ref{2-5}) and (\ref{2-11}), 
we further evaluate the staggered magnetization as 
\begin{widetext}
\begin{eqnarray}
\label{3d-7}
\vec M_l &\equiv& S\langle \vec n_l \rangle = 
S \sum_r U_{rl}\langle \vec m_r\rangle   
=\left\{
\begin{array}{cc}
\frac{2}{N+1}\sum_{r=1}^N\delta_{r,{\rm odd}}\sin\left(\frac{rl\pi}{N+1}\right)
\left[\tan\left(\frac{r\pi}{2(N+1)}\right)\right]^{-1}\left(\frac{\xi_r}{a}\right)^2
 \frac{\vec H}{J}   &   ({\rm ladders})  \\
\left(\frac{\xi_0}{a}\right)^2\frac{\vec H}{J}  & ({\rm tubes}:\, N\geq 3)
\end{array}
\right.,\,\,\,\,\,\,\,
\end{eqnarray}
\end{widetext}
where $\vec M_l$ is parallel to the field $\vec H$ like the single-chain
case [see Eq.~(\ref{2-11})], namely $\vec M_l=(0,0,M_l)$.
The staggered magnetizations are independent of the chain index $l$ in
the tubes: $M_l=M_s$.
On the other hand, the $l$-dependence of the magnetizations clearly 
exists in the ladders. We emphasize that these inhomogeneous distribution
of the staggered magnetization cannot be predicted by the standard NLSM
scheme, which assumes a short-range AF or FM order to arise for the rung
direction. From 
$\sin(\frac{rl \pi }{N+1})=\sin(\frac{r(N+1-l)\pi}{N+1})$, 
we find that $M_l=M_{N+1-l}$ is realized in the ladders.
The results $M_l=M_s$ in tubes and $M_l=M_{N+1-l}$ in ladders indicate
that the present NLSM plus SPA scheme preserves the translational
symmetry along the rung in tubes, and the reflection one 
about the plane containing the central axis of the ladder, in the
$[\pi,0]$-field case [refer to the
discussion about the validity of Eqs.~(\ref{3a-1}) and (\ref{3a-2}) 
in Sec.~\ref{sec3a}].

\begin{figure}
\scalebox{0.38}{\includegraphics{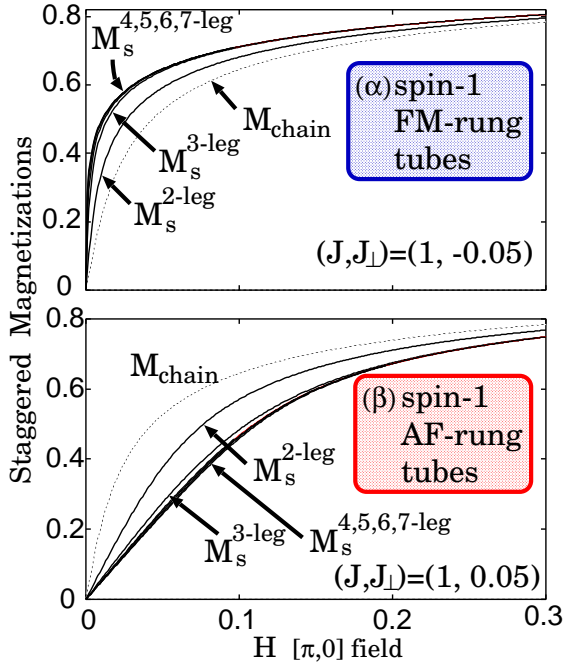}}
\caption{\label{fig12}Staggered magnetizations of $N$-leg
 spin-$1$ FM- or AF-rung tubes with the $[\pi,0]$ field, $M_s^{\text{N-leg}}$. 
The dotted curve $M_{\rm chain}$ stands for the staggered magnetization 
$m_s^z$ of the single AF chain with the staggered field 
(see Fig.~\ref{fig1}). The 2-leg tube means the 2-leg ladder. 
A relation $M_s^{\text{p-leg}}(H)<M_s^{\text{(p+1)-leg}}(H)$ 
[$M_s^{\text{p-leg}}(H)>M_s^{\text{(p+1)-leg}}(H)$] is realized for 
the FM [AF]-rung case. 
Magnetizations $M_s^{\text{4,5,6,7-leg}}$ almost overlap.}
\end{figure}

We will investigate the magnetizations $\vec M_l$ and the magnon modes
in detail, below.

\subsubsection{\label{sec3d-1}Staggered Magnetizations}
\begin{figure}
\scalebox{0.5}{\includegraphics{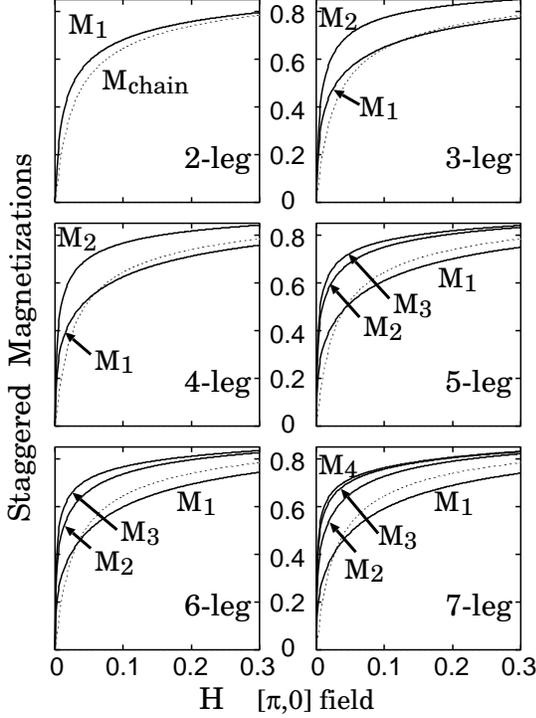}}
\caption{\label{fig13}Staggered magnetizations of $N$-leg 
spin-$1$ FM-rung ladders with the $[\pi,0]$ field and $J_\perp/J=-0.05$. 
For $l< [\frac{N}{2}]$, $M_{l+1}(H)>M_l(H)$ is realized. 
Similarly to Fig.~\ref{fig12}, the symbol 
$M_{\rm chain}$ means $m_s^z$.} 
\end{figure}

\begin{figure}
\scalebox{1.0}{\includegraphics{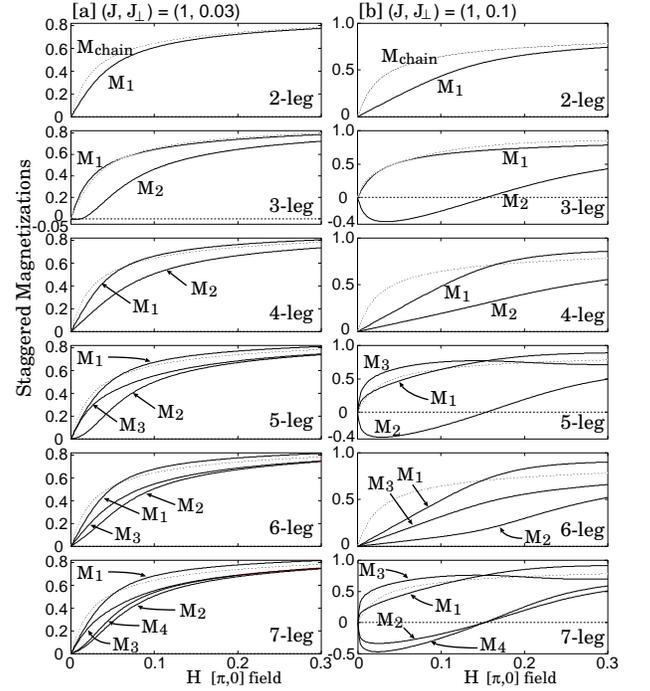}}
\caption{\label{fig14} Staggered magnetizations $M_l$ of $N$-leg
 spin-$1$ AF-rung ladders with the $[\pi,0]$ field.} 
\end{figure}

If $\lambda_{\rm sp}$ is fixed by the SPE, 
the magnetizations $\vec M_l$ are also done in Eq.~(\ref{3d-7}). 
Figures~\ref{fig12}-\ref{fig14}
display them in zero temperature. 
The FM rung coupling and the $[\pi,0]$ field are 
cooperative each other, while the AF rung coupling competes with the
field. A consequence is, as shown in Fig.~\ref{fig12}, that 
for FM (AF)-rung tubes, $M_s$ are always larger (smaller) than that of
the single AF chain with a staggered field. In addition, 
the growth of $N$ gradually enhances (reduces) the magnetization $M_s$
in FM (AF)-rung tubes. 
The magnetization profile such as the panel ($\alpha$) in
Fig.~\ref{fig12} is also expected 
in nonfrustrated (FM-rung) ladders. Actually, as expected,
Fig.~\ref{fig13} indicates that for the small-field regime $H\alt 0.05$,
$M_l$ tend to increase together the growth of $N$. 
It further explains that the more the $l$th chain approaches the
center of the ladder, the larger its magnetization $M_l$ becomes. 
This is understood from the consideration that 
the chains near the center are more subject to the FM-rung 
correlation effects than ones near the edge. 
On the other hand, one can extract following two unexpected features
in Fig.~\ref{fig13}. (\rnum{1}) The maximum magnetization in 
$M_l$ of the $N$-leg ladder, $M_{\rm max}^N$, is slightly larger 
than that of $(N+1)$-leg ladder for the regime $H\agt 0.05$ and $N\geq 3$. 
(\rnum{2}) The edge magnetization $M_{1,N}$ is smaller than that of 
the single chain in the same regime. 
Since the $[\pi,0]$ field and FM rung coupling, which is absent in the
single chains, must cooperatively enhance the growth of $M_l$, these two
results are expected to be incorrect.  
Because 
in the spin-1 AF chain, the staggered magnetization obtained by the SPA 
is almost consistent with the DMRG data within 
$0\leq \frac{H}{J}\leq 0.5$,~\cite{Erco} 
these unexpected results would
mainly originate from the averaging of constraints, Eq.~(\ref{3a-2}).
The approximation (\ref{3a-2}) would prevent $M_l$ from increasing in
the regime $H\agt 0.05$. 
The true magnetization curves are expected to have a less 
$l$-dependence so that the edge magnetization $M_{1,N}$ is always larger
than that of the single chain, $M_{\rm chain}$ in Fig.~\ref{fig13}. 
Moreover, $M_{\rm max}^{N+1}>M_{\rm max}^{N(\geq 3)}$ 
must hold for all region $0\leq H< \infty$ in the FM-rung ladders.

Figure~\ref{fig14} provides the staggered magnetizations in
the frustrated AF-rung ladders with the $[\pi,0]$ field.
The left panel [a] insists that the staggered magnetizations tend to 
point to the same direction as the $[\pi,0]$ field, 
when the AF rung coupling is small enough; $\frac{J_\perp}{J} \alt 0.03$. 
The edge magnetization $M_{1,N}$ increases most rapidly 
in such a weak rung-coupling region 
since the edge chain receives the competitive AF rung coupling from only
one side, unlike other chains. The rapid growth of $M_{1,N}$ and
the AF rung coupling would make the growths of $M_{l=\text{even}}$ slower. 
While, the panel [b] contains the following interesting phenomena: 
when the $[\pi,0]$ field and the rung coupling are sufficiently small 
and large, respectively ($\frac{H}{J} \alt 0.1$ and 
$\frac{J_\perp}{J} \agt 0.03$) in the odd-leg tubes,
the field induces the staggered magnetization pointing to the opposite
direction to it in the even-$l$ chains. This result is unique for the
ladders, and does not appear in AF-rung tubes (see Fig.~\ref{fig12}).
Such a magnetization configuration staggered along the rung does not also 
occur in even-leg ladders, because the configuration 
cannot be compatible with the edge magnetization 
turning to the $[\pi,0]$ field. One may call the result in
the panel [b] as an even-odd property in the ladders with the
$[\pi,0]$ field. Although the panel [b] further implies the simultaneous
crossings of $(M_1,M_3)$ and $(M_2,M_4,\text{zero-magnetization line})$, 
they might be a coincidence depending on the approximation (\ref{3a-2}).
From the discussion about Fig.~\ref{fig13}, 
Fig.~\ref{fig14} might be also less accurate for the regime $H\agt 0.05$.

\subsubsection{\label{sec3d-2}Magnon Dispersions}
We next study the magnon dispersions of the $[\pi,0]$-field case. 
Following the functional derivative technique such as Eq.~(\ref{2-6}), 
we can represent $G_r^{T(L)}$ as 
\begin{eqnarray}
\label{3d-8}
G_r^{T(L)}({\bf x}-{\bf x'}) = S^2G_r^0({\bf x}-{\bf x'})
+2S^2\int d{\bf y}\bigg[\nonumber\\
G_r^0({\bf x}-{\bf y})G_r^0({\bf y}-{\bf z})
J_r^{x(z)}({\bf z})
\Big(i\frac{\delta\lambda({\bf y})}{\delta J_r^{x(z)}({\bf x'})}\Big)\bigg].
\end{eqnarray}
In the calculation of $\delta\lambda/\delta J_r^\alpha$, we suppose that
each $J_r^\alpha$ is an independent external field.
The relation $J_r^{x,y}=0$ leads to $G_r^T({\bf x})=S^2G_r^0({\bf x})$. 
The Fourier transformation of $G_r^T$ is 
\begin{eqnarray}
\label{3d-9}
\tilde G_r^T({\bf k}) &=& \frac{S^2 gc}{\omega_n^2+\epsilon_r(k)^2}.
\end{eqnarray}
Therefore, we can immediately conclude that the dispersion of
the $r$-th transverse mode $\epsilon_r^T(k)$ is given by $\epsilon_r(k)$. 
Similarly to the single-chain case, this mode has the 
double degeneracy corresponding to the $x$ and $y$ components.
Furthermore, like the no-field case, 
the tubes exhibit the four-fold degeneracy 
$\epsilon_r=\epsilon_{-r}$ except for $r=0$ and $\frac{N}{2}$ modes.
The SPE~(\ref{3d-6}) tells us that 
$\xi^{-2}=-\frac{2ig\lambda_{\rm sp}}{c}$ is enhanced by 
the field $H$ (or $J_r^z$). The transverse bands $\epsilon_r^T(k)$ 
and gaps $\Delta_r^T\equiv\epsilon_r^T(0)$
hence rise monotonically with $H$ increasing.

Although the estimation of the longitudinal mode $\epsilon_r^L(k)$ is rather
complicated due to the presence of $J_r^z$, it is possible through the
application of the method of deriving $\epsilon_L(k)$ in the single chain
with a staggered field. The trivial relation 
$\frac{\delta}{\delta J_r^z}(\frac{\delta
S_{\rm E}^{[\pi,0]}}{\delta\lambda})=0$ is available as the integral equation
determining $\delta \lambda/\delta J_r^z$ in Eq.~(\ref{3d-8}). 
Imitating Eqs.~(\ref{2-6'})-(\ref{2-8}), we can transform it as follows:
\begin{eqnarray}
\label{3d-10}
\int d{\bf y} I({\bf x}-{\bf y})
\Big(i\frac{\delta\lambda({\bf y})}{\delta J_r^z({\bf x'})}\Big)
&=& -2 {\cal M}_r G_r^0({\bf x}-{\bf x'}),\,\,\,\,\,
\end{eqnarray}
where 
\begin{subequations}
\label{3d-11}
\begin{eqnarray}
{\cal M}_r &=& S \langle m_r^z \rangle = S^2 \tilde G_r^0({\bf 0})H/a,
\label{3d-11-1}\\
I({\bf x}-{\bf y}) &=& \sum_r \Big[6\Gamma_r({\bf x}-{\bf y})\nonumber\\
&&+({\cal M}_r/S)^2 G_r^0({\bf x}-{\bf y})\Big],\label{3d-11-2}\\
\Gamma_r({\bf x}-{\bf y})&=& G_r^0({\bf x}-{\bf y})G_r^0({\bf y}-{\bf x}).
\label{3d-11-3}
\end{eqnarray}
\end{subequations}
Equations~(\ref{3d-8}) and (\ref{3d-10}) lead to the following
expression for the Fourier transformation of $G_r^L$:  
\begin{eqnarray}
\label{3d-12}
\tilde G_r^L({\bf k}) &=& \tilde G_r^T({\bf k})
\frac{\sum_p 3\tilde \Gamma_p({\bf k})
+\sum_{p\neq r}2 \tilde {\cal M}_p^2\tilde G_p({\bf k})}
{\sum_p[3\tilde \Gamma_p({\bf k})+2 \tilde {\cal M}_p^2\tilde G_p({\bf k})]},
\,\,\,\,\,\,\,\,\,\,\,\,
\end{eqnarray}
where $\tilde {\cal M}_r={\cal M}_r/S=\langle m_r^z \rangle$. 
The longitudinal mode $\epsilon_r^L(k)$ can be fixed by the pole
structures of the real-time retarded Green's function 
$\tilde G_r^L(\bar{\bf k})\equiv
\tilde G_r^L({\bf k})|_{\omega_n\to -iz}$, 
where $\bar{\bf k}=(z,k)$, $z=\omega+i\eta$, and $\eta\to +0$. 
Here, as in Eq.~(\ref{2-14}), let us introduce new symbols,
\begin{subequations}
\label{3d-13}
\begin{eqnarray}
G_r(z)&=&\tilde G_r^L(\bar{\bf k})/(S^2gc),\label{3d-13-1}\\
\Gamma_r(z)&=& 3 \tilde \Gamma_p(\bar{\bf k})/(2gc),\label{3d-13-2}\\
\Gamma_{\text{tot}}(z)&=&\sum_r\Gamma_r(z),\label{3d-13-3}\\
{\cal F}^{1(2)}(z)&=& \mathfrak{Re}(\mathfrak{Im}){\cal F}(z),\label{3d-13-4}
\end{eqnarray}
\end{subequations}
where we omit the subscript $k$, and ${\cal F}(z)$ is an arbitrary
function of $\bar{\bf k}$.
In terms of these symbols, we obtain the simplified expression of 
the real-time Green's function,
\begin{eqnarray}
\label{3d-14}
G_r(z)&=&\frac{\Gamma_{\text{tot}}(z)
+\sum_{p\neq r}\frac{\tilde{\cal M}_p^2}{\epsilon_p(k)^2-z^2}}
{(\epsilon_r(k)^2-z^2)\big[\Gamma_{\text{tot}}(z)
+\sum_p\frac{\tilde{\cal M}_p^2}{\epsilon_p(k)^2-z^2}\big]}.\,\,\,\,\,\,\,\,\,\,
\end{eqnarray}
Analyzing Eq.~(\ref{3d-14}), one can find the longitudinal magnon
dispersions.

[{\bf Tubes}] 
The calculation of $G_r(z)$ in the tubes is easier than that in the
ladders, because the tubes take $J_r^z\propto \delta_{r,0}$ and 
$\tilde{\cal M}_r\propto \delta_{r,0}$.
These properties bring 
\begin{eqnarray}
\label{3d-15}
G_{r\neq 0}(z)&=&\frac{1}{\epsilon_r(k)^2-z^2}
\propto\tilde G_r^T(\bar{\bf k}).
\end{eqnarray}
The longitudinal dispersion $\epsilon_r^L(k)$, thus, is equivalent to the
transverse one $\epsilon_r^T(k)$ for $r\neq 0$. Namely, in the tubes
with the $[\pi,0]$ field, the $r$th magnon mode
is triply degenerate like the zero-field case (Fig.~\ref{fig4}), 
except for the $0$th mode. Of course, there exists the additional
degeneracy $\epsilon_r=\epsilon_{-r}$. On the other hand, the $0$th mode
Green's function is written as 
\begin{eqnarray}
\label{3d-16}
G_0(z)&=&\frac{\Gamma_{\text{tot}}(z)}
{(\epsilon_0(k)^2-z^2)\Gamma_{\text{tot}}(z)+\tilde {\cal M}_0^2}.
\end{eqnarray}
The form of $G_0(z)$ is quite similar to that of $G(z)$ in Eq.~(\ref{2-16}).
Moreover, at $T=0$, $\Gamma_r^{1,2}(z)$ have the same form as 
$\Gamma^{1,2}$ fixed by Eq.~(\ref{2-15}). 
Therefore, following the calculation from Eq.~(\ref{2-17})
to Eq.~(\ref{2-20}), we can achieve, at $T=0$,
\begin{eqnarray}
\label{3d-17}
\epsilon_0^L(k)^2 &=& \epsilon_0^T(k)^2 + 
\tilde{\cal M}_0^2/\Gamma_{\rm{tot}}^1(\epsilon_0^L(k),k),
\end{eqnarray}
where we restore the subscript $k$ in $\Gamma_{\rm{tot}}^1$.
Note that in the derivation of Eq.~(\ref{3d-17}), we assume 
$\epsilon_0^L(k)<2\epsilon_{\text{min}}(k/2)$, where 
$\epsilon_{\text{min}}(k)$ is defined as the minimum of all the 
transverse dispersions $\epsilon_r(k)$ (see Fig.~\ref{fig4}).
Because the inequality $\Gamma_{\rm{tot}}^1(\epsilon_0^L(k),k)>0$ 
is realized under $\epsilon_0^L(k)<2\epsilon_{\text{min}}(k/2)$, Eq.~(\ref{3d-17})
explains that the $0$th longitudinal band $\epsilon_0^L(k)$ is always
larger than the transverse one $\epsilon_0^T(k)$. 
Employing the explicit form of $\Gamma_r^1(\epsilon_0^L(0),0)$ 
[see Eq.~(\ref{2-19'})], we can calculate the longitudinal gap 
$\Delta_0^L\equiv\epsilon_0^L(0)$ as follows:
\begin{eqnarray}
\label{3d-18}
{\Delta_0^L}^2&=&{\Delta_0^T}^2 + \frac{2\tilde{\cal M}_0^2}{3g}
\Big[\sum_r{\Delta_r^T}^{-2}K(\Delta_0^L/\Delta_r^T)\Big]^{-1},\,\,\,\,\,
\end{eqnarray}
where 
\begin{eqnarray}
\label{3d-19}
K(x)&=&\frac{1}{2\pi x\sqrt{1-x^2/4}}\arctan\left(\frac{x}{2\sqrt{1-x^2/4}}\right),
\,\,\,\,\,\,
\end{eqnarray}
and we, of course, assumed that the minimum point of the band 
$\epsilon_0^L(k)$ is located at $k=0$. The SPE~(\ref{3d-6}) and 
Eqs.~(\ref{3d-15})-(\ref{3d-18}) enable us to know all the magnon band
structures in the tubes ($N\geq 3$) with the $[\pi,0]$ field at $T=0$.
\begin{figure}
\scalebox{0.4}{\includegraphics{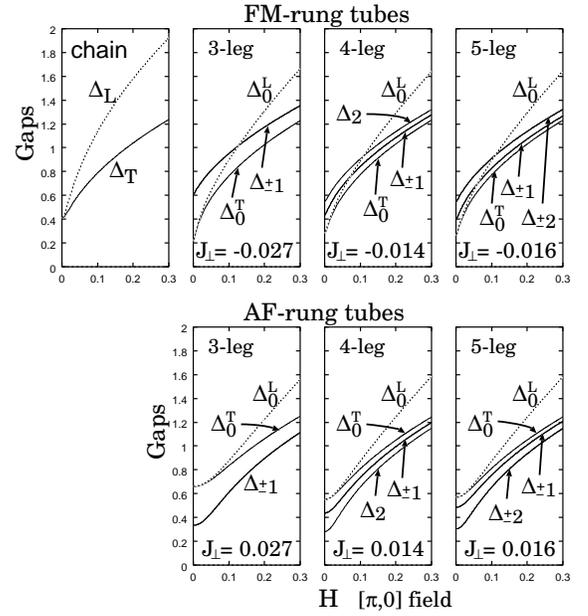}}
\caption{\label{fig15}Transverse and longitudinal gaps of $N$-leg 
spin-$1$ tubes ($N\geq 3$) with the $[\pi,0]$ field and $J=1$. 
As a comparison, we also draw the gaps of the spin-$1$ chain with 
staggered field (see Fig.~\ref{fig2}) in the left upper panel.}
\end{figure}
Figure~\ref{fig15} shows the gaps $\Delta_r^{T,L}$ at $T=0$. 
In this figure, we see that $\Delta_r^T=\Delta_r^L\equiv\Delta_r$ holds for
$r\neq 0$. Since the strong rung coupling destroys the
condition $\epsilon_0^L(k)<2\epsilon_{\text{min}}(k/2)$, our scope in
Fig.~\ref{fig15} is restricted to the extremely weak rung-coupling
regime. We, however, believe that the gap behavior in Fig.~\ref{fig15}
is robust even with a moderately strong rung coupling. 
The lowest (highest) band $\epsilon_0$ is split by the field 
in the FM (AF)-rung tubes. Thus, there are the magnon-band crossings 
only in the FM-rung tubes. 
The manner of the lowest-band splitting goes for 
that of the single chain with the staggered field (see the upper
panels in Fig.~\ref{fig15}, and Fig.~\ref{fig2}). It implies 
that an $N$-leg spin-$S$ FM-rung tube has the same low-energy properties
as the spin-$(N\times S)$ single chain even for the weak rung-coupling
regime. The growths of gaps in the AF-rung case
are slightly slower than those in the FM-rung case. It must reflect 
the frustration between the rung coupling and the field.
As already predicted, one can verify from Fig.~\ref{fig15} that 
all the gaps monotonically grow up with $H$ increasing. 
We, thus, conclude that in tubes
the $[\pi,0]$ field induces no critical phenomena at least in the weak
rung-coupling regime, irrespective of the presence of the frustration.
Although the SPA predicts that only the degeneracy of the bands
$\epsilon_0$ is lifted by the field, 
actually the other bands are also expected to more or less split
(because any mechanisms preserving the triple degeneracy of magnon bands
are not found in the $[\pi,0]$-field case).

[{\bf 2-leg ladder}] Let us next 
investigate the longitudinal magnons for the ladders. 
For the 2-leg case, the estimation of $G_r$ is as simple as
that in the tubes because of $J_2^z={\cal M}_2=0$, which leads to
$G_2(z)\propto G_2^T(\bar{\bf k})$. Therefore, the dispersion of the 
longitudinal mode $\epsilon_2^L(k)$ is identical with $\epsilon_2^T(k)$.
On the other hand, $G_1$ is written as 
\begin{eqnarray}
\label{3d-20}
G_1(z)&=&\frac{\Gamma_{\text{tot}}(z)}
{(\epsilon_1(k)^2-z^2)\Gamma_{\text{tot}}(z)+\tilde {\cal M}_1^2}.
\end{eqnarray}
Because the form of $G_1$ is same as Eq.~(\ref{3d-16}), the process
from Eq.~(\ref{3d-16}) to Eq.~(\ref{3d-19}) can be directly applied 
as a way determining the longitudinal dispersion $\epsilon_1^L(k)$.
As a result, at $T=0$, we obtain two equations,
\begin{subequations}
\label{3d-21}
\begin{eqnarray}
\epsilon_1^L(k)^2 &=& \epsilon_1^T(k)^2 + 
\tilde{\cal M}_1^2/\Gamma_{\rm{tot}}^1(\epsilon_1^L(k),k),\label{3d-21-1}\\
{\Delta_1^L}^2&=&{\Delta_1^T}^2 + \frac{2\tilde{\cal M}_1^2}{3g}
\Big[\sum_{r=1}^2{\Delta_r^T}^{-2}K(\Delta_1^L/\Delta_r^T)\Big]^{-1},
\label{3d-21-2}\,\,\,\,\,\,\,\,\,\,\,\,\,\,
\end{eqnarray}
\end{subequations}
under the condition $\epsilon_1^L(k)<2\epsilon_{\rm{min}}(k/2)$.
Equation~(\ref{3d-21-1}) indicates $[2\epsilon_{\rm{min}}(k/2)>]$ 
$\epsilon_1^L(k)>\epsilon_1^T(k)$.

[{\bf 3-leg ladder}] 
Like $G_2$ in the 2-leg ladder, ${\cal M}_2=0$ leads to 
$\epsilon_2^L(k)=\epsilon_2^T(k)$ in the 3-leg ladder.
On the other hand, $G_{1,3}$ are 
more complicated than the Green's functions in the 2-leg case.
After a simple calculation, they are represented as 
\begin{eqnarray}
\label{3d-22}
G_{1,3}(z)&=&\frac{{\cal C}_{1,3}(\omega)+i{\cal D}_{1,3}(\omega)}
{{\cal A}(\omega)+i{\cal B}(\omega)},
\end{eqnarray}
where 
\begin{subequations}
\label{3d-23}
\begin{eqnarray}
{\cal A}&=&E_1E_3\Gamma_{\rm{tot}}^1(z)
+\tilde{\cal M}_1^2E_3+\tilde{\cal M}_3^2E_1\nonumber\\
&&+2\eta\omega(E_1+E_3)\Gamma_{\rm{tot}}^2(z)
-4\eta^2\omega^2\Gamma_{\rm{tot}}^1(z),\label{3d-23-1}\,\,\,\,\,\,\\
{\cal B}&=&E_1E_3\Gamma_{\rm{tot}}^2(z)-2\eta\omega(E_1+E_3)\Gamma_{\rm{tot}}^1(z)
\nonumber\\
&&-2\eta\omega(\tilde{\cal M}_1^2+\tilde{\cal M}_3^2)
-4\eta^2\omega^2\Gamma_{\rm{tot}}^2(z),\label{3d-23-2}\\
{\cal C}_{1(3)}&=& E_{3(1)}\Gamma_{\rm{tot}}^1(z)+2\eta\omega\Gamma_{\rm{tot}}^2(z)
+\tilde{\cal M}_{3(1)}^2,\label{3d-23-3}\\
{\cal D}_{1(3)}&=& E_{3(1)}\Gamma_{\rm{tot}}^2(z)
-2\eta\omega\Gamma_{\rm{tot}}^1(z),\label{3d-23-4}\\
E_r&=&\epsilon_r(k)^2-\omega^2+\eta^2.\label{3d-23-5}
\end{eqnarray}
\end{subequations}
Under the condition $|\omega|<2\epsilon_{\rm{min}}(k/2)$ and $T=0$, in
which $\Gamma_{\rm{tot}}^1>0$ and $\Gamma_{\rm{tot}}^2=0$,
$G_{1,3}$ are fairly simplified. In that case, the explicit forms of their
imaginary part are
\begin{widetext}
\begin{eqnarray}
\label{3d-23}
G_{1,3}^2(z)=\frac{2\eta\omega \Big[
(E_3\Gamma_{\rm{tot}}^1+\tilde{\cal M}_3)^2
+\tilde{\cal M}_1^2\tilde{\cal M}_3^2  +{\cal E}_{1,3} \Big]}
{\Big[E_1E_3\Gamma_{\rm{tot}}^1+\tilde{\cal M}_1^2E_3+\tilde{\cal M}_3^2E_1
-4\eta^2\omega^2\Gamma_{\rm{tot}}^1\Big]^2
+4\eta^2\omega^2\Big[(E_1+E_3)\Gamma_{\rm{tot}}^1
+\tilde{\cal M}_1^2+\tilde{\cal M}_3^2\Big]^2},
\end{eqnarray}
where ${\cal E}_1=4\eta^2\omega^2(\Gamma_{\rm{tot}}^1)^2$ and 
${\cal E}_3=4\eta^2\omega^2$.
At the limit $\eta\to +0$ [see Eq.~(\ref{2-18})], 
both $G_1^2$ and $G_3^2$ take the same pole structure as follows:
\begin{eqnarray}
\label{3d-24}
\lim_{\eta\to +0}G_{1,3}^2 \propto \delta(g(\omega)),&& 
g(\omega) = \frac{E_1E_3\Gamma_{\rm{tot}}^1(\omega)
+\tilde{\cal M}_1^2E_3+\tilde{\cal M}_3^2E_1}
{(E_1+E_3)\Gamma_{\rm{tot}}^1(\omega)
+\tilde{\cal M}_1^2+\tilde{\cal M}_3^2}.
\end{eqnarray}
From the solution of $g(\omega)=0$, 
we obtain the following two longitudinal bands $\epsilon_\pm^L(k)$:
\begin{eqnarray}
\label{3d-25}
\epsilon_\pm^L(k)^2 = \frac{1}{2}\Big[\epsilon_1(k)^2+\epsilon_3(k)^2\Big]
+\frac{1}{2\Gamma_{\rm{tot}}^1}(\tilde{\cal M}_1^2+\tilde{\cal M}_3^2)
\hspace{8.3cm}\nonumber\\
\pm \frac{1}{2\Gamma_{\rm{tot}}^1}
\sqrt{
\Big[(\epsilon_1(k)^2+\epsilon_3(k)^2)\Gamma_{\rm{tot}}^1
+\tilde{\cal M}_1^2+\tilde{\cal M}_3^2\Big]^2
-4\Gamma_{\rm{tot}}^1
\Big[\epsilon_1(k)^2\epsilon_3(k)^2\Gamma_{\rm{tot}}^1
+\tilde{\cal M}_1^2\epsilon_3(k)^2 
+\tilde{\cal M}_3^2\epsilon_1(k)^2\Big] 
},
\end{eqnarray}
\end{widetext}
where $\Gamma_{\rm{tot}}^1$ means
$\Gamma_{\rm{tot}}^1(\epsilon_\pm^L(k),k)$.
From this result, it is clear that at least within the SPA scheme, the
$[\pi,0]$ field engenders the hybridization between two magnon bands
$\epsilon_1$ and $\epsilon_3$ in the 3-leg ladders.
Provided that the minimum of $\epsilon_\pm^L(k)$ is in $k=0$, 
the gap $\Delta_\pm^L=\epsilon_\pm^L(0)$ can be determined by the
replacement $(\epsilon_\pm^L,\epsilon_1,\epsilon_3)
\to(\Delta_\pm^L,\Delta_1,\Delta_3)$ in Eq.~(\ref{3d-25}). 
In the zero-field limit $H\to 0$, where $\tilde{\cal M}_{1,3}\to 0$, 
Eq.~(\ref{3d-25}) reduces to 
\begin{eqnarray}
\label{3d-26}
\epsilon_\pm^L(k)^2 &\approx&\frac{1}{2}
\Big[\epsilon_1(k)^2+\epsilon_3(k)^2  
\pm \big|\epsilon_1(k)^2-\epsilon_3(k)^2\big|\Big].\,\,\,\, 
\end{eqnarray}
This result and the inequality $\epsilon_1(k)<\epsilon_3(k)$ 
[$\epsilon_1(k)>\epsilon_3(k)$] in the FM [AF]-rung ladder reveal that 
$\epsilon_+^L$ and $\epsilon_-^L$ are, respectively, split from
$\epsilon_3$ and $\epsilon_1$ ($\epsilon_1$ and $\epsilon_3$) in the 
FM (AF)-rung ladder. 
Therefore, $\epsilon_+^L$ and $\epsilon_-^L$ should be rewritten as 
$\epsilon_3^L$ and $\epsilon_1^L$ ($\epsilon_1^L$ and $\epsilon_3^L$)
for the FM (AF)-rung ladder. The gaps $\Delta_\pm^L$ also can be redefined.

[{\bf 4-leg and higher-leg ladders}]
The logic calculating the longitudinal dispersions in the 2- and 3-leg
ladders is successful even for the 4-leg ladders.
We mention only the results of them. 
The identities ${\cal M}_{2,4}=0$ leads to 
$\epsilon_{2,4}^L(k)=\epsilon_{2,4}^T(k)$. 
While, $G_{1,3}(z)$ take the same form as Eq.~(\ref{3d-23}) under the
condition $|\omega|<2\epsilon_{\rm{min}}(k/2)$, except that 
$\Gamma_{\rm{tot}}^1=\sum_{r=1}^3\Gamma_r^1$ and $k_r=\frac{r\pi}{4}$
are replaced with $\sum_{r=1}^4\Gamma_r^1$ and $k_r=\frac{r\pi}{5}$, 
respectively. Therefore, $\epsilon_{1,3}^L(k)$ can be fixed 
like Eq.~(\ref{3d-25}).
As easily expected, the evaluations of the magnon dispersions in the 
higher-leg ladders demand the more complicated analyses. Here we do not
perform them. In principle, one can study all the longitudinal bands 
using Eq.~(\ref{3d-14}).

\begin{figure}
\scalebox{0.4}{\includegraphics{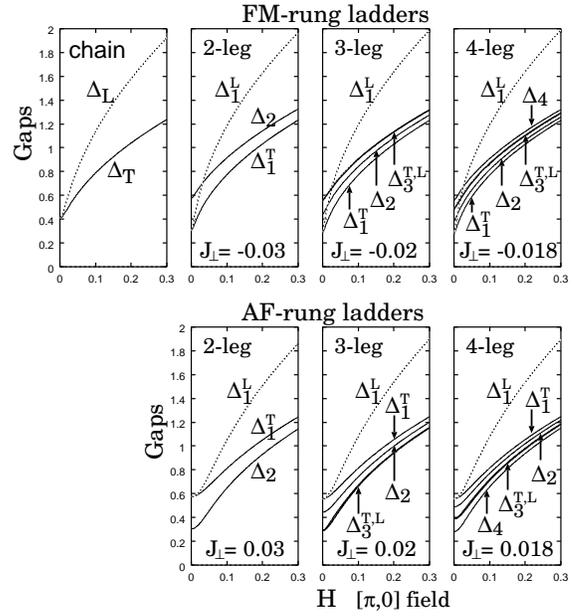}}
\caption{\label{fig16}Transverse and longitudinal gaps of $N$-leg 
spin-$1$ ladders with the $[\pi,0]$ field and $J=1$. The gaps
 $\Delta_3^T$ and $\Delta_3^L$ almost overlap.}
\end{figure}

We show the gaps of 2, 3, and 4-leg ladders in Fig.~\ref{fig16}. 
The gap behavior is much similar to that of the tubes in
Fig.~\ref{fig15}: the highest (lowest) band is largely split into
the doubly degenerate transverse bands and the single longitudinal band
in the FM (AF)-rung ladders.
The splitting of $\epsilon_3(k)$ is considerably small. Namely, 
$\Delta_3^L$ and $\Delta_3^T$ almost overlap. 
Observing carefully the numerical data of 3-leg and 4-leg FM-rung
ladders, we see that $\Delta_3^L$ ($\Delta_3^T$) is a little larger than 
$\Delta_3^T$ ($\Delta_3^L$) for the case $\Delta_1^L<\Delta_3^T$ 
($\Delta_1^L>\Delta_3^T$). Even though the band crossings in FM-rung
tubes (Fig.~\ref{fig15}) are allowed from the translational symmetry
along the rung, the ladders do not possess such a symmetry. 
Therefore, the level crossing in Fig.~\ref{fig16} might, in fact, be an 
avoided crossing. Moreover, more quantitative analyses would lift the
remaining triple degeneracy of the bands $\epsilon_{1,2}$ in 
Fig.~\ref{fig16}. 

Summarizing all the discussions about the magnon dispersions, we can
conclude that the $[\pi,0]$ field engenders the monotonic raise of
all magnon bands, and cannot induce any critical phenomena
at least for the weak rung-coupling regime. While, we have already
predicted that the other staggered field, the $[0,\pi]$ field, induces
the gap reduction (Fig.~\ref{fig11}). Therefore, 
our results in the $[0,\pi]$, $[\pi,0]$, and $[\pi,\pi]$-field cases
suggest that the spatial direction with the staggered component of the fields
essentially affects the low-energy excitations of the spin ladders and
tubes.

Finally, we notice again that through the transformation 
$\vec n_{l=\rm{even}}\to -\vec n_{l=\rm{even}}$, all the results in 
Figs.~\ref{fig12}-\ref{fig16} can be interpreted as those of the
$[\pi,\pi]$-field case. In addition, note again that odd-leg tubes are
absent in the $[\pi,\pi]$-field case.
The $[\pi,\pi]$ field competes with the FM rung coupling. 
This frustration, of course, can induce the even-odd property as in
Fig.~\ref{fig14}. The tubes with a $[\pi,\pi]$ field do not have the
one-site translational symmetry along the rung, and do only the two-site
one. Therefore, the band degeneracy caused from the translational symmetry
should partially vanish in the $[\pi,\pi]$-field case. However, due to the 
symmetry restoration via the mapping 
$\vec n_{l=\rm{even}}\to -\vec n_{l=\rm{even}}$, which would be
valid only in the low-energy limit, 
the tubes with a $[\pi,\pi]$ field take the same bands as those of the 
tubes with a $[0,\pi]$ field within our strategy.


\section{\label{sec5}Summary and Discussions}
We provided a systematic analysis for the low-energy properties
of $N$-leg integer-spin ladders and tubes~(\ref{1-1}) 
with several kinds of external fields (\ref{1-2}) within the NLSM
and SPA framework. Our results would be reliable for the weak
rung-coupling, weak external-field, and small-$N$ cases. Furthermore, 
we expect that several results are robust even for the strong
rung-coupling, strong external-field, and large-$N$ cases. 
Although we concentrated on only the zero-temperature case, the SPA
strategy used here, of course, can be applied to the 
low-temperature case. 

Our results are summarized as follows.
(\rnum{1}) In the no-field case, we derived the magnon band structure
in Fig.~\ref{fig4}, and predicted a new even-odd nature: for AF-rung
tubes, only the odd-leg tubes possess the sixfold degenerate magnon 
band as the lowest one. The sixfold degeneracy is not a merely
approximate result, and is protected by the translational symmetry
along the rung. Several SPA results were compared with the 
QMC data in Figs.~\ref{fig9} and \ref{fig10}. 
(\rnum{2}) In the $[0,0]$-field case, we predicted another even-odd
nature: when the field is sufficiently strong and a finite uniform
magnetization emerges, the GS of odd-leg AF-rung
tubes becomes a $c=1+1$ massless state (two-component TLL), 
while a standard TLL state with $c=1$ appears in other systems. 
Generally, Zamolodchikov $c$ theorem~\cite{Zamol} prefers
the emergence of a $c=1$ state to that of higher-$c$ ones in 1D U(1)-symmetric
systems. However, we predicted, using the GL and bosonization analyses, 
that the translational symmetry along the rung and the reflection one
(Fig.~\ref{fig3}) in the frustrated tubes make the $c=1+1$ state stabilized. 
Inversely, once these symmetries are broken down 
(for example, due to an inhomogeneous rung coupling), the
$c=1+1$ state would disappear and a $c=1$ state emerges instead. 
Regarding the case where the uniform field is further strong, 
we also predicted that the above $c=1+1$ state is taken over by 
a $c=1$ one when the second lowest magnons are condensed. At the
transition from the $c=1+1$ state to the $c=1$ one, one could see a new  
cusp structure, which does not accompany the divergence of the
susceptibilities, in the magnetization curve. 
Furthermore, the validity of our GL theory was briefly discussed.
(\rnum{3}) In the $[0,\pi]$-field case, the
SPA analysis suggested that the lowest doubly degenerate bands go down
with the field increasing, in all systems. From this, one may think that 
a $c=1+1$ state is also possible in the $[0,\pi]$-field case. However, 
it is doubtful since we were not able to find
any symmetries leading to the degeneracy of the lowest two bands.
We thus anticipate that the above double degeneracy is an approximate
result as the $[0,\pi]$ field is small enough.
(\rnum{4}) In the $[\pi,0]$ and $[\pi,\pi]$-field cases, 
we analyzed the magnetizations and the
magnon dispersions (Figs.~\ref{fig12}-\ref{fig16}). The inhomogeneous
magnetization in the ladders were predicted. Moreover, it was shown that 
the $[\pi,0]$ and $[\pi,\pi]$ field do not induce any critical
phenomena at least for the weak rung-coupling regime.    
This is in contrast with the the gap reduction by the $[0,\pi]$
field.

The new even-odd nature and the quantum phase transition between two
critical phases in the $[0,0]$-field case 
are most fascinating among all the results. However, one
has to remember that our
NLSM strategy is originally based on the case without external fields. 
Therefore, within such a strategy, 
one can not essentially provide a quantitative prediction for
the case where the uniform field is so large that magnons are 
condensed. We will revisit the 
magnon-condensed state in the frustrated tubes using other methods
elsewhere. Furthermore, we will discuss half-integer-spin ladders and
tubes in the near future.

Besides our frustrated spin tube, (as we already stated) other
mechanisms generating the magnetization cusp have been 
known.~\cite{Fra-Kre,Kawa1,Kawa2,Oku1,Oku2,Yamamoto} 
However, such mechanisms usually require artificial or fine-tuned
interactions in the models. On the other hand, the structure of spin
tubes is quite simple, and it was shown in Sec.~\ref{sec3b-3} 
that the cusp in the tube is stable against some perturbations. 
Thus, we think that our scenario of the cusp
has a higher possibility of realization compared with other ones.

Our previous work,~\cite{Masa} 
based on the perturbation theory and bosonization techniques, 
shows that the 2-leg
spin-$S$ AF-rung ladder with the $[\pi,0]$ field has $2S$ critical
curves in the sufficiently strong rung-coupling regime, and they vanish
in the weak rung-coupling one. This prediction is consistent with our
analysis for the weak rung-coupling case. Both studies, however, cannot
explain how $2S$ critical curves fade away.

It is worth noticing that all staggered ($[0,\pi]$, $[\pi,0]$, and
$[\pi,\pi]$) fields generally make triply degenerate spin-1 magnon
bands split
into the doubly degenerate transverse modes and single longitudinal
one within our SPA framework. 
It has already known~\cite{O-A,A-O,Ess,E-T} that the same type of the 
band splitting appears in the spin-$\frac{1}{2}$ AF chain 
with the staggered field, when the field is sufficiently small: 
the effective theory of such a spin-$\frac{1}{2}$ 
chain is a sine-Gordon model, which low-energy spectrum consists of 
the massive soliton, the antisoliton (these two are degenerate) 
and the breather (bound state of the soliton and antisoliton). 
Therefore, the band splitting of two and single ones may be a universal 
feature in 1D AF spin systems with an alternating field around the
isotropic [SU(2)] point.

\begin{acknowledgments}
First of all, the author wishes to thank Masaki Oshikawa for critical 
reading this manuscript, 
relevant advice and valuable discussions. He gratefully thanks Ian 
Affleck for the discussion about the magnon condensation in the 
frustrated tubes. Furthermore, he is grateful to Munehisa Matsumoto 
and Kouichi Okunishi for giving the QMC data of the spin-1 ladders 
(Figs.~\ref{fig9} and \ref{fig10}) and useful comments on the 
magnetization cusp, respectively.
This work was partially supported by a 21st Century COE Program at
Tokyo Tech ``Nanometer-Scale Quantum Physics'' by the
Ministry of Education, Culture, Sports, Science and Technology.
\end{acknowledgments}

\appendix
\section{Some results of simple matrices}
\label{app1}
Here, we write down some results of simple eigenvalue
problems, which are used in Sec.~\ref{sec3}.  

Let us define the following two $N\times N$ Hermitian matrices 
appearing in Sec.~\ref{sec3}:
\begin{subequations}
\label{ap1-1}
\begin{eqnarray}
{\cal A}&=& 
\left(
\begin{array}{cccccccc}
A_1        & A_2     &            &         &          \\
A_2        & A_1     &   \ddots   &         &          \\
           & \ddots  &   \ddots   &  \ddots &          \\
           &         &   \ddots   &  \ddots &  A_2     \\
           &         &            & A_2     &  A_1     \\
\end{array}
\right),\label{ap1-1-1}\\
{\cal B}&=& 
\left(
\begin{array}{cccccccc}
B_1        & B_2     &            &         &  B_2     \\
B_2        & B_1     &   \ddots   &         &          \\
           & \ddots  &   \ddots   &  \ddots &          \\
           &         &   \ddots   &  \ddots &  B_2     \\
   B_2     &         &            &   B_2   &  B_1     \\
\end{array}
\right).\label{ap1-1-2}
\end{eqnarray}
\end{subequations}
Eigenvalues ${\cal A}_m$ and corresponding eigenvectors 
$\vec a_m$ of $\cal A$ are given by 
\begin{subequations}
\label{ap1-2}
\begin{eqnarray}
{\cal A}_m &=& A_1+ 2A_2\cos k_m , \,\,\,{\rm where}
\,\,\,k_m=\frac{m\pi}{N+1},\label{ap1-2-1}\\
\vec a_m&=&\sqrt{\frac{2}{N+1}}\,\,{}^T\Big(\sin k_m,\sin 2k_m,\dots,\sin Nk_m\Big),
\,\,\,\,\,\,\,\,\,\,\,\,\,\,\,\,\,\,\label{ap1-2-2}
\end{eqnarray}
\end{subequations}
where $m=1,\dots, N$ and $\vec a_m^2=1$. Similarly, eigenvalues 
${\cal B}_n$ and eigenvectors $\vec b_n$ of ${\cal B}$ ($N\geq 3$) are 
\begin{subequations}
\label{ap1-3}
\begin{eqnarray}
{\cal B}_n &=& B_1+ 2B_2\cos k_n , \,\,\,{\rm where}
\,\,\,k_n=\frac{2n\pi}{N},\label{ap1-3-1}\\
\vec b_n&=&\sqrt{\frac{2}{N}}\,\,{}^T\Big(\sin(k_n+\pi/4),\sin(2k_n+\pi/4),
\nonumber\\
&& \hspace{2cm}\dots,\sin(Nk_n+\pi/4)\Big),\label{ap1-3-2}
\end{eqnarray}
\end{subequations}
where $n=q,\dots, N-1+q$ ($q\in \mathbb{Z}$) and $\vec b_n^2=1$.

We next introduce a matrix 
\begin{eqnarray}
\label{ap1-5}
C &=& 
\left(
\begin{array}{ccc}
C_{11}        & C_{12}        \\
C_{21}        & C_{22}       \\
\end{array}
\right),
\end{eqnarray}
where $C_{11}$ is a normal matrix. The determinant of the matrix~(\ref{ap1-5}) 
satisfies the following well-known formula:
\begin{eqnarray}
\label{ap1-4}
\det|C|&=& \det |C_{11}|\times \det|C_{22}-C_{21}C_{11}^{-1}C_{12}|.
\end{eqnarray}

\section{Single chains with the staggered field}
\label{app2}
We give a short review of the Green's function
method for integer-spin chains with the staggered field, which was
discussed in Ref.~\onlinecite{Erco}.

We start from the NLSM coupling a general external field $\vec J({\bf x})$, 
which Euclidean action is 
\begin{eqnarray}
\label{2-1}
S_{\rm E}[\vec n,\lambda:\vec J]&=&\int d{\bf x}\left[{\cal L}_{\rm E}
-S\vec J \cdot \vec n \right],\,\,\,\,\,\,
\end{eqnarray}
where ${\cal L}_{\rm E}$ is same as Eq.~(\ref{2a-5-3}). 
The effective theory for the
staggered-field case corresponds to $\vec J=\vec H/a=(0,0,H/a)$. 
Here, let us introduce a Green's function 
$G^0({\bf x},{\bf x'})$ as
\begin{eqnarray}
\label{2-2}
-\frac{1}{gc}\Big[\partial_\tau^2+c^2\partial_x^2+2igc \lambda({\bf x})\Big]
G^0({\bf x},{\bf x'})=\delta^2 ({\bf x}-{\bf x'}).\,\,\,\,\,
\end{eqnarray}
After integrating out $\vec n$, the action becomes 
\begin{eqnarray}
\label{2-3}
S_{\rm E}[\lambda: \vec J]=\frac{3}{2}{\rm Tr}\left[\ln G^0({\bf x},{\bf y})\right]
\hspace{3cm}\nonumber\\
-\frac{S^2}{2}\int d{\bf x}d{\bf y}\vec J({\bf x})\cdot G^0({\bf x},{\bf y})
\vec J({\bf y})
+i\int d{\bf x}\lambda({\bf x}).
\end{eqnarray}
The SPE $\delta S_{\rm E}[\lambda:\vec J]/\delta \lambda|_{\vec J:{\rm
fixed}}=0$ is evaluated as
\begin{eqnarray}
\label{2-4}
3G^0({\bf x},{\bf x})\hspace{4.8cm}&&\nonumber\\
+S^2\int d{\bf y}d{\bf z}G^0({\bf y},{\bf x})G^0({\bf x},{\bf z})
\vec J({\bf y})\cdot\vec J({\bf z}) &=& 1,\,\,\,\,\,
\end{eqnarray}
which determines the saddle-point value $\lambda_{\rm sp}({\bf x})$.
One can represent several quantities using $G^0$ within the above SPA
scheme. The staggered magnetization is
\begin{eqnarray}
\label{2-5}
m_{\rm s}^\alpha &\equiv& S\langle n^\alpha({\bf x})\rangle 
=\frac{\delta \ln Z}{\delta J^\alpha({\bf x})}
\approx - \frac{\delta S_{\rm E}[\lambda_{\rm sp}:\vec J]}
{\delta J^\alpha({\bf x})}\nonumber\\
&=& \frac{S^2}{2}\int d{\bf y}[G^0({\bf x},{\bf y})+G^0({\bf y},{\bf x})]
J^\alpha({\bf y}).
\end{eqnarray}
The excitation structures are estimated from the singularities of 
real-time connected Green's functions. 
They are associated with imaginary-time (Matsubara) connected Green's
functions through analytical continuation. The latter is 
\begin{eqnarray}
\label{2-6}
G_c^{\alpha\beta}({\bf x},{\bf x'})&=& 
S^2\langle {\cal T}_\tau n({\bf x})^\alpha n({\bf x'})^\beta\rangle_c
\nonumber\\
&\equiv& S^2\big[
\langle {\cal T}_\tau n({\bf x})^\alpha n({\bf x'})^\beta\rangle
-\langle n({\bf x})^\alpha\rangle \langle n({\bf x'})^\beta\rangle\big]
\nonumber\\
&=&\frac{\delta^2 \ln Z}{\delta J^\alpha({\bf x})\delta J^\beta({\bf x'})}
\approx -\frac{\delta^2 S_{\rm E}[\lambda_{\rm sp}:\vec J]}
{\delta J^\alpha({\bf x})\delta J^\beta({\bf x'})}\nonumber\\
&=& S^2 \Big[G^0({\bf x},{\bf x'})+G^0({\bf x'},{\bf x})\Big]
\delta_{\alpha\beta}/2  \nonumber\\
&&+S^2\int d{\bf y}d{\bf z}\Big[G^0({\bf x},{\bf z})G^0({\bf z},{\bf y})\\
&&+G^0({\bf y},{\bf z})G^0({\bf z},{\bf x})\Big]J^\alpha({\bf y})
\left(i\frac{\delta\lambda_{\rm sp}({\bf z})}{\delta J^\beta({\bf x'})}\right),
\nonumber
\end{eqnarray}
where the functional derivative $\delta^2/\delta A\delta B$ means that
first $\delta/\delta A$ is performed, and then $\delta/\delta B$ is done.
The final term $\delta\lambda_{\rm sp}/\delta J^\beta$
can be determined by the following trivial equation:  
\begin{eqnarray}
\label{2-6'}
0=\frac{\delta}{\delta J^\alpha({\bf x'})}
\Big(\frac{\delta S_{\rm E}}{\delta \lambda({\bf x})}\Big|_{\vec J}\Big)
=\int d{\bf y}\bigg[
\frac{\delta^2 S_{\rm E}}{\delta\lambda({\bf x})\delta\lambda({\bf y})}
\Big |_{\vec J}\nonumber\\
\times\frac{\delta\lambda({\bf y})}{\delta J^\alpha({\bf x'})}
+\frac{\delta}{\delta J^\alpha({\bf x'})}
\Big(\frac{\delta S_{\rm E}}{\delta\lambda({\bf x})}\Big |_{\vec J}\Big)
\Big |_\lambda\bigg], 
\end{eqnarray}
where $\delta/\delta\lambda |_{\vec J}$ is the functional derivative
under the condition that $\vec J$ is fixed, and 
$\delta/\delta J^\alpha |_\lambda$ in the final term means the
derivative with respect to the ``explicit'' $J^\alpha$-dependence 
of $\delta S_{\rm E}/\delta\lambda |_{\vec J}$. 
Through an easy calculation, Eq.~(\ref{2-6'}) becomes
\begin{eqnarray}
\label{2-7}
\int d{\bf y} H({\bf x},{\bf y})
\left(i\frac{\delta\lambda_{\rm sp}({\bf y})}{\delta J^\beta({\bf x'})}\right)
=-S^2\int d{\bf y}J^\beta({\bf y})   \nonumber\\
\times\Big[G^0({\bf y},{\bf x})G^0({\bf x},{\bf x'})
+G^0({\bf x'},{\bf x})G^0({\bf x},{\bf y})\Big],
\end{eqnarray}
where 
\begin{eqnarray}
\label{2-8}
H({\bf x},{\bf y})&\equiv&
\frac{\delta^2 S_{\rm E}}{\delta\lambda({\bf x})\delta\lambda({\bf y})}
\Big |_{\vec J}\nonumber\\
&=& 6\Gamma({\bf x},{\bf y})+2S^2\int d{\bf z}d{\bf w} 
\vec J({\bf z})\cdot \vec J({\bf w}) \nonumber\\
&&\times\Big[G^0({\bf z},{\bf y})G^0({\bf y},{\bf x})G^0({\bf x},{\bf w})
+\nonumber\\
&&G^0({\bf z},{\bf x})G^0({\bf x},{\bf y})G^0({\bf y},{\bf w})\Big],\nonumber\\
\Gamma({\bf x},{\bf y})&=& G^0({\bf x},{\bf y})G^0({\bf y},{\bf x}).
\end{eqnarray}

Let us apply the above results to our staggered-field case, in which
$\vec J=\vec H/a$. We assume that 
$\lambda_{\rm sp}$ is independent of $\bf x$.
It leads to the relation $G^0({\bf x}, {\bf x'})=G^0({\bf x}-{\bf x'})$. 
The Fourier transformation of $G_0({\bf x})$, therefore, can be defined as
\begin{eqnarray}
\label{2-9}
\tilde{G}^0({\bf k})=\int d{\bf x}e^{-i{\bf k}{\bf x}}G^0({\bf x})
=\frac{gc}{\omega_n^2+c^2k^2+c^2\xi^{-2}},
\end{eqnarray}
where ${\bf k}{\bf x}=kx-\omega_n\tau$.
From Eqs.~(\ref{2-4}) and (\ref{2-9}), the SPE is calculated as 
\begin{eqnarray}
\label{2-10}
\frac{3gc}{2\pi}\int_0^\Lambda \frac{dk}{\epsilon(k)}
\coth\Big(\frac{\beta}{2}\epsilon(k)\Big)
&=& 1-\left(\frac{SgH}{ca}\right)^2\xi^4.\,\,\,\,\,\,\,\,\,\,\,\,\,\,\,\,
\end{eqnarray}
The final term denotes the deviation from the SPE~(\ref{2a-9}) of the 
no-field case. From Eq.~(\ref{2-5}), the staggered
magnetization is 
\begin{eqnarray}
\label{2-11}
m_{\rm s}^\alpha   &=&   S^2\tilde{G}^0({\bf 0})\frac{H^\alpha}{a}
=\left(\frac{\xi}{a}\right)^2\frac{H^\alpha}{J}.
\end{eqnarray}
This result indicates that the staggered
magnetization is parallel to $\vec H$, namely $\vec m_s=(0,0,m_s)$. 
From Eqs.~(\ref{2-6})-(\ref{2-9}), 
the Fourier components of the connected Green's functions are estimated 
as follows:
\begin{subequations}
\label{2-12}
\begin{eqnarray}
\tilde{G}_c^{xx,yy}({\bf k})&=&S^2\tilde{G}^0({\bf k})
=\frac{S^2gc}{\omega_n^2+\epsilon(k)^2},
\label{2-12-1}\\
\tilde{G}_c^{zz}({\bf k})&=&\tilde{G}_c^{xx}({\bf k})
\frac{3\tilde{\Gamma}({\bf k})}
{3\tilde{\Gamma}({\bf k})+2(m_{\rm s}/S)^2\tilde{G}^0({\bf k})},
\,\,\,\,\,\,\,\,\,\,\,\,\,\,\,\,\,\,\,\,\label{2-12-2}
\end{eqnarray}
\end{subequations}
where $\tilde{\Gamma}({\bf k})=\frac{1}{L\beta}\sum_{\bf p}
\tilde{G}^0(\frac{1}{2}[{\bf k}+{\bf p}])\tilde{G}^0
(\frac{1}{2}[{\bf k}-{\bf p}])$ is the Fourier transformation of
$\Gamma({\bf x})$. Other Green's functions all vanish. 
In order to know the excitation spectrums, let us investigate 
the Fourier components of real-time Green's functions, 
$\tilde{G}_c^{\alpha\alpha}(\bar{\bf k})\equiv 
\tilde{G}_c^{\alpha\alpha}({\bf k})|_{\omega_n\rightarrow
-iz}$, where $\bar{\bf k}=(z,k)$ and $z=\omega+i\eta$. 
We refer the excitation modes
determined from the poles of $\tilde{G}_c^{xx,yy}(\bar{\bf k})
\equiv\tilde{G}_c^T(\bar{\bf k})$ 
[$\tilde{G}_c^{zz}(\bar{\bf k})\equiv\tilde{G}_c^L(\bar{\bf k})$] 
to transverse [longitudinal] magnon modes.
From Eq.~(\ref{2-12-1}), the imaginary part of 
$\tilde{G}_c^T(\bar{\bf k})$ is 
\begin{eqnarray}
\label{2-13}
\mathfrak{Im}\tilde{G}_c^T(\bar{\bf k})&=&\frac{\pi S^2 gc}{2\epsilon(k)}
\Big[\delta(\omega-\epsilon(k))-\delta(\omega+\epsilon(k))\Big].
\,\,\,\,\,\,\,\,\,\,\,\,\,\,\,\,
\end{eqnarray}
The delta-function singularity means that the transverse modes are 
exhausted by the single-magnon excitations with the dispersion 
$\epsilon_T(k)=\epsilon(k)$. The transverse
gap is defined as $\Delta_T=\epsilon_T(0)$. 
The mode $\epsilon_T(k)$ is doubly
degenerate correspondingly to $x$ and $y$ directions, 
in the present SPA scheme.

The singularity structure of $\tilde{G}_c^L$ is much more
involved than that of $\tilde{G}_c^T$. Here, we show only its results. 
For convenience, we introduce several new quantities,
\begin{subequations}
\label{2-14}
\begin{eqnarray}
G(\omega)=\tilde{G}_c^L(\bar{\bf k})/(S^2gc),\label{2-14-1}\\  
G^{1(2)}(\omega)=\mathfrak{Re}(\mathfrak{Im})G(\omega),
\label{2-14-2}\\  
\Gamma^{1(2)}(\omega)=3[\mathfrak{Re}(\mathfrak{Im})
\tilde{\Gamma}(\bar{\bf k})]/(2gc),
\label{2-14-3}\\
M=m_{\rm s}/S,\,\,\,\,\epsilon=\epsilon(k),\label{2-14-4}
\end{eqnarray}
\end{subequations}
where we omit the indication of the wave number $k$, and 
$\tilde{\Gamma}(\bar{\bf k})\equiv 
\tilde{\Gamma}({\bf k})|_{\omega_n\rightarrow -iz}$.
At $T=0$, using $\tilde{G}^0$, we calculate $\Gamma^{1,2}$ as
\begin{subequations}
\label{2-15}
\begin{eqnarray}
\mathfrak{Im}\tilde{\Gamma}(\bar{\bf k})&=&\frac{g^2}{4A(k,p')}
\Theta_{\rm s}\left(\omega^2-4\epsilon(k/2)^2\right){\rm sgn}(\omega),\label{2-15-1}\\
\mathfrak{Re}\tilde{\Gamma}(\bar{\bf k})&=&
\int_{2\epsilon(\frac{k}{2})}^\infty 
\frac{dz}{\pi}\, 2z\,\mathfrak{Im}\tilde{\Gamma}(k,z){\cal P}\left(\frac{1}{z^2-\omega^2}\right),
\,\,\,\,\,\,\,\,\,\,\,\,\,\,\,\,\,\,\,\,\label{2-15-2}
\end{eqnarray}
\end{subequations}
where $A(k,p')=|p'\epsilon(p'+k)+(p'+k)\epsilon(p')|$ 
[$p'=\frac{\omega}{2c}\sqrt{(\omega^2-4\epsilon(k/2)^2)/(\omega^2-c^2k^2)}-k/2$],
$\Theta_{\rm s}$ is the Heaviside's step function, 
${\rm sgn}(\omega)$ is the sign function,
and ${\cal P}$ means the Cauchy principal part. To derive Eq.~(\ref{2-15-2}), 
we used the Kramers-Kronig relation.
Using the new symbols~(\ref{2-14}), Eq.~(\ref{2-12-2}) is simplified as 
\begin{eqnarray}
\label{2-16}
G(\omega)=\frac{\Gamma(\omega)}{\Gamma(\omega)(\epsilon^2-(\omega+i\eta)^2)+M^2}.
\end{eqnarray}
Thus, $G^{1,2}(\omega)$ are written as   
\begin{subequations}
\label{2-17}
\begin{eqnarray}
G^1(\omega)&=&\frac{\alpha(\omega)\Gamma^1(\omega)-\beta(\omega)\Gamma^2(\omega)}
{\alpha(\omega)^2+\beta(\omega)^2},\label{2-17-1}\\
G^2(\omega)&=&\frac{\beta(\omega)\Gamma^1(\omega)+\alpha(\omega)\Gamma^2(\omega)}
{\alpha(\omega)^2+\beta(\omega)^2},\label{2-17-2}
\end{eqnarray}
\end{subequations}
where $\alpha(\omega)=(\epsilon^2-\omega^2+\eta^2)\Gamma^1(\omega)
+2\eta\omega\Gamma^2(\omega)+M^2$ and 
$\beta(\omega)=2\eta\omega\Gamma^1(\omega)
-(\epsilon^2-\omega^2+\eta^2)\Gamma^2(\omega)$.
The pole structure of $G^2$ gives the longitudinal mode $\epsilon_L(k)$.
For $|\omega|<2\epsilon(k/2)$, in which
$\Gamma^2=0$, we have 
\begin{eqnarray}
\label{2-18}
G^2(\omega)&=&\frac{2\eta\omega}
{[\epsilon^2-\omega^2+\eta^2-M^2/\Gamma^1(\omega)]^2
+4\eta^2\omega^2}\nonumber\\
&\overset{\eta\to +0}{\rightarrow}&\pi\,{\rm sgn}(\omega)\,\delta[f(\omega)], 
\end{eqnarray}
where $f(\omega)=\omega^2-\epsilon^2-M^2/\Gamma^1(\omega)$.
From this, one sees that under the condition $\epsilon_L(k)<2\epsilon(k/2)$,
the longitudinal mode satisfies $f(\epsilon_L(k))=0$, i.e.,
\begin{eqnarray}
\label{2-19}
\epsilon_L(k)^2 &=& \epsilon_T(k)^2+M^2/\Gamma^1(\epsilon_L(k)). 
\end{eqnarray}
If the lowest excitation of the longitudinal mode is located in $k=0$, 
the longitudinal gap is defined by $\Delta_L=\epsilon_L(0)$.
Under the condition $\Delta_L<2 \Delta_T$, one can
easily perform the integral in $\Gamma^1(\Delta_L)|_{k=0}$. 
As a result, its explicit form becomes
\begin{eqnarray}
\label{2-19'}
\Gamma^1(\Delta_L)|_{k=0}&=& 
\frac{3g\,\,\arctan\left(\frac{y}{\sqrt{4-y^2}}\right)}
{2\pi\Delta_T^2 y\sqrt{4-y^2}},
\end{eqnarray}
where $y=\Delta_L/\Delta_T$. From Eqs.~(\ref{2-19}) and (\ref{2-19'}), 
we can arrive in the equation fixing $\Delta_L$,~\cite{note1}
\begin{eqnarray}
\label{2-20}
y^2 &=& 1+\frac{4m_{\rm s}^2 y\sqrt{1-y^2/4}}{3S\left[1-\frac{2}{\pi}
\arctan\Big(\frac{2}{y}\sqrt{1-y^2/4}\Big)\right]},\,\,\,\,\,\, 
\end{eqnarray}
at $T=0$.
On the other hand, $G^2$ does not have any
singularities for $|\omega|>2\epsilon(k/2)$, in the SPA scheme.

\bibliography{apssamp}

\begin{thebibliography}{l}
\bibitem{NLS-Hal}F. D. M. Haldane, Phys. Rev. Lett. {\bf 50}, 1153
	(1983); Phys. Lett. {\bf 93A}, 464 (1983). 
\bibitem{LSM}E. Lieb, T. D. Schulz and D. C. Mattis, Ann. Phys (N.Y)
	{\bf 16}, 407 (1961).
\bibitem{Aff-Lieb}I. Affleck and E. H. Lieb, Lett. Math. Phys. {\bf 12},
	57 (1986).
\bibitem{Gogo}A. O. Gogolin, A. A. Nersesyan and A. M. Tsvelik, {\em
	Bosonization and Strongly Correlated Systems} (Cambridge Univ.
	Press, Cambridge, England, 1998).
\bibitem{CFT}P. D. Francesco, P. Mathieu and D. S\'en\'echal, {\em
	Conformal\ Field Theory} (Springer-Verlag, New York, 1997).
\bibitem{Dag}E. Dagotto, J. Riera and D. Scalapino, Phys. Rev. B {\bf
	45}, R5744 (1992).
\bibitem{White}S. R. White, R. M. Noack and D. J. Scalapino,
	Phys. Rev. Lett. {\bf 73}, 886 (1994).
\bibitem{Dag2}E. Dagotto and T. M. Rice, Science {\bf 271}, 618 (1996).
\bibitem{Fri}B. Frischmuth, S. Haas, G. Sierra and T. M. Rice,
	Phys. Rev. B {\bf 55}, R3340 (1997).
\bibitem{Rojo}A. G. Rojo, Phys. Rev. B {\bf 53}, 9172 (1996).
\bibitem{Sier}G. Sierra, J. Phys. A {\bf 29}, 3299 (1996);
	also see cond-mat/9610057.
\bibitem{Dell}S. Dell'Aringa, E. Ercolessi, G. Morandi, P. Pieri and
	M. Roncaglia, Phys. Rev. Lett. {\bf 78}, 2457 (1997).
\bibitem{Hiroi}Z. Hiroi, M. Azuma, M. Takano and Y. Bando, J. Solid
	State Chem. {\bf 95} 230 (1991).
\bibitem{Azuma}M. Azuma, Z. Hiroi, M. Takano, K. Ishida and Y. Kitaoka,
	Phys. Rev. Lett. {\bf 73}, 3463 (1994).
\bibitem{Chab}G. Chabuoussant, M. -H. Julien, Y. Fagot-Revurat, M. Hanson, 
	L. P. L\'evy, C. Berthier, M. Hovati\'c and O. Piovesana,
	Eur. Phys. J. B {\bf 6}, 167 (1998).
\bibitem{DT-TTF}D. Ar\v con, A. Lappas, S. Margadonna, K. Prassides,
	E. Ribera, J. Veciana, C. Rovira, R. T. Henriques and
	M. Almeida, Phys. Rev. B {\bf 60}, 4191 (1999).
\bibitem{Nojiri}J. Schnack, H. Nojiri, P. K\"ogerler, G. J. T. Cooper
	and L. Cronin, Phys. Rev. B {\bf 70}, 174420 (2004). 
\bibitem{Mill}P. Millet, J. Y. Henry, F. Mila and J. Galy, J. Solid
	State Chem. {\bf 147}, 676 (1999).
\bibitem{Shul}H. J. Schulz, cond-mat/9605075.
\bibitem{Taka}K. Kawano and M. Takahashi, J. Phys. Soc. Jpn. {\bf 66},
	4001 (1997).
\bibitem{Cab}D. C. Cabra, A. Honecker and P. Pujol, Phys. Rev. B {\bf
	58}, 6241 (1998).
\bibitem{Ichi}I. Ichinose and Y. Kayama, Nucl. Phys. B {\bf 522}, 569
	(1998).
\bibitem{Cit}R. Citro, E. Orignac, N. Andrei, C. Itoi and S. Qin,
	J. Phys.: Condens. Matter {\bf 12}, 3041 (2000).
\bibitem{Cit2}E. Orignac, R. Citro and N. Andrei, Phys. Rev. B {\bf 61},
	11533 (2000). 
\bibitem{oddlegs1/2}A. L\"uscher, R. M. Noack, G. Misguich, V. N. Kotov and
	F. Mila,   Phys. Rev. B {\bf 70}, 060405(R) (2004).
\bibitem{Aff-Lec}I. Affleck, in {\em Fields, Strings and Critical
	Phenomena}, edited by E. Br\'ezin and J. Zinn-Justin, p. 564 
	(North-Holland, Amsterdam, 1990).
\bibitem{Man}E. Manousakis, Rev. Mod. Phys. {\bf 63}, 1 (1991).
\bibitem{Frad}E. Fradkin, {\em Field Theories of Condensed Matter Systems}
	(Addison-wesley, 1991).
\bibitem{Auer}A. Auerbach, {\em Interacting Electrons and Quantum
	Magnetism} (Springer-Verlag, New York, 1994).


\bibitem{Sene}D. S\'en\'echal, Phys. Rev. B {\bf 52}, 15319 (1995).

\bibitem{O-A}M. Oshikawa and I. Affleck, Phys. Rev. Lett. {\bf 79}, 2883 (1997).
\bibitem{A-O}I. Affleck and M. Oshikawa, Phys. Rev. B {\bf 60}, 1038
	(1999).
\bibitem{E-T}F. H. L. Essler and A. M. Tsvelik, Phys. Rev. B {\bf 57},
	10592 (1998).
\bibitem{Ess}F. H. L. Essler, Phys. Rev. B {\bf 59}, 14376 (1999). 
\bibitem{Nomu}M. Tsukano and K. Nomura, J. Phys. Soc. Jpn. {\bf 67},
	302 (1998). 
\bibitem{Z-R-M}A. Zheludev, E. Ressouche, S. Maslov, T. Yokoo,
	S. Raymond and J. Akimitsu, Phys. Rev. Lett. {\bf 80}, 3630
	(1998).
\bibitem{Mas-Zhe}S. Maslov and A. Zheludev, Phys. Rev. B {\bf 57}, 68
	(1998); Phys. Rev. Lett. {\bf 80}, 5786 (1998).
\bibitem{Lou}J. Lou, X. Dai, S. Qin, Z. Su and L. Yu, Phys. Rev. B {\bf
	60}, 52 (1999).
\bibitem{Erco}E. Ercolessi, G. Morandi, P. Pieri and M. Roncaglia,
	Phys. Rev. B {\bf 62}, 14860 (2000). 
\bibitem{Erco2}E. Ercolessi, G. Morandi, P. Pieri and M. Roncaglia,
	Europhys. Lett. {\bf 52}, 434 (2000).
\bibitem{Cap}M. Capone and S. Caprara, Phys. Rev. B {\bf 64}, 184418
	(2001).
\bibitem{Wang}Y. -J. Wang, F. H. L. Essler, M. Fabrizio and
	A. A. Nersesyan, Phys. Rev. B {\bf 66}, 024412 (2002).
\bibitem{M-O}M. Sato and M. Oshikawa, Phys. Rev. B {\bf 69}, 054406
	(2004).
\bibitem{Masa}M. Sato, Phys. Rev. B {\bf 71}, 024402 (2005).
\bibitem{Cu-Den}D. C. Dender, P. R. Hammar, D. H. Reich, C. Broholm and
	G. Aeppli, Phys. Rev. Lett. {\bf 79}, 1750 (1997).
\bibitem{Cu-A1}T. Asano, H. Nojiri, Y. Inagaki, J. P. Boucher,
	T. Sakon, Y. Ajiro and M. Motokawa, Phys. Rev. Lett. {\bf 84},
	5880 (2000).
\bibitem{Cu-A2}T. Asano, H. Nojiri, W. Higemoto, A. Koda, R. Kadono
	and Y. Ajiro, J. Phys. Soc. Jpn. {\bf 72}, 594 (2002).
\bibitem{PMCu-ex}R. Feyerherm, S. Abens, D. G\"unther, T. Ishida,
	M. Mei{\ss}ner, M. Meschke, T. Nogami and M. Steiner, J. Phys.:
	Condens. Matter {\bf 12}, 8495 (2000).
\bibitem{YbAs-O}M. Oshikawa, K. Ueda, H. Aoki, A. Ochiai and M. Kohgi,
	J. Phys. Soc. Jpn. {\bf 68}, 3181 (1999).
\bibitem{YbAs-S}H. Shiba, K. Ueda and O. Sakai, J. Phys. Soc. Jpn. {\bf
	69}, 1493 (2000).
\bibitem{Yb4As3}M. Kohgi, K. Iwasa, J. M. Mignot, B. Fak, P. Gegenwart,
	M. Lang, A. Ochiai, H. Aoki and T. Suzuki, Phys. Rev. Lett. {\bf
	86}, 2439 (2001).




\bibitem{Ven}L. C. Venuti, E. Ercolessi, G. Morandi, P. Pieri and
	M. Roncaglia, Int. J. Mod. Phys. B {\bf 16}, 1363 (2002).


\bibitem{Zam}A. Zamolodchikov and A. B. Zamolodchikov, Ann. Phys. {\bf
	120}, 253 (1979).

\bibitem{Mahan}For example see section 3.5 in G. D. Mahan, 
	{\em Many-Particle Physics 3rd ed} (Plenum Press, New York and
	London, 2000).

\bibitem{Todo-Kato}S. Todo and K. Kato, Phys. Rev. Lett. {\bf 87},
	047203 (2001).
\bibitem{Qin}S. Qin, X. Wang, and L. Yu, Phys. Rev. {\bf B} 56, R14251
	(1997).
\bibitem{S-A}E. S. S\o rensen and I. Affleck, Phys. Rev. Lett. {\bf 71},
	1633 (1993).
\bibitem{Yama}S. Yamamoto, Phys. Rev. B {\bf 51}, 16128 (1995).

\bibitem{H-A}M. D. P. Horton and I. Affleck, Phys. Rev. B {\bf 60},
	11891 (1999).
\bibitem{Ess2}F. H. L. Essler, Phys. Rev. B {\bf 62}, 3264 (2000).







\bibitem{note0}Authors in Ref.~\onlinecite{Erco} insist that for the
	case $J\gg H$, the gap $\Delta_L$ derived from a single-mode
	approximation is more accurate rather than that of the Green's
	function method. However, we guess that they take a
	careless mistake in the calculation of $\Delta_L$, and their
	statement is incorrect. See Eq.~(\ref{2-20}) and our
	comment~\onlinecite{note1}. 





\bibitem{Allen}D. Allen and D. S\'en\'echal, Phys. Rev. B {\bf 61},
	12134 (2000).
\bibitem{Todo}S. Todo, M. Matsumoto, C. Yasuda and H. Takayama,
	Phys. Rev. B {\bf 64}, 224412 (2001).


\bibitem{Tsv}A. M. Tsvelik, Phys. Rev. B {\bf 42}, 10499 (1990).
\bibitem{Kita-Nomu}A. Kitazawa and K. Nomura, Phys. Rev. B {\bf 59},
	11358 (1999).

\bibitem{MMcom}M. Matsumoto (private communication, 2004).

\bibitem{S-T}T. Sakai and M. Takahashi, Phys. Rev. B {\bf 42}, R1090
	(1990); J. Phys. Soc. Jpn. {\bf 60}, 760 (1991); {\em ibid},
	3615 (1991); Phys. Rev. B {\bf 43}, 13383 (1991).
 
\bibitem{Aff-magcond}I. Affleck, Phys. Rev. B {\bf 43}, 3215 (1991).
\bibitem{Aff-bose}I. Affleck, Phys. Rev. B {\bf 41}, 6697 (1990). 
\bibitem{Kon}R. M. Konik and P. Fendley, Phys. Rev. B {\bf 66}, 144416
	(2002).
\bibitem{Fath}G. F\'ath, Phys. Rev. B {\bf 68}, 134445 (2003).



\bibitem{Hal2}F. D. M. Haldane, Phys. Rev. Lett. 47, 1840 (1981).
\bibitem{Cabra}D. C. Cabra and A. Honecker and P. Pujol, Phys. Rev. B
	{\bf 58}, 6241 (1998). 
\bibitem{Shank}R. Shankar, Acta Phys. Polo. {\bf 26}, 1835 (1995).
\bibitem{Delft}J. v. Delft and H. Schoeller, Ann. Phys. {\bf 4}, 225
	(1998).
\bibitem{Selton}D. G. Shelton and A. A. Nersesyan and A. M. Tsvelik,
	Phys. Rev. B {\bf 53}, 8521 (1996).
\bibitem{Fu-Zh}A. Furusaki and S. C. Zhang, Phys. Rev. B {\bf 60}, 1175
	(1999). 
\bibitem{relevant}In addition to the vertex operators, derivative type
	operators such as $\partial_x\phi_q$ can exist as the relevant
	or marginal terms in the effective theory. However, the relevant
	first-order derivative terms can be absorbed into the Gaussian
	free-boson part ${\cal L}_{\rm E}$ in Eq.~(\ref{3b-2-6}), and the
	marginal second-order derivative ones just change the parameters
	$K$ and $v$.    
\bibitem{OYA}M. Oshikawa, M. Yamanaka and I. Affleck, 
	Phys. Rev. Lett. {\bf 78}, 1984 (1997).
\bibitem{Cardy}J. L. Cardy, {\em Scaling and Renormalization in
	Statistical Physics} 
	(Cambridge Univ. Press, Cambridge, England, 1996).


\bibitem{Fra-Kre}H. Frahm and V. E Korepin, Phys. Rev. B {\bf 42}, 10553
	(1990); {\em ibid}. {\bf 43}, 5653 (1991). 

\bibitem{Kawa1}M. Fujii, S. Fujimoto and N. Kawakami,
	J. Phys. Soc. Jpn. {\bf 65}, 2381 (1996).
\bibitem{Kawa2}A. Kawaguchi, T. Fujii and N. Kawakami, Phys. Rev. B {\bf
	63}, 144413 (2001).

\bibitem{Oku1}K. Okunishi, Y. Hieida and Y. Akutsu, Phys. Rev. B {\bf
	60}, R6953 (1999).
\bibitem{Oku2}K. Okunishi and T. Tonegawa, J. Phys. Soc. Jpn. {\bf 72},
	479 (2003).

\bibitem{Yamamoto}T. Yamamoto, R. Manago, Y. Mori and C. Ishii,
	J. Phys. Soc. Jpn. {\bf 72}, 3204 (2003). 










\bibitem{Wein}For example, see S. Weinberg, {\em Quantum Theory of
	Fields Vol.1} (Cambridge Univ. Press. 1995).







\bibitem{Zamol}A. B. Zamolodchikov, JETP Lett. {\bf 43}, 730 (1986).





\bibitem{note1}Equation (\ref{2-20}) was first derived in
	Ref.~\onlinecite{Erco}. However, we think that Eq.~(88) in
	Ref.~\onlinecite{Erco}, which corresponds to Eq.~(\ref{2-20}),
	contains a careless mistake: the factor $\frac{g^2c}{2}$ in
	Eq.~(83) and the numerator of the second term of
	the right-hand side in Eq.~(88) should be replaced with $g^2c$ and  
	$4m_s^2 S^{-1}\sqrt{1-y^2/4}$, respectively.


\end{thebibliography}

\end{document}